\documentclass[1p]{elsarticle}
\pdfoutput=1
\usepackage{epsfig}
\usepackage{hyperref}
\usepackage{verbatim}
\usepackage{bm}
\usepackage[titletoc]{appendix}
\usepackage{multirow}

\def\Journal#1#2#3#4{{#1} {#2} (#4) #3 }

\def\NPA{{\em Nucl.\ Phys.} A}

\def\PLB{{\em Phys.\ Lett.} B}

\def\PRL{\em Phys.\ Rev.\ Lett.}

\def\PREP{\em Phys.\ Rep.}

\def\PRD{{\em Phys.\ Rev.} D}
\def\PRC{{\em Phys.\ Rev.} C}

\newcommand{\be}{\begin{equation}}
\newcommand{\ee}{\end{equation}}
\newcommand{\bea}{\begin{eqnarray}}
\newcommand{\eea}{\end{eqnarray}}

\begin{document}

\title{Heavy-Quark QCD Exotica}

\author[asu]{Richard F. Lebed\corref{cor}}
\ead{richard.lebed@asu.edu}

\author[iu]{Ryan E. Mitchell}
\ead{remitche@indiana.edu}

\author[pitt]{Eric S. Swanson}
\ead{swansone@pitt.edu}

\cortext[cor]{Corresponding author}

\address[asu]{Department of Physics, Arizona State University, Tempe,
Arizona 85287-1504, USA}

\address[iu]{Department of Physics, Indiana University, Bloomington,
Indiana 47405, USA}

\address[pitt]{Department of Physics and Astronomy, University of
Pittsburgh, Pennsylvania 15260, USA}

\begin{abstract}
  This review presents an overview of the remarkable progress in the
  field of heavy-quark exotic hadrons over the past 15 years.  It
  seeks to be pedagogical rather than exhaustive, summarizing both the
  progress and specific results of experimental discoveries, and the
  variety of theoretical approaches designed to explain these new
  states.
\end{abstract}
\maketitle
\tableofcontents

\section{Introduction} \label{sec:Intro}

\subsection{Overview}

When one considers all the elementary particles discovered in the past
two decades, the Higgs boson rightly takes center stage as the most
significant example, not only by virtue of its eminent role of
completing the Standard Model, but also in capturing the attention of
both the scientific community and the public.  However, most of the
particles discovered in the current era were almost completely
unexpected, and they seem to be broadly interrelated.  Quite
remarkably, though, no scientific consensus has yet emerged to explain
all of them by means of a single, universal theoretical principle.
These particles are, of course, the 30 or so observed candidate {\it
exotic hadrons\/} ({\it i.e.}, ones that do not fit into the paradigms
of either $q\bar q$ bosonic {\it mesons\/} or $qqq$ fermionic {\it
baryons} [$q$ being a generic quark]).  Most of these states have
masses in the same region as conventional charmonium states (hadrons
consistent with a $c\bar c$ bound-state structure), and indeed have so
far never been observed to decay into hadrons not containing
charm---and hence are called {\it charmoniumlike}).  Almost all of the
other candidate exotic hadrons appear in the $b\bar b$ bound-state
sector and are {\it bottomoniumlike}.  In addition, a few states have
recently been observed in the lighter-quark sectors that are good
candidate exotics; however, the most unambiguous candidates for exotic
hadrons observed to date appear in the $c\bar c$ and $b\bar b$
sectors, and their formation, properties, and structure comprise the
subject of this review.

Quantum Chromodynamics (QCD) has long been the accepted quantum field
theory of the strong nuclear force responsible for holding atomic
nuclei together.  The three interaction charges, called {\it colors},
are carried by elementary strongly interacting spin-$\frac 1 2$
particles called {\it quarks}, which interact through the exchange of
massless force-carrying {\it gauge bosons\/} called {\it gluons}.  In
these respects, QCD is very similar to Quantum Electrodynamics (QED),
the quantum field theory of electricity and magnetism, for which the
interaction charge is simply electric charge, and the gauge bosons are
photons.  However, QCD is much more complicated in several ways: The
presence of three distinct color charges and the non-Abelian nature of
the gauge group means that the gluons themselves carry color and
therefore can interact with other gluons (in contrast to
charge-neutral photons).  More significantly, no particle---quark or
gluon---carrying a net color charge has ever been experimentally
isolated, a phenomenon called {\it color confinement}.  The contrast
with QED could not be more pronounced, in which it is effortless to
produce free charged particles, such as electron beams.  Instead,
colored particles are only found in color-neutral compounds called
{\it hadrons}, and until very recently, it was possible to classify
every known hadron as a conventional meson or baryon.

The mathematical structure of the {\it gauge symmetry\/} describing
the 3 color charges is that of the group SU(3), which produces an
easily enumerated list of possible color-singlet combinations, and
hence of possible hadron structures.  The rule is very simple: Any net
number of quarks ($N_q$ quarks minus $N_{\bar q}$ antiquarks) that is
divisible by 3, plus any number $N_g$ of valence gluons (except for a
single gluon with no quarks) can form a color singlet.  Mesons have
$N_q = N_{\bar q} = 1$, $N_g = 0$, while baryons have $N_q = 3$,
$N_{\bar q} = 0$, $N_g = 0$ (and antibaryons of course swap $N_q
\leftrightarrow N_{\bar q}$).  ``Valence'' here refers just to those
gluons---or, if one prefers, the total gluonic field of the
hadron---affecting the overall spin ($J$), parity ($P$), and charge
conjugation parity ($C$) quantum numbers $J^{PC}$ of the hadron,
because real QCD is so strong that innumerable gluons (not to mention
virtual $q\bar q$ {\it sea quark\/} pairs) are constantly created and
destroyed in any hadron.  Then a hadron with valence structure $q\bar
q g$ ($N_q = N_{\bar q} = 1$, $N_g \ge 1$) is called a {\it hybrid\/}
meson, and a hadron with valence gluons ($N_g \ge 2$) but no valence
quarks is called a {\it glueball}.  Multiquark hadrons are also
possible, the smallest two options being $q\bar q q \bar q$ {\it
  tetraquarks\/} and $qqqq\bar q$ {\it pentaquarks}.  All of these
hadrons except conventional mesons and baryons are called exotics.

The possibility of exotic hadrons was anticipated even before the
advent of the color charge degree of freedom or of QCD, in the seminal
works by Gell-Mann~\cite{GellMann:1964nj} (which introduced quarks)
and Zweig~\cite{Zweig:1981pd,Zweig:1964jf} (which introduced the
equivalent ``aces'').  The development of the quark model provides a
special intellectual resonance with the current state of affairs in
exotics studies: The quark model was developed as a simple paradigm
that brought order to the confusing proliferation of hadrons
discovered in the 1950s and 1960s, and the current era awaits the
development of an analogous resolution for exotics.  As of the time of
this writing, several paradigms or physical pictures have been
developed to understand the known exotics, such as {\it hadronic
molecules}, {\it diquarks}, and so on, and we will discuss the
successes and shortcomings of each of these pictures in detail in this
review.  But one should be under no illusion that any single one of
these pictures has yet accommodated all of the experimental data on
masses, production mechanisms, decay modes, and decay rates of the
exotics.  Of course, it is likely that no single picture will suffice
to explain all of the new data.

Even so, some interesting patterns have begun to emerge, and we will
offer some opinions on the directions that the data and theory appear
to be heading.

Speaking of the data, the very first charmoniumlike exotic was
discovered by the Belle collaboration at KEK in
2003~\cite{Choi:2003ue}, and although a great deal is now known about
this state, it is still called $X(3872)$ to indicate its fundamentally
unknown nature.  It was just the first of many unforeseen states to be
observed in subsequent years, some at multiple facilities and some (so
far) only at one.  Several of the most important experiments for
uncovering these exotics are still in operation ({\it e.g.}, BESIII,
LHCb, CMS), some are undergoing upgrades for future runs ({\it e.g.},
JLab12, Belle~II), and yet others are in the planning/development
stages ({\it e.g.}, ${\bar{\rm P}}$ANDA).  The full story of exotics
is an ongoing one: It is a rich, data-driven field producing numerous
new and surprising results every year, in which novel theoretical
ideas compete and are continually tested.

Our purpose in this review is to summarize and clarify the field of
heavy-quark QCD exotics for non-experts.  It is not meant to be
completely exhaustive; dozens of experimental papers and on the order
of 1000 theoretical papers have already contributed to the study of
these states.  Nor is it the first review dedicated to heavy-quark
exotics, which date back as far as 2006~\cite{Swanson:2006st,
Zhu:2007wz,Godfrey:2008nc,Drenska:2010kg,Pakhlova:2010zza,
Yuan:2014rta,Esposito:2014rxa,Olsen:2014qna,Chen:2016qju,
Ali:2016gli}.  In addition, several previous reviews whose subject is
heavy quarkonium, or exotics in general, discuss
exotics~\cite{Brambilla:2004wf,
Eichten:2007qx,Klempt:2007cp,Voloshin:2007dx,Brambilla:2010cs,
Bodwin:2013nua}.  We intend to paint a detailed picture, accessible to
non-expert researchers, on the discovery history, current status, and
future prospects of these remarkable states.

The remainder of Sec.~\ref{sec:Intro} presents the key theoretical
underpinnings of the states and an overview of the methods used to
analyze them; Sec.~\ref{sec:Expt} provides a semi-historical account
(organized by physical process) of the key experimental findings on
the exotics; Sec.~\ref{sec:Theory} describes the leading theoretical
pictures for the exotics in greater detail; Sec.~\ref{sec:Prospects}
offers a discussion of the possible future directions for exotics
studies; and Sec.~\ref{sec:Concl} concludes.
Appendix~\ref{app:states} summarizes the exotic candidates
individually.

\subsection{Distinguishing Conventional from Exotic Hadrons}

In order to substantiate a claim that an exotic state has been
observed, one must first ask two questions: how does one know that a
state has been observed, and second, how does one know that it is
exotic?

First, with regard to observation, a sufficiently long-lived charged
state leaves a measurably long track in a detector, while a
sufficiently long-lived neutral state leaves a measurable gap between
its production point and its decay via charged particles or observed
absorption in a calorimeter.  Kinematical reconstruction is then used
to identify the energy, momentum, and ultimately mass, of the
particle.  A short-lived particle is identified via the energy-time
uncertainty principle as a resonant peak in the production amplitude,
its mean lifetime given as the reciprocal of the full width at half
maximum, $\Gamma$.  The idealized form representing such a resonant
state is the {\it Breit-Wigner amplitude},
\begin{equation} \label{eq:BW}
f(s) = \frac{\Gamma/2}{M-\sqrt{s}-i\Gamma/2} \, ,
\end{equation}
where $\sqrt{s} = E_{\rm CM}$ is the center-of-momentum (CM) frame
energy and $M$ is the resonant-state mass parameter (to be specific,
this is a Lorentz-invariant expression of the nonrelativistic form).
The absolute square $|f(s)|^2$, which is the quantity observed in a
scattering cross section, gives a Lorentzian peak at $\sqrt{s} = M$ of
the same type that is familiar from multiple branches of physics.
However, in a physical situation one finds that the parameter $\Gamma$
can assume an energy dependence, that closely spaced resonances do not
assume the simple form of a sum of Breit-Wigners~\cite{Guo:2016bjq},
and that other amplitude effects, such as the opening of thresholds
for the formation of on-shell particles, can severely obscure the
idealized form of Eq.~(\ref{eq:BW}).  One cannot assume that every
bump in a cross section corresponds to a new resonance.  Fortunately,
Eq.~(\ref{eq:BW}) also presents one additional handle for discerning
resonant behavior, through the phase $\delta = \arg f(s)$ (the phase
shift of scattering theory).  In the neighborhood of $\sqrt{s} = M$,
$\delta$ increases rapidly from 0 to $\pi$, passing through $\frac \pi
2$ precisely at $\sqrt{s} = M$.  The amplitude $f(s)$ exhibits a
counterclockwise ``looping'' behavior in the Argand plane, which is
taken as the standard indicator of resonant behavior, idealized
Breit-Wigner or not.

Second, the procedure for deciding whether a state is conventional or
exotic can be carried out at several levels.  The simplest is by means
of $J^{PC}$ quantum numbers.  In a nonrelativistic quark model, a
$q\bar q$ meson with total spin angular momentum $S$ and relative
orbital angular momentum $L$ has $P = (-1)^{L+1}$ and $C =
(-1)^{L+S}$.  The case $J^{PC} = 0^{--}$ and the series $J^{P = (-1)^J
\! , \, C = (-1)^{J+1}} \in \{ 0^{+-}$, $1^{-+}, \ldots \}$, cannot be
reached for any values of $S$ and $L$, and are therefore manifestly
exotic.  No manifestly exotic baryon $J^{PC}$ value occurs, however.

The known electric charges of quarks (or alternately, their isospin
[for $u$, $d$] and other quark flavor [$s$, $c$, or $b$]) content
provide another signal for exotics.  A bosonic hadron with charge
$+2$, for example, cannot be formed as a $q\bar q$ state, and
therefore is at minimum a tetraquark.  Charmoniumlike states with
nonzero charge are necessarily exotic, since they must contain more
valence quarks than just the (neutral) pair $c\bar c$; for example,
the $Z_c(3900)^+$, which was the first state of this type with its
discovery confirmed (in 2013)~\cite{Ablikim:2013mio,Liu:2013dau}, is
believed to be a $c\bar c u\bar d$ state.

More typically, however, an exotic candidate carries the same $J^{PC}$
and charge as some conventional state.  In such circumstances, one
expects the states to mix quantum-mechanically, making identification
even trickier.  However, even in those cases, one can make headway.
If one has a principle for deciding how many states of a given
$J^{PC}$ should occur in a certain mass range and finds
extras---so-called {\it supernumerary\/} states---then one can be sure
that the set of these states contains an exotic component; we shall
see such examples in the $J^{PC} = 1^{--}$ charmoniumlike sector.
Second, a state may have the same $J^{PC}$ and approximately the
expected mass of a conventional state, but it may be difficult to
produce in the expected way, or it may decay into unexpected channels
or have suppressed decay rates into the expected channels.  For
instance, the $X(3872)$ has the same $J^{PC} = 1^{++}$ as a yet-unseen
$c\bar c$ state, $\chi_{c1}(2P)$, but its behavior is inconsistent
with expectations for the $\chi_{c1}(2P)$ in a variety of ways.  In
particular, its width is quite small, $\Gamma < 1.2$~MeV\@.  The
$\chi_{c1}(1P)$ state for which $\chi_{c1}(2P)$ is a radial excitation
is well known, and despite being hundreds of MeV lighter and therefore
having much less available phase space for decays, it has a width only
slightly smaller (840~keV) than the $X(3872)$ upper bound.

\subsection{Heavy Quarkonium}

In order to predict a spectrum of hadronic bound states, one must make
certain assumptions about quark and gluon interactions to solve
their equation of motion.  The complexity of the strong interaction,
both in its magnitude and in the intricacies of gluon self-coupling
and sea-quark production, makes for a problem that cannot be
completely solved analytically, and even reliable numerical solutions
(in the form of {\it lattice QCD\/} simulations) remain difficult to
achieve.  These effects are particularly prominent in the light-quark
($u$, $d$, $s$) sector, in which the quarks are manifestly
relativistic.  However, in the heavy-quark ($c$, $b$)
sector\footnote{$t$ quarks are so heavy that they decay weakly to $b$
quarks long before they could form hadrons.}, one can consider the
quarks to be slowly moving color sources interacting with a background
field of gluons and sea quarks---a {\it Born-Oppenheimer\/} scale
separation.  The typical energies associated with light quarks and
glue are given by the QCD scale $\Lambda_{\rm QCD} = O(200~{\rm
MeV})$, which is larger than $m_{u,d,s}$ but much smaller than
$m_{c,b}$.  The production of $c\bar c$, $b\bar b$ sea-quark pairs in
hadrons is also known to be suppressed, which justifies treating the
$c\bar c$ in charmoniumlike states (or $b\bar b$ in bottomoniumlike
states) to be valence quarks.

One can thus obtain a substantial simplification by treating the heavy
quarks as interacting nonrelativistically via a chosen potential
$V(r)$.  The best known such example is the {\it Cornell
potential}~\cite{Eichten:1978tg,Eichten:1979ms} for heavy quarkonium,
\begin{equation} \label{eq:Cornell}
V(r) = -\frac{\kappa}{r} + br \, ,
\end{equation}
where the first term represents the short-distance Coulomb-like
one-gluon exchange interaction, while the second represents the
confinement potential, ever-increasing with separation.  The Cornell
model has been enhanced over the years, its most thorough application
being Ref.~\cite{Barnes:2005pb}.  One may then solve the
Schr\"{o}dinger equation with the heavy-quark pair interacting through
$V(r)$ just as one does for $e^- p$ in hydrogen, thus obtaining a
complete spectrum of conventional $c\bar c$ or $b\bar b$ states.  The
results are quite impressive: 14 $c \bar c$ and 17 $b \bar b$ states
predicted by this quark-potential model have already been observed.
Several of these states lie above the threshold for producing
open-flavor heavy hadrons ({\it i.e.}, $c\bar q + \bar c q$ for $c
\bar c$), which is the analogue to the ionization threshold for
hydrogen; in other words, confinement allows for the production of
prominent above-threshold resonances, which have been experimentally
seen.  Solving the Schr\"{o}dinger equation also produces specific
eigenstate wave functions, which can then be used to predict hadronic
and radiative transition rates, and in turn provide comparisons to
data in order to confirm a state's conventional quarkonium status.

\subsection{Quarkoniumlike Exotics}

\begin{figure}
\begin{center}
\includegraphics[width=21.5cm]{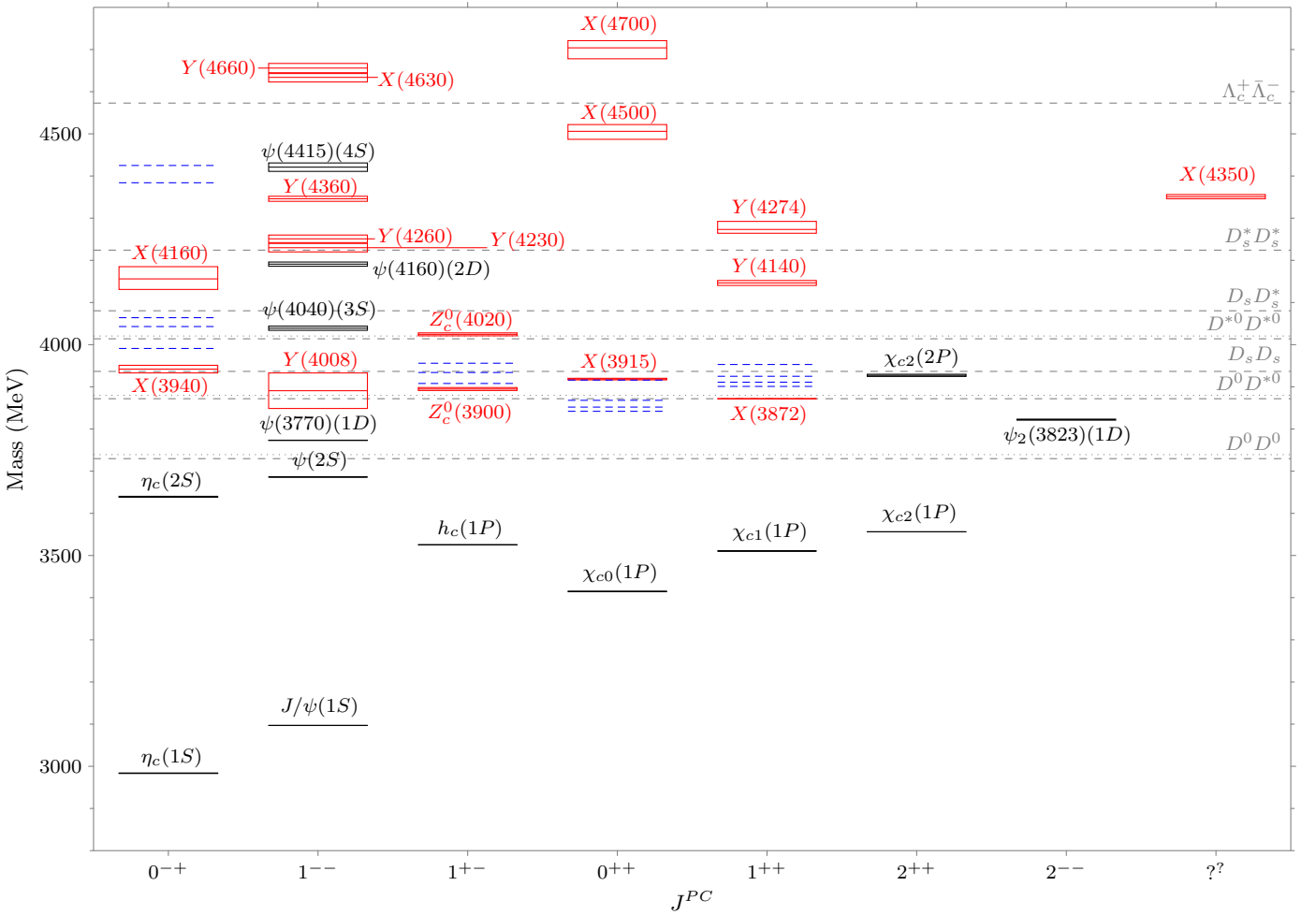}
\end{center}
\vspace{-14.3cm}
\caption{Level diagram for the neutral $c\bar c$ sector.  Conventional,
  observed $c\bar c$ states are solid (black) lines labeled by Greek
  letters, the lowest predicted yet-unobserved conventional $c\bar c$
  states are labeled with dashed (blue) lines (the clusters indicating
  predictions of several variant model calculations), and the solid
  (red) lines labeled by $X$, $Y$, or $Z$ indicate exotic
  charmoniumlike candidates.  Each measured state mass, including its
  central value and uncertainty, is presented as a rectangle (lines
  simply indicating very thin rectangles).  Relevant thresholds are
  given by gray dashed lines; if a gray dotted line is nearby, it
  indicates the threshold isospin partner to the labeled dashed line.
  In some cases, likely quantum numbers have been assigned to states
  for which some uncertainty remains; this is the case, for example,
  for the $X(3940)$ and $X(4160)$, which have been studied as
  $\eta_c(3S)$, $\eta_c(4S)$ candidates.  The actual known quantum
  numbers are listed in
  Table~\ref{tab:XYZbyMass}.\label{fig:ccneutral}}
\end{figure}

\begin{figure}
\begin{center}
\includegraphics[width=21.5cm]{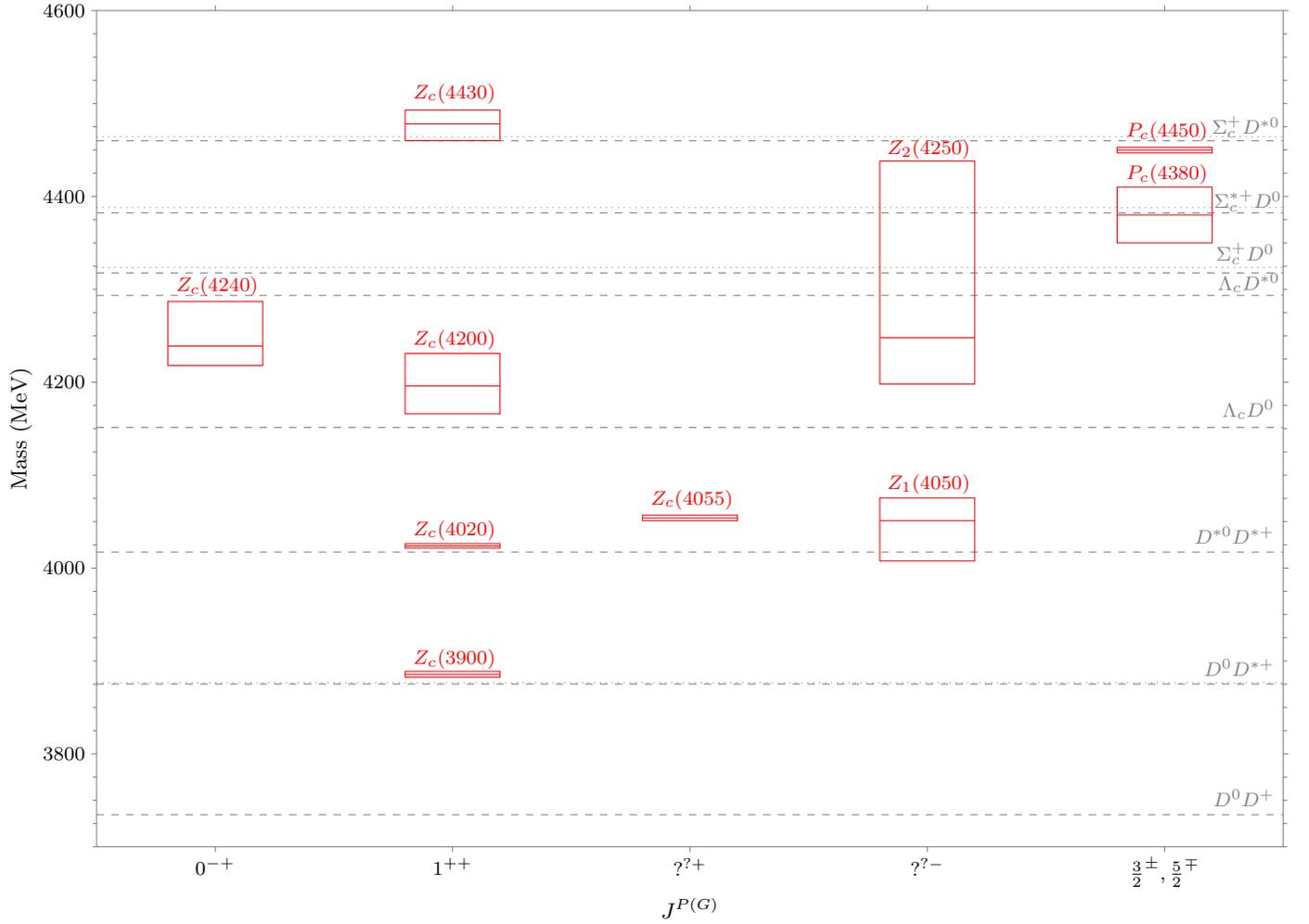}
\end{center}
\vspace{-14.3cm}
\caption{Charged charmoniumlike states, both bosonic and fermionic.
Each measured state mass, including its central value and uncertainty,
is presented as a rectangle.  Relevant thresholds are given by gray
dashed lines; if a gray dotted line is nearby, it indicates the
threshold isospin partner to the labeled dashed
line.\label{fig:cccharged}}
\end{figure}

The level diagram for neutral $c\bar c$-containing states is presented
in Fig.~\ref{fig:ccneutral}.  Conventional $c\bar c$ states are solid
(black) lines labeled by Greek-letter particle names, while the lowest
predicted but yet-unobserved $c\bar c$ states are represented by
dashed (blue) lines (clusters indicating the predictions of variant
models).  The stunning result of the years since 2003 is the
observation of all the levels marked in red and labeled as $X$, $Y$,
or $Z$, the exotic candidates.  The states of the charged
charmoniumlike sector, which as noted above are manifestly exotic, are
presented in Fig.~\ref{fig:cccharged}.  These figures should be
considered a snapshot in time, as some of the states may disappear
under closer scrutiny, while additional ones will doubtlessly be
discovered.  In the higher-energy (and thus not quite as easily
accessible through current experiments) $b\bar b$ sector, 2 neutral
and 2 charged exotic candidates have been observed.

The naming scheme currently in use for the new states is still not
entirely settled.  The labels currently employed are $X, Y, Z$, and
$P_c$.  The original $X(3872)$ was first seen~\cite{Choi:2003ue} as a
$J/\psi \, \pi^+ \pi^-$ resonance in the decay $B^{\pm} \to K^{\pm}
J/\psi \, \pi^+ \pi^- $, and therefore $X$ has generically been used
to denote neutral resonances appearing in $B$-meson decays.  However,
not only has $X(3872)$ been since observed in other processes (such as
at hadron colliders~\cite{Acosta:2003zx,Chatrchyan:2013cld}), but also
some states appearing in $B$ decays [{\it e.g.}, $Y(4140)$] are
labeled as $Y$, and some states appearing so far only in other
production processes [{\it e.g.}, $X(4350)$ in $\gamma \gamma$ fusion]
are labeled as $X$.

The first new state observed in the {\it initial-state radiation\/}
(ISR) process $e^+ e^- \to \gamma_{\rm ISR} Y$ pioneered at the BaBar
experiment at SLAC was seen in 2005 and named
$Y(4260)$~\cite{Aubert:2005rm}.  Currently, all such states produced
this way are labeled $Y$, but as noted above, so are a few others.

The first charged charmonium-like state was observed by Belle in
2008~\cite{Choi:2007wga} and was named $Z_c(4430)$.  Since then, the
label $Z$ has been used for all charged quarkoniumlike bosons, but
since all such states are expected to have neutral isospin partners
({\it e.g.}, $Z_c(3900)^+$ has an observed $Z_c(3900)^0$
partner~\cite{Xiao:2013iha,Ablikim:2015tbp,Ablikim:2015gda}), the
label now has come to mean exotic candidates in an isospin multiplet
that has a charged member.

Bottomoniumlike states are labeled with a $b$ subscript ($Y_b$,
$Z_b$).

The Particle Data Group (PDG)~\cite{Agashe:2014kda} currently avoids
the naming ambiguities in the meson sector simply by calling all
charmoniumlike bosons $X$.

Only the label $P_c$ is, as yet, completely unambiguous.  It is used
for baryonic charmoniumlike exotics (first observed in 2015 at
LHCb~\cite{Aaij:2015tga}).  Certainly, a clearer naming scheme for the
new states is highly desirable.

\subsection{Overview: Models and Pictures for Exotics}

The large amount of data on candidate exotic states has spawned a
number of potential theoretical explanations that are summarized here
and described in greater detail in the following sections.  It should
be emphasized that no single picture naturally accommodates all
the observed states; some might turn out to be of a molecular nature,
some of a hybrid nature, and so on, or a given state could easily be a
quantum-mechanical mixture of more than one type.  Furthermore,
distinctions between some of the pictures can sometimes blur,
depending upon dynamical assumptions.

\subsubsection{Molecular Picture} \label{sec:Molecular}

Upon first glance, the most obvious interpretation of a tetraquark
(pentaquark) state is that of a hadronic molecule of two mesons (one
meson and one baryon).  In this picture, each component meson is bound
internally by strong QCD color forces, while the mesons bind to each
other by means of a much weaker color-neutral residual QCD force, the
analogue of the van der Waals attraction in chemistry.  Molecules
formed of separate color-neutral hadrons are of course plentiful in
nature---after all, all atomic nuclei beyond hydrogen have this
structure.  Supporting the molecular interpretation is a mathematical
fact of color algebra: A 2-quark 2-antiquark system can be assembled
into a color singlet in only two independent ways; in the case of
$c\bar c q^\prime \! \bar q$, the combinations are $(c\bar c)(q^\prime
\! \bar q)$ and $(c\bar q)(\bar c q^\prime)$, {\it i.e.}, either the
color structure of charmonium plus a light meson, or a pair of
open-charm mesons (with analogous results for pentaquarks).  Indeed,
the original proposal of molecules formed from charmed-meson pairs is
almost as old as QCD itself~\cite{Voloshin:1976ap,DeRujula:1976zlg}.
Similar comments hold for pentaquarks, charmed or
not~\cite{Hogaasen:1978jw,Chan:1977st}.  Of course, this result alone
does not necessarily imply that the quarks segregate themselves inside
the tetraquark to resemble two separate hadrons.  The color-singlet
pairs can be completely delocalized within the tetraquark in a variety
of interesting ways, the specifics of which define the other physical
pictures to be described below.

The plausibility of the molecular picture is greatly substantiated by
two further facts.  First, a number of exotic candidates lie
remarkably close to two-meson thresholds.  The most impressive example
is provided by the original exotic candidate, $X(3872)$, whose mass
obeys $m_{X(3872)} - m_{D^{*0}} - m_{D^0} = +0.01 \pm 0.18$~MeV\@.
This value suggests a state {\em at least 10 times\/} more weakly
bound than a deuteron, which itself is already considered a weakly
bound hadronic molecule.  Second, the absence of evidence for
near-degenerate quartets of exotic candidates containing $u\bar u$,
$u\bar d$, $d\bar u$, and $d\bar d$ forming $I=0$ and $I=1$ isospin
multiplets, despite a dedicated search~\cite{Aubert:2004zr} in this
energy region, suggests a preference for binding in certain isospin
channels.  Such a result is natural when one supposes that the
necessary molecular binding is the result of meson exchanges, and
recalls that the lightest mesons $\pi$ have $I=1$.  Indeed, a
significant part of the binding of the $I=0$ deuteron is accomplished
through $\pi$ exchange.  Moreover, no prominent resonant structure
seems to occur at the $D^0 \bar D^0$ threshold~\cite{Abe:2007sya},
which is consistent with a molecular picture in which the binding is
accomplished through the exchange of $J^P = 0^-$ mesons like $\pi$,
since the invariance of strong interactions under rotations plus
parity forbids a three-pseudoscalar coupling such as $D^0 \bar D^0
\pi$.

Not every exotic candidate lies just below a suitable threshold,
however.  For example, the $Z_c(3900)^+$ lies about 20~MeV above the
nearest ($D \bar D^*$) threshold and dominantly decays through this
mode.  In this case, one faces the awkward problem of attempting to
form a bound state out of components into which it can freely decay.
It is also worth remembering that the deuteron, unlike all charm
molecules, faces no such stability issues.

In addition, the production rate of $X(3872)$ at high-energy collider
experiments in the primary interaction region (``prompt'' production)
is observed to be comparable to that of ordinary charmonium
states~\cite{Acosta:2003zx,Chatrchyan:2013cld}.  If $X(3872)$ is a
purely molecular $D^0 \bar D^{*0} + {\bar D}^0 D^{*0}$ state
(subsequently we write just the first term, the charge conjugate being
understood), then presumably the two mesons form in the collision
first, and they must furthermore possess a sufficiently small relative
momentum---a rare occurrence in a high-energy collision---to allow
their subsequent coalescence into a molecule.  The large measured
prompt production rate of $X(3872)$ argues against this state having a
purely molecular nature~\cite{Bignamini:2009sk,Bignamini:2009fn}.

\subsubsection{Hadrocharmonium Picture}

In the {\it hadrocharmonium} (or more generally, {\it
  hadroquarkonium}) picture for multiquark exotics, the heavy-quark
pair $Q\bar Q$ forms a compact core about which the light $q\bar q$ or
$qqq$ forms a quantum-mechanical
cloud~\cite{Voloshin:2007dx,Dubynskiy:2008mq}.  The $Q\bar Q$ in the
simplest variant of the picture forms a color singlet, in which case
the light-quark cloud does as well, and their mutual binding again
occurs through weak color van der Waals forces, like in molecular
models.  Alternately, both the core and the cloud can occur in the
color-adjoint representation, thereby creating a much stronger mutual
binding, but this configuration lies outside the original
hadrocharmonium proposal.

The hadrocharmonium picture was originally motivated by the strong
preference of several exotics to decay to conventional charmonium
[{\it e.g.}, $J/\psi$, $\psi(2S)$, and $\chi$] rather than to heavy
open-flavor hadrons ($D$, $D^*$), which can naturally be viewed as the
dissociation of the charmonium core from the light cloud.  If the
dynamics is such that the heavy degrees of freedom largely decouple
from the light ones, as expected from {\it heavy-quark spin
symmetry\/}~\cite{Manohar:2000dt}, then the spin and wave function of
the $Q\bar Q$ within the exotic should leave an imprint on the
final-state quarkonium.  For example, the preference of $Z_c(4430)$ to
decay to $\psi(2S)$ rather than the $1S$ state
$J/\psi$~\cite{Chilikin:2014bkk} may indicate a radially excited
$c\bar c$ core in $Z_c(4430)$.  Decays of a particular exotic state
into final states with more than one $c\bar c$ spin [{\it e.g.}, both
$J/\psi$ ($s_{c\bar c} = 1$) and $h_c$ ($s_{c\bar c} = 0$)] need not
violate heavy-quark spin symmetry, as long as the exotic contains an
admixture of the two spin states~\cite{Li:2013ssa}.

On the other hand, hadroquarkonium also has conceptual drawbacks.  If
the forces holding the state together are sufficiently weak, it is
unclear why the system would persist long enough to be identified as a
distinct state.  If the forces are sufficiently strong, it is unclear
why the quarkonium and light components would remain largely
decoupled, rather than immediately rearranging into two heavy
open-flavor hadrons.  Indeed, the exotics that decay prominently into
open-flavor pairs, such as $X(3872)$ into $D^0 {\bar D}^{*0}$, do not
appear to admit a satisfactory hadrocharmonium interpretation.

\subsubsection{Diquark Picture}

The binding of color-singlet hadrons through the triplet ({\bf
3})-antitriplet ($\bar{\bf 3}$) $q\bar q$ combination is so familiar
from the ubiquity of mesons that it is easy to forget certain other
color combinations are also attractive.  Of course, the baryon $qqq$
combination also forms color-singlet hadrons, and since each of the
quarks transforms as a {\bf 3}, each complementary quark pair must
form a color-conjugate $\bar{\bf 3}$.  The {\it diquark\/} combination
${\bf 3} \otimes {\bf 3} \to \bar{\bf 3}$ must therefore be attractive
in order for the baryon to be bound.  Indeed, it is straightforward to
compute the color dependence of the short-distance coupling of
particles in SU(3)$_c$ representations $R_1$ and $R_2$, whose color
generators appearing at the interaction vertices are $T^a_1$ and
$T^a_2$, respectively, to the product representation $R$.  The trick
is precisely the one used to compute spin-spin couplings for SU(2):
${\bf s}_1 \! \cdot {\bf s}_2 = \frac 1 2 [({\bf s}_1 \! + {\bf
s}_2)^2 - {\bf s}_1^2 - {\bf s}_2^2]$.  For an arbitrary group, the
squares define the quadratic Casimirs $C_2(R) \equiv T^a_R T^a_R$,
\begin{equation} \label{eq:CasimirDef}
{\cal C}(R,R_1,R_2) \equiv C_2 (R) - C_2 (R_1) - C_2 (R_2) \, .
\end{equation}
From Eq.~(\ref{eq:CasimirDef}), one may compute the relative size of
short-distance color couplings for all $qq$ or $q \bar q$ systems:
\begin{equation} \label{eq:CasimirNums}
{\cal C}(R,R_1,R_2) =  \frac 1 3 (-8,-4,+2,+1) \ {\rm for} \ R =
({\bf 1},\bar {\bf 3},{\bf 6},{\bf 8}) \, ,
\end{equation}
respectively.  Unsurprisingly, the most attractive coupling is that of
the color-singlet $q\bar q$ combination.  However, the diquark
$\bar{\bf 3}$ coupling is also quite large, being half as strong at
short distance, while the two repulsive couplings are rather weaker.

Diquarks therefore provide a promising potential source of
substructure in hadronic physics, and have long been studied as
such~\cite{Anselmino:1992vg}, particularly in the baryon sector.  In
the tetraquark sector, the structure is that of a bound state of a
diquark and antidiquark, although such systems are often confusingly
dubbed ``tetraquark models'' in the literature.  Originally suggested
for the light-quark scalar mesons $a_0(980)$ and
$f_0(980)$~\cite{Jaffe:1976ig}, the diquark picture was applied to the
charmoniumlike exotics in Ref.~\cite{Maiani:2004vq} by means of a
constituent-quark Hamiltonian including spin couplings between the
quarks.  Due to the equal importance of each colored quark in
determining the structure of the full state (as opposed to molecular
models, in which the $q\bar q$ pairs are first combined into color
singlets), diquark models predict a rich spectrum of states.  Indeed,
when the pattern of observed exotics became more apparent in the past
few years, the original model of~\cite{Maiani:2004vq} was found to
predict too many states, and was modified through the {\it Ansatz\/}
that the only significant spin couplings are the ones within each
diquark~\cite{Maiani:2014aja}.  In such models, the $Y$ exotics are
understood as $L = 1$ orbital excitations of lower states such as
$X(3872)$ (which is supported by possible observation of the decay
$Y(4260) \to \gamma X(3872)$~\cite{Ablikim:2013dyn}), while the
lighter $Z$ states [$Z_c(3900)$, $Z_c(4020)$] are related to the
$X(3872)$ through different diquark spins and relative orientations,
and $Z_c(4430)$ is a radial excitation of the $Z_c(3900)$.  The $P_c$
states can be considered analogously~\cite{Maiani:2015vwa}.  Even with
these successes, the diquark model of~\cite{Maiani:2014aja} still
predicts many more states than have yet been observed.

Additionally, a Hamiltonian treatment suggests an (approximately)
common rest frame for all the components, since it implies a single,
shared time coordinate.  If the diquark-antidiquark $(cq)(\bar c \bar
q)$ pair form a relatively static molecule, then the question of
stability again arises: Why should the system not simply reorganize
itself into the more tightly bound open-flavor heavy-meson molecule
$(c\bar q)(\bar c q)$?  The {\it dynamical diquark
picture\/}~\cite{Brodsky:2014xia} addresses this question by proposing
that the diquark-antidiquark pair are created with a large relative
momentum and, if below the threshold for creating extra $q\bar q$
pairs, can only hadronize through the long-distance tails of meson
wave functions stretching between the quarks and antiquarks, providing
an explanation of exotics' relatively small widths.  The large diquark
pair separation also gives a natural explanation for the spin {\it
Ansatz\/} of Ref.~\cite{Maiani:2014aja}.  The pentaquarks $P_c$ can be
constructed by an extension of the dynamical diquark
attraction~\cite{Lebed:2015tna} to the color-triplet attraction of
{\it triquarks}, $\bar c_{\bar{\bf 3}} (qq)_{\bar{\bf 3}} \to [\bar c
(qq)]_{\bf 3}$, thus describing the $P_c$ as diquark-triquark states.
However, the dynamical diquark picture has not yet been developed as a
fully predictive model.

\subsubsection{Hybrids}

The previous sections have focused on quark dynamics for conventional
and novel hadrons.  As noted, an alternative way to construct novel
hadrons is by admitting explicit gluonic degrees of freedom, in
addition to quarks, in the state. These states are called {\it
  hybrids}~\cite{Meyer:2015eta}, while states that are dominated by
gluonic degrees of freedom are called {\it glueballs}.

The history of the development the quark model and QCD illustrates
that discovering explicit nonperturbative glue can be difficult.  As
stated above, all the well-established mesons have $J^{PC}$ equal to
$0^{-+}$, $0^{++}$, $1^{++}$, {\it etc.}, which can be created with
$q\bar q$ pairs in a given orbital momentum state.  The $q\bar q$
picture is supported further by the absence of mesons with isospin or
strangeness greater than unity.  Thus it appears that quarks are
spin-$\frac 1 2$ entities, while the spectrum ordering suggests that
energy eigenvalues increase with orbital angular momentum.  In this
way, the simple quark model of mesons (and baryons) was partly
motivated by the {\it absence\/} of ``exotic'' hadrons such as
multiquark or gluonic states.  It is therefore perhaps no surprise
that QCD exotics have been difficult to observe.

The simplest explanation for this absence is that gluonic degrees of
freedom are somewhat ``stiff'' and therefore difficult to excite.  Of
course, with the increasing energy, luminosity, and capabilities of
modern accelerators and detectors, one might hope that this impediment
can be overcome.

Historically, the nonperturbative gluonic degrees of freedom have been
analyzed in the context of two broad ideas: They are some sort of {\it
string\/} or {\it flux tube}, or they manifest as an effective
constituent confined by a {\it bag\/} or
potential~\cite{Barnes:1977lta,Jaffe:1975fd}.  Alternatively,
nonperturbative glue can be thought of as either collective, nonlocal
degrees of freedom, or as a local {\it quasiparticle\/} degree of
freedom.  More recently, lattice gauge theory computations have
provided compelling evidence that nonperturbative gluons are
effectively chromomagnetic quasiparticles of quantum numbers $J^{PC} =
1^{+-}$~\cite{Dudek:2011bn} with an excitation energy of approximately
1 GeV\@.  Thus, the lightest charmonium hybrid multiplet is expected
near 4180 MeV, with quantum numbers $J^{PC} = (0,1,2)^{-+}$ and
$1^{--}$.  In addition, effective field theories have been developed
for hybrids that place on a rigorous footing some of the lore of the
field~\cite{Berwein:2015vca}.


Hybrid mesons are every bit as ``hadronic'' as conventional mesons,
and thus convincingly identifying them will rely on developing a
robust and reliable model of their spectrum and production and decay
characteristics. This effort profits greatly from recent algorithmic
and computational advances in lattice gauge field theory.  These
developments, coupled with the nascent GlueX experiment and the
forthcoming $\bar{\rm P}$ANDA experiment, provide much hope for
dramatic progress in the subfield.

\subsubsection{Kinematical Effects}

As stated above, many of the $XY \! Z$ states lie near threshold and
are therefore naturally associated with weakly bound molecular
interpretations.  Intriguingly, several of the new states lie just
{\em above\/} threshold: $Z_c(3900)$ [$D\bar D^*$], $Z_c(4020)$
[$D^*\bar D^*$], $Z_b(10610)$ [$B\bar B^*$], and $Z_b(10650)$
[$B^*\bar B^*$].  This fact strongly suggests that these experimental
enhancements may be due to threshold rescattering rather than
quark-level dynamics.

That something nontrivial can happen at a threshold can be seen with
the following two-channel nonrelativistic example.  Consider $a\to a$
and $a\to b$ scattering (the letters refer to channels) described by
the $S$~matrix:
\begin{equation}
  S = \left(\begin{array}{cc}
      \sqrt{1-\rho^2}\, {e}^{2i \delta_a} & i \rho \,
      {e}^{i (\delta_a + \delta_b)} \\
      i\rho \, {e}^{i (\delta_a + \delta_b)} & \sqrt{1-\rho^2}\,
      {e}^{2 i \delta_b} \end{array} \right).
\end{equation}
Near an $s$-wave threshold at $E=E_0$, $\rho^2 \approx 2 c k$, where
$c$ is a constant, and
\begin{equation}
k^2 = 2 \mu_b (E-E_0).
\end{equation}
Under these conditions,
\begin{equation}
  \sigma( a \to a) \approx \frac{4\pi}{2\mu_a E} \left| \frac{(1-ck)\,
      {e}^{2 i \delta_a} -1 }{2i}\right|^2 \approx
  \frac{4\pi}{2\mu_a E} (1-ck) \, \sin^2\delta_a.
\end{equation}
As $E_0$ is approached from above, $\sigma(E)$ is well behaved, but
$d\sigma/dE \to -\infty$, indicating a slope discontinuity.
Continuing $\sigma(E)$ below $E_0$ shows that this discontinuity can
appear as a cusp. This effect was first pointed out by Wigner in
1948~\cite{Wigner:1948zz} and was studied further by Baz' and
Okun~\cite{BO} and Nauenberg and Pais~\cite{NP} in the late 1950s.


\begin{figure}[t]
\begin{center}
\includegraphics[width=6cm,angle=0]{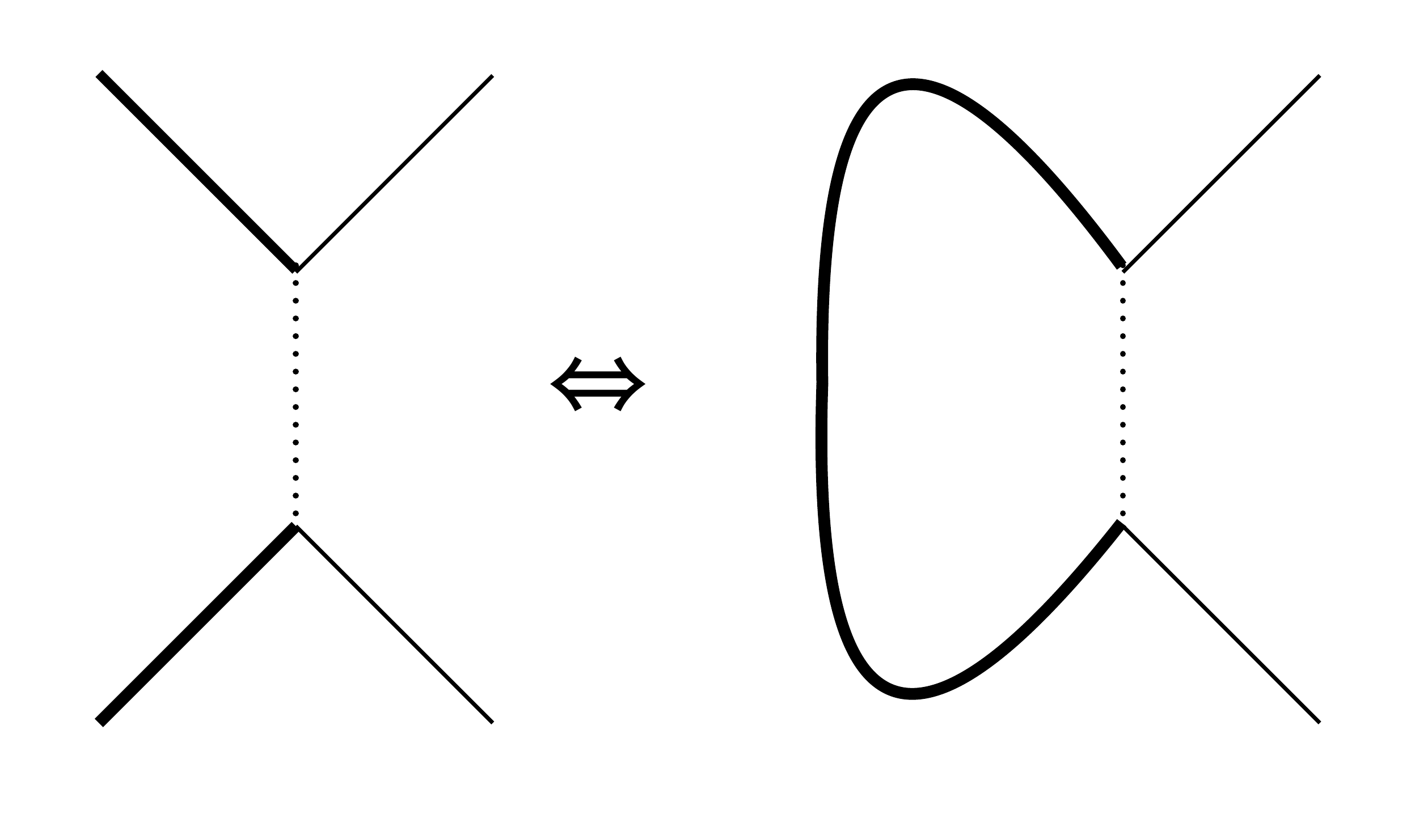}
\end{center}
\caption{The relationship of scattering and self-energy amplitudes.}
\label{fig:cut-prop}
\end{figure}

Two-particle scattering can be mapped to a two-point function by
cutting a propagator (Fig.~\ref{fig:cut-prop}).  Thus, the
opening-channel singularity is related to the self-energy threshold
singularity.  Because of these connections, terms such as ``threshold
effect'', ``rescattering effect'', and ``cusp effect'' all refer to
similar dynamics, and tend to be used interchangeably.

\begin{figure}[t]
\begin{center}
  \includegraphics[width=8cm,angle=0]{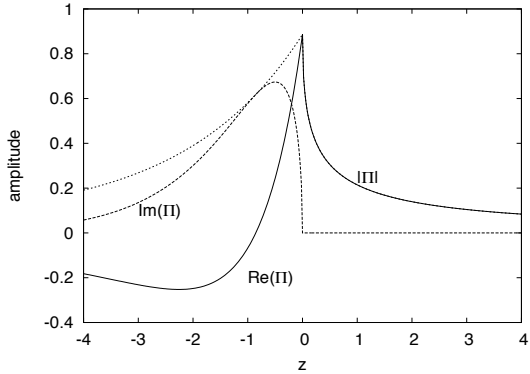}
\end{center}
\caption{Self-energy $\Pi(z)$ vs.\ $z$.  $|\Pi|$ for $z < 0$ is given
by the dotted line.}
\label{fig:bubble}
\end{figure}

We illustrate the two-point function behavior with a simple
nonrelativistic expression for the self-energy of a scalar particle
coupled to an intermediate state $AB$, dropping overall coefficients:
\begin{equation}
  \Pi(s) = \int\frac{d^3q}{(2\pi)^3} \,
  \frac{\exp{(-2q^2/\beta^2)}}{\sqrt{s} - m_A - m_B - q^2/
    (2\mu^{\vphantom\dagger}_{AB})
    + i \epsilon} \, .
\end{equation}
A phenomenological exponential ``form factor'' with scale $\beta$ has
been included in the expression to account for the spatial extent of
the hadrons in the process.  The integral can be evaluated in closed
form and is given by
\begin{equation} \label{eq:Pi_closed}
  \Pi(s) = - \frac{\mu{\vphantom\dagger}_{AB}\beta}{(2\pi)^{3/2}}
  \, \left[ 1 -
    \sqrt{\pi z}\exp(z) {\rm erfc}(\sqrt{z})\right] \, ,
\end{equation}
with
\begin{equation}
z = \frac{4 \mu_{AB}}{\beta^2}(m_A + m_B - \sqrt{s}).
\end{equation}
The behavior of the self-energy [in units of the prefactor of
Eq.~(\ref{eq:Pi_closed})] is shown in Fig.~\ref{fig:bubble}.  The
imaginary part of the amplitude is zero for positive $z$ ({\it i.e.},
below threshold) and turns on rapidly once the intermediate-state
threshold is crossed.  The real part of the amplitude also exhibits
singular behavior near threshold [as required by the complex
analyticity of $\Pi(s)$], and it is no surprise that the rate can
display large enhancements just above threshold.  Furthermore, the
``cusp'' produces phase motion that is similar to that of a
Breit-Wigner amplitude (but differs in that the motion follows the
real axis until threshold is reached).

%
%
T{\"o}rnqvist~\cite{Tornqvist:1995kr} and
Bugg~\cite{Bugg:2008wu,Bugg:2011jr} have stressed the importance of
this simple phenomenon for interpreting hadronic reactions for many
years, highlighting, among other effects, the mechanism by which
resonances are ``attracted'' to threshold cusps.

\subsection{Theoretical Techniques}

\subsubsection{Quark Potential Models}

Attempts to understand hadrons with potential models date from shortly
after quarks were introduced by Gell-Mann~\cite{GellMann:1964nj} and
Zweig~\cite{Zweig:1981pd,Zweig:1964jf}, and were based on a
simple-minded extension of the quantum-number coupling techniques of
nuclear physics.  Early problems with the apparent lack of free quarks
and fermionic quark statistics were obviated by the introduction of
QCD (or, more accurately, were replaced with the color-confinement
problem), which in turn led to a renaissance of the
field~\cite{DeRujula:1975qlm,Isgur:1977ef}.  In the modern treatment,
the application of simple quantum-mechanical models to the structure
of hadrons is underpinned by potential nonrelativistic QCD (pNRQCD),
which builds a systematic description of low-lying heavy hadrons by
integrating out ({\it {\`a} la\/} Wilson) a series of large energy
scales in QCD~\cite{Brambilla:2004jw}.

Typical quark potential models of heavy quarkonia assume a central
confinement potential, often taken to be linear, a Coulombic
interaction [or both, as in the Cornell potential of
Eq.~(\ref{eq:Cornell})], and supplemental QCD-motivated spin-dependent
interactions.  The agreement with the established charmonium and
bottomonium spectra can be startling, and it is in fact difficult to
understand why it continues to work for states above the open-flavor
threshold.  For example, a four-parameter nonrelativistic
model~\cite{Barnes:2005pb} obtains the masses of the 12 charmonium
masses known at the time with an average error of 0.26\%.

The application of quark potential models to highly excited heavy
hadrons and light hadrons is a {\em model\/} of QCD, in the sense that
it is not justified by an expansion in an identifiable small
parameter.  Nevertheless, in the same way that Landau's {\it
quasiparticles\/} emerge from strongly interacting fermionic systems,
the hope is that many phenomena can be subsumed into model parameters.
We remark that the potential concept is often stated to be
inapplicable in the light-quark regime.  However, the QCD Hamiltonian
can be defined in Coulomb gauge, which induces an explicit
instantaneous interaction.  The presence of light quarks can greatly
modify this interaction, but does not eliminate it.

Important conceptual problems nevertheless do occur once light quarks
are included.  For example, light quarks permit transitions between
Fock sectors, and these transitions are expected to play a large role
in hadronic properties high in the spectrum.  Gluonic degrees of
freedom must also have an impact once excitations of the order of the
effective gluon mass are reached (around 1~GeV).  Finally, chiral
symmetry breaking is the dominant dynamical feature of QCD in the very
low-energy regime, but is impossible to incorporate into simple
nonrelativistic potential models.

We remark that all of these problems can be overcome with
generalizations of the potential model approach: coupled-channel
models can be used to model transitions, gluonic degrees of freedom
can be explicitly included, relativistic kinematics and dynamics can
be assumed, and chiral symmetry breaking can be incorporated with
simple many-body physics techniques.

\subsubsection{Meson Exchange Models} \label{sec:Yukawa}

The idea that meson exchange is relevant to nuclear structure dates to
a seminal paper by Yukawa in 1935~\cite{Yukawa:1935xg}.  In just over
a decade, Yukawa's suggestion was confirmed with the discovery of the
pion.  By modern standards, this saga is something of a fluke: Baryons
are made of quarks, and quarks strongly interact via gluon exchange.
Had Yukawa known this, he would have arrived at an infinite-ranged
$NN$ interaction---which he knew could not be right, because real
nuclei have a finite size.  Alternatively, had he known of color
charge and confinement as well, he would have arrived at an
interaction range that is too small.  In fact, obtaining the correct
interaction requires spontaneous chiral symmetry breaking {\em and\/}
the correct amount of explicit chiral symmetry breaking in order to
achieve a pion of the ``correct'' physical mass.  This circumstance
conspires to make the pion unnaturally light, and therefore ubiquitous
in hadronic interactions.

Yukawa's idea has been taken to its limit by the
Nijmegen~\cite{Stoks:1994wp}, Bonn~\cite{Machleidt:2000ge}, and
Argonne~\cite{Wiringa:1994wb} groups, who have built extensive models
of nucleon-nucleon interactions based upon the exchange of many
mesons.  Of course QCD is a theory of quarks and gluons, and
presumably the short-range nucleon-nucleon interaction is dominated by
these degrees of freedom.  Since mesons couple to quarks, it is also
natural to consider meson-exchange contributions to the interquark
interaction, and a phenomenology of the baryon spectrum has been based
on this idea~\cite{Glozman:1995fu}.  We add, however, that this idea
has been heavily criticized as inapplicable to mesons and incompatible
with baryon phenomenology~\cite{Isgur:1999jv}.

Since pion exchange provides an essential source of binding for the
deuteron, it might be relevant to other hadronic interactions.  The
first to treat this idea seriously was T{\"o}rnqvist, who found many
possible bound-states of combinations of $D$, $D^*$, $B$, or $B^*$
mesons~\cite{Tornqvist:1993ng}.  The idea has found many applications
to novel hadrons, especially those that couple to heavy-meson pairs in
an $s$ wave and have masses just below the decay-channel threshold.

In spite of the enthusiasm for meson-exchange dynamics, several
conceptual difficulties bedevil the field.  The standard approach is
to consider the nonrelativistic limit of pion (or other meson)
exchange for the process $AB \to AB$.  For pion exchange, the resulting
scattering amplitude is of the form
\[
{\cal M} \propto \frac{({\bm \sigma}_A\cdot {\bm q}) \,
({\bm \sigma}_B \cdot {\bm q})} {{\bm q}^2 + \mu^2} \,
{\bm \tau}_A\cdot {\bm \tau}_B \, ,
\]
where $\mu^2 = m_\pi^2 - (m_A-m_B)^2$.  Here ${\bm q}$ is the momentum
exchange in the process, and ${\bm \sigma}$ and ${\bm \tau}$ refer to
spin and isospin.  Fourier transforming this expression yields a
central potential with a delta function and a tensor function that is
not an admissable quantum-mechanical interaction because of its
singular nature.  These problems are addressed by introducing a
regulator that modifies the interaction at short distance.  This
modification can be drastic: It is typically of the opposite sign to
the central Yukawa potential and very strong at the origin.  In the
case of the deuteron, the regulator ``core'' is useful because it
matches expectations for the repulsive $NN$ core interaction.
However, in general the regulator dependence is arbitrary and cannot
be expected to match to reality.  This problem is especially visible
when repulsive Yukawa interactions are considered, since these
correspond to an (unphysical) attractive core.

An additional problem arises when the masses of hadrons $A$ and $B$
are not equal.  If the difference is large enough, $\mu$ and the
interaction become imaginary.  Such a result may be a reasonable
analytic continuation for the amplitude, but it implies that the
system should be considered as a three-body problem ($A$-$B$-$\pi$) to
capture the essential dynamics.

\subsubsection{Heavy Quark Symmetry}

Were quarks degenerate in mass and electric charge, they would give
rise to hadron multiplets degenerate in masses and couplings.  This
effect is well illustrated by the phenomenon of isospin symmetry,
since the few-MeV difference between $m_u$ and $m_d$ and the
QED-induced energies are small compared to the strong-interaction
energy scale $\Lambda_{\rm QCD}$\@.  In the opposite limit $\epsilon_Q
\equiv \Lambda_{\rm QCD}/m_Q \to 0$, different heavy quark flavors $Q$
become interchangeable static sources of color charge.  Moreover,
since the spin-dependent interactions of these quarks are suppressed
by powers of $\epsilon_Q$, one finds a near degeneracy between
different spin states of heavy quarks (as seen, {\it e.g.}, from the
relative smallness of the difference $m_{D^*} - m_D$ compared to its
average, since the $D$ and $D^*$ differ only by the flip of the
$c$-quark spin).  Taken together, this {\it heavy-quark spin-flavor
  symmetry}~\cite{Isgur:1989vq,Isgur:1989ed} gives rise to the
powerful {\it heavy-quark effective
  theory}~\cite{Grinstein:1990mj,Georgi:1990um}, which has
$\epsilon_Q$ as its expansion parameter; for a review,
see~\cite{Manohar:2000dt}.

Heavy-quark spin symmetry (HQSS) also plays an important role in the
spectra and decays of the multiquark exotics, generally producing
nearly degenerate multiplets of different spin, analogous to the
$D$-$D^*$ pair.  Each theoretical picture produces a distinctive set
of constraints on the spectrum, once HQSS is
imposed~\cite{Cleven:2015era}.  For example, the spin of the heavy
$c\bar c$ pair in hadrocharmonium should be a conserved quantum
number, so that if the $Y(4260)$ is a $J^{PC} = 1^{--}$
hadrocharmonium state, it should have a slightly lighter $0^{-+}$
partner, while molecular models based upon single-pion exchange should
have multiplets in which a given spin state has either isospin 0 or 1
(but not both), and static diquark models should produce multiplets
with dozens of states.

From the perspective of heavy-quark physics, the chief difference
between the $c\bar c$ and $b\bar b$ spectra lies in the fact that
$\epsilon_c \simeq 3\epsilon_b$, leading to the open-flavor threshold
$(Q\bar q)(\bar Q q)$ falling in a different location with respect to
the conventional quarkonium states, depending upon the quark flavor.
For example, the $D\bar D$ threshold occurs only slightly above the
$\psi(2S)$, while the $B\bar B$ threshold occurs slightly below the
$\Upsilon(4S)$, a fact used to great effect at the $B$ factory
experiments BaBar and Belle.  The location of exotics with respect to
open-flavor thresholds are similarly expected to be flavor dependent.

\subsubsection{Chiral Unitary Models}

Effective Lagrangians can also be formulated in terms of certain
hadronic degrees of freedom by exploiting the {\it chiral symmetry\/}
of the QCD Lagrangian under transformations $q \to \exp(i\theta
\gamma_5) q$ for massless quarks $q$, which is broken spontaneously by
quantum effects and leads, by means of {\it Goldstone's theorem}, to
the appearance of a multiplet of massless $J^P = 0^-$ mesons.  In the
real world, the masses $m_u$, $m_d$ (and to a lesser extent, $m_s$)
are small but nonzero, leading to the lightness of pions (and to a
lesser extent, $K$ and $\eta$).  An expansion in inverse powers of the
scale $\Lambda_\chi \simeq 1$~GeV of chiral symmetry breaking, or more
accurately, in powers of the typical momenta $p$ of physical hadronic
processes in the combination $\epsilon_\chi \equiv p/\Lambda_\chi$,
leads to the rather successful {\it chiral perturbation theory\/}
($\chi$PT)~\cite{Ecker:1994gg}.

However, $\chi$PT is only valid for $p \ll \Lambda_\chi$.  As $p$
approaches $\Lambda_\chi$, the number of terms in the effective
Lagrangian contributing significantly to the process increases rapidly
(just like the number of terms of a Taylor series needed for the
accurate representation of a function near its radius of convergence),
degrading the predictive power of $\chi$PT.  The key physical
ingredient one can use to extend the range of usefulness of such
calculations is the unitarity of the scattering
matrix~\cite{Oller:2000ma}.

One such unitarization approach, called the Inverse Amplitude Method
(IAM), uses $\chi$PT to fix constants that appear in a dispersion
relation for the physical amplitude (which, by construction, acts as a
Pad{\'e} resummation of the perturbative series and satisfies
unitarity).  The amplitude, when re-expanded, not only reproduces the
low-energy input of $\chi$PT, but can generate nonperturbative
resonant poles as well.  The IAM was first described for elastic
scattering in~\cite{Truong:1988zp} and for coupled-channel systems
in~\cite{Oller:1997ng}.

The other common approach~\cite{Oller:1998zr} uses the $N/D$
method~\cite{Chew:1960iv} and allows one to incorporate explicitly the
existence of known resonant poles.  Here, the numerator $N$ and
denominator $D$ functions for a partial-wave amplitude are separately
defined so as to isolate the contributions of branch cuts in various
regions of the complex momentum plane corresponding to scattering
processes and their crossed channels.  Such an approach is
advantageous in that it allows one to probe whether a given resonance
has an existence independent of its couplings to other hadrons, or
only appears as the dynamical effect of the rescattering of lighter
hadrons.  This distinction is especially interesting for multiquark
exotics, where even the most basic questions of their structure remain
unanswered.

\subsubsection{QCD Sum Rules} \label{sec:QCDsumrules}

The {\it operator product expansion\/} (OPE) of a two-point
correlation Green's function at some momentum transfer $q^2$,
\begin{equation}
\Pi(q^2) \equiv i \int d^4 x \, e^{iqx} \left< 0 \left| T J(x)
J^\dagger(0) \right| 0 \right> \, ,
\end{equation}
for a quark current $J$ of some chosen quantum numbers, forms the
starting point of the {\it QCD sum rule\/}
method~\cite{Shifman:1978bx}.  One writes $\Pi(q)$ in two ways: as a
sum of Wilson coefficients $C_n(q^2)$ times vacuum expectation values
of local operators $\hat O_n$ that are expressed in terms of the
fundamental quark and gluon degrees of freedom (the operator product
side), and as a dispersion integral over the imaginary part of
$\Pi(q)$, which (reminiscent of the optical theorem) can in turn can
be written in terms of the hadronic spectral density function
$\rho(s)$, a function of measurable masses $M$, decay constants $f$,
and continuum-state form-factor contributions for all hadronic states
that can couple to $J$:
\begin{equation}
\Pi(q) = \sum_n C_n \left< \hat O_n \right> = \int ds \,
\frac{\rho(s)}{s-q^2+i\epsilon} \, ,
\end{equation}
where
\begin{equation}
\rho(s) = \sum_n \delta(s-M_n^2) \left< 0
\left| J^{\vphantom\dagger} \right| n \right> \left< n \left|
J^\dagger \right| 0 \right> + \rho_{\rm cont} = \sum_n f_n^2 \delta
(s-M_n^2) + \rho_{\rm cont} \, .
\end{equation}

Choosing a value $s = s_0$ above which continuum and higher resonance
($n>0$) contributions are expected to dominate, and performing an
integral transform on both sides of the equation that gives extra
weight to the lower-energy contributions where the lightest resonance
occurs (a {\it Borel transformation\/} with mass parameter $M^2$), one
obtains a result for the lowest resonance mass $M_0$:
\begin{equation}
M_0^2 = \frac{\int_0^{s_0} ds \, e^{-s/M^2} s \, \rho^{\rm OPE}(s)}
{\int_0^{s_0} ds \, e^{-s/M^2} \rho^{\rm OPE} (s)}
\, .
\end{equation}
The quantity $\rho^{\rm OPE} (s)$ here is the spectral function
computed from the OPE.

A good deal of artistry is needed to achieve successful application of
QCD sum rules.  For example, one can obtain numerically stable results
only by a careful choice of the current $J$, the operators $\hat O_n$
to include, and the values of $s_0$ and $M$.  Early applications of
QCD sum rules to the charmoniumlike exotics are reviewed
in~\cite{Nielsen:2009uh}, and more recent ones in~\cite{Chen:2016qju}.

\subsubsection{Lattice QCD} \label{sec:lattice}

In 1974 Ken Wilson examined the strong-coupling behavior of QCD by
discretizing the theory on a spacetime grid (called a {\it lattice\/}
in the community)~\cite{Wilson:1974sk}.  It was soon realized that
this approach provides a representation of QCD (really, a
regularization) suitable for carrying out Monte Carlo
simulations~\cite{Creutz:1979zg,Creutz:1980zw}, thereby spawning the
discipline of {\it lattice gauge field theory}.  The intervening four
decades have seen tremendous advances in computational abilities, with
commensurate improvements in the quality of lattice calculations.  At
the same time, software and algorithms have progressed from an era
where a single investigator could write a complete simulation in a few
days, to one in which suites of sophisticated code are maintained by
large collaborations.

Lattice gauge field theory computations are typically set up with
scalar and spinor fields on lattice sites, and gauge fields appear on
{\it links}, which connect two spatially separated sites, $\bm{x}$ and
$\bm{x}+ \bm{\mu}$.  Link variables map to the gauge field $A_\mu({\bm
x})$ in the continuum limit, and provide a convenient way to maintain
QCD gauge invariance.  Monte Carlo computations of observables then
amount to executing Markov-chain processes that iteratively
equilibrate to the normalized exponentiated Euclidean action $S_E$,
\[
\frac{\exp(-S_E)}{\int D\phi \exp(-S_E[\phi])} \, .
\]
It is necessary that this factor defines a real probability density,
and thus Grassmann-valued fields (spinors corresponding to dynamical
fermions) must be integrated out explicitly.  This integration gives
rise to a determinant that must be included in the Markov process,
which unfortunately introduces substantial numerical noise in the
computation.  Because of this limitation, early calculations either
were performed in the pure gauge theory, or simply ignored the
determinant (the {\it quenched approximation}).  This impediment has
been overcome in the last few years, and all modern lattice
computations are now performed with dynamical quarks of varying types.

Measuring correlation functions permits the extraction of particle
masses, via  expressions like
\begin{equation}
\langle T \phi(x)\phi(0)\rangle  = \frac{\int D\phi\, \phi(x)\phi(0)
\, \exp(-S_E)}{\int D\phi\, \exp(-S_E)} =
\sum_n |\langle n|\phi(0)|0\rangle |^2\, \exp[-(E_n-E_0) x].
\label{eq:vev}
\end{equation}
Similarly, measuring three-point or higher correlation functions
permits the extraction of hadronic couplings.  The matrix elements in
Eq.~(\ref{eq:vev}) also provide information on the structure of
states, although the values obtained in a simulation must be
interpreted with care, since they depend upon the regularization
scale.

The accomplishments of the field are impressive; among them are the
establishment of color confinement, a precision computation of the
pure-glue spectrum, the computation of the proton mass---along with
the masses of other low-lying mesons and baryons in each flavor
sector---and a convincing demonstration of the independent existence
of hybrid mesons.  Recently, the resonance structure of hadrons has
started to be explored, the extraction of scattering parameters for
simple processes has been achieved, and beyond-Standard Model physics
is being explored.

We remark that hadronic properties that feature ``unnatural'' scales
({\it i.e.}, scales much smaller than the scale $\Lambda_{\rm QCD}
\sim 200$~MeV), such as in nuclear physics ({\it e.g.}, the deuteron
binding energy 2.2~MeV), or the properties of weakly bound exotic
states, remain stubbornly out of reach.

\subsubsection{Born-Oppenheimer Approximation}
\label{sect:bo}

The presence of heavy quarks in many of the exotic hadrons suggests
that the Born-Oppenheimer approximation is a useful tool in the study
of these systems.  This approach was introduced by Born and
Oppenheimer in 1927~\cite{Born:1927boa} in an effort to understand
atomic binding in molecules. The method relies on the large ratio of
electron to nuclear masses to separate their temporal scales. Thus,
electron motion can be considered in the potential created by static
nuclear Coulombic sources. The energy of these systems can then be
traced as a function of the positions of the nuclei, thereby
generating {\it Born-Oppenheimer potentials}. Finally, masses can be
obtained by studying nuclear dynamics in the Born-Oppenheimer
potentials.

\begin{figure}[t]
\begin{center}
\includegraphics[width=8cm,angle=0]{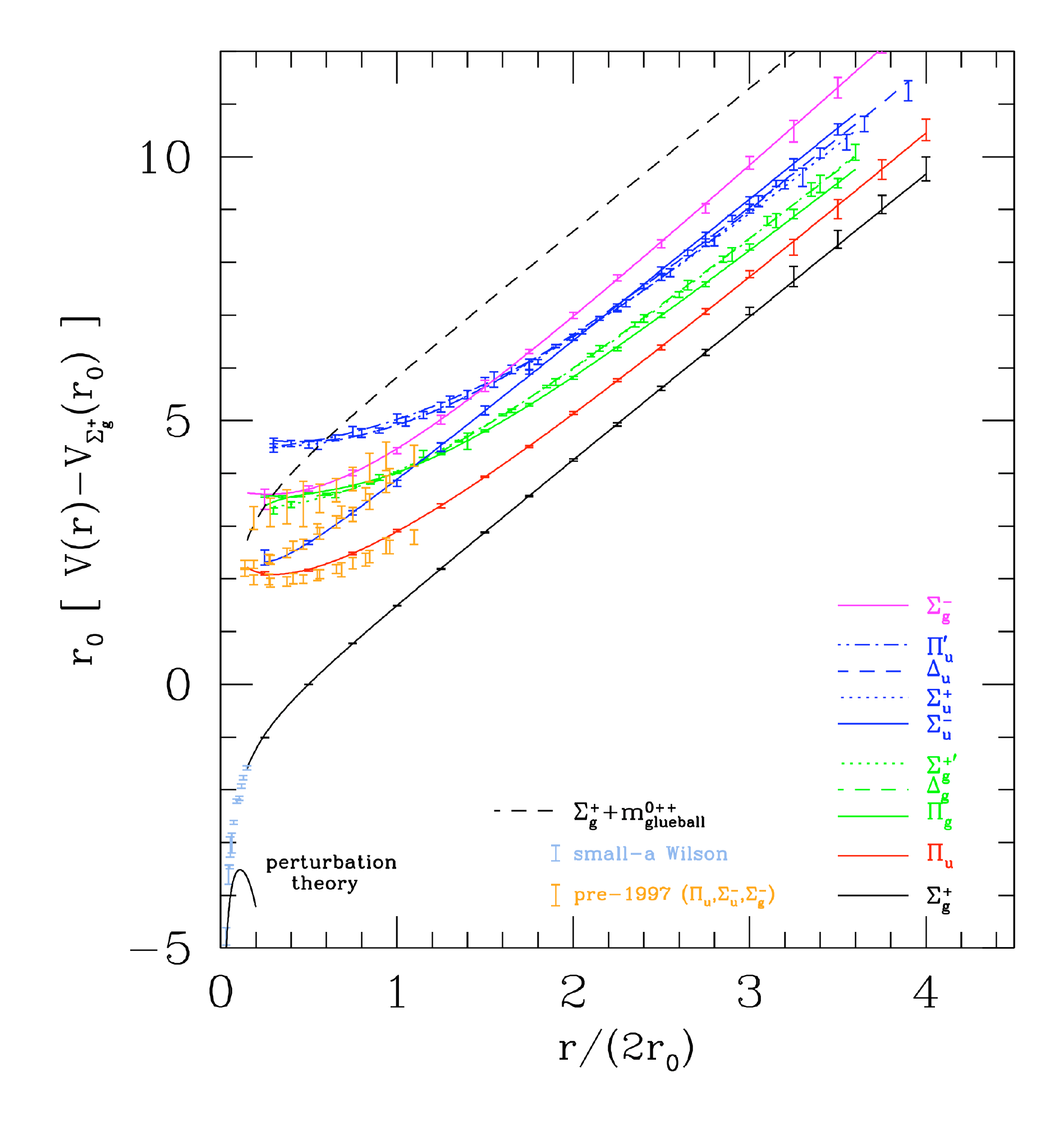}
\end{center}
\caption{Lattice adiabatic hybrid potentials. The curves are labeled
  with diatomic quantum numbers $\Lambda^Y_\eta$, where $\Lambda$ is
  the projection of the gluonic angular momentum on the
  quark-antiquark axis, $Y$ represents parity under reflection through
  a plane containing this axis, and $\eta$ is the product of gluonic
  parity (through the midpoint of the quark-antiquark pair) and charge
  conjugation. The quantity $r_0$ is approximately $0.5$~fm.
  Figure courtesy of C. Morningstar.}
\label{fig:adia}
\end{figure}

The first lattice gauge theory computation of the Born-Oppenheimer
potentials for meson was made in 1983 by Griffiths {\it et
  al.}~\cite{Griffiths:1983ah}. In the static limit, the quark and
antiquark serve as a color source and sink, and the gluonic field
arranges itself into configurations described by the quantum numbers
of diatomic molecules. These potentials were traced in great detail in
the quenched approximation by Juge {\it et al.}~\cite{Juge:1999ie},
and are displayed in Fig.~\ref{fig:adia}.

Once nontrivial quark dynamics are permitted, the resulting mesons are
interpreted as hybrids with their gluonic degrees of freedom in the
appropriate representation. These energies were compared to the
corresponding meson masses in a full lattice calculation by Juge {\it
et al.}, with agreement at the 10\% level. It is thus likely that the
Born-Oppenheimer approximation is a useful guide to properties of
(spin-averaged) heavy-quark hadrons.  Subsequent work has extended the
method to heavy baryons~\cite{Ichie:2002mi,Takahashi:2002bw,
Okiharu:2003vt}, heavy four-quark systems~\cite{Alexandrou:2001ip,
Green:1995df,Pennanen:1998nu}, and pentaquark
systems~\cite{Cardoso:2012rb,Cardoso:2011fq}.  The use of
Born-Oppenheimer techniques for the $XY \! Z$ mesons is discussed in
Ref.~\cite{Braaten:2014qka}.

\section{Experimental Foundations} \label{sec:Expt}

\subsection{Historical Sketch and Overview}
\label{sec:expintro}

\begin{figure}[htb]
\begin{center}
\includegraphics*[width= 1.0\columnwidth]{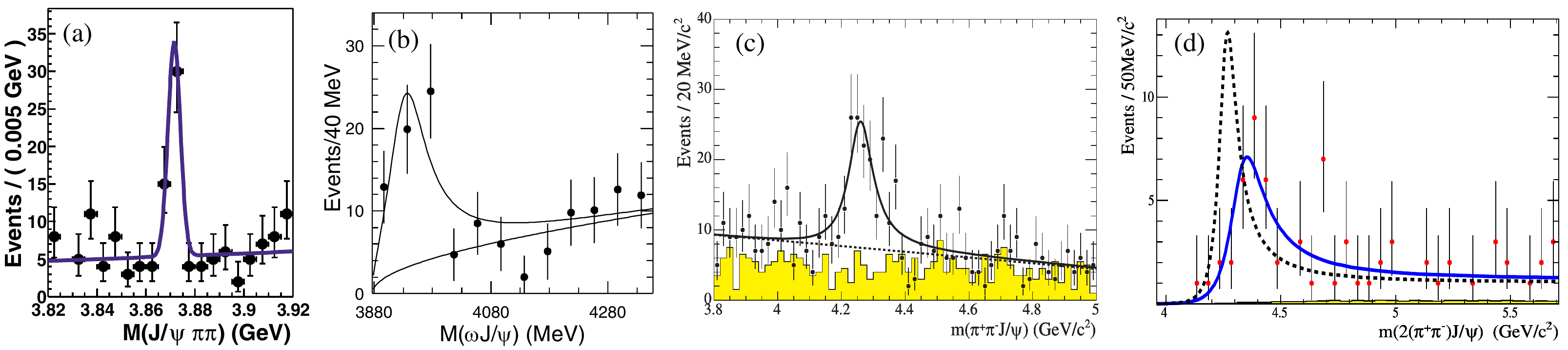}
\end{center}
\caption{\label{fig:discoveries} Earliest observations of the $XY
  \! Z$.  (a)~The $X(3872)$ was discovered in $B\to K X(3872)$ with
  $X(3872)\to\pi^+\pi^-J/\psi$~\cite{Choi:2003ue}.  The
  $\pi^+\pi^-J/\psi$ mass spectrum is shown (from~\cite{Choi:2003ue}). 
  (b)~The $Y(3940)$ was discovered in $B\to
  K Y(3940)$ with $Y(3940)\to\omega J/\psi$~\cite{Abe:2004zs}, as part
  of a search for $X(3872)\to\omega J/\psi$. The $\omega J/\psi$ mass
  spectrum is shown~(from~\cite{Abe:2004zs}). (c)~The
  $Y(4260)$ was discovered in $e^+e^- \to Y(4260)$ with
  $Y(4260)\to\pi^+\pi^-J/\psi$~\cite{Aubert:2005rm}, following a
  search for $e^+e^-\to X(3872)$.  The $\pi^+\pi^-J/\psi$ mass
  spectrum is shown, along with the background estimation from
  $J/\psi$ sidebands~(from~\cite{Aubert:2005rm}).  (d)~The
  $Y(4360)$ was discovered in $e^+e^- \to Y(4360)$ with
  $Y(4360)\to\pi^+\pi^-\psi(2S)$~\cite{Aubert:2007zz}, as part of a
  search for $Y(4260)\to\pi^+\pi^-\psi(2S)$.  The $\pi^+\pi^-\psi(2S)$
  mass spectrum is shown~(from~\cite{Aubert:2007zz}).  The
  solid curve is the $Y(4360)$ and the dotted curve is how the
  $Y(4260)$ would appear if it decayed to $\pi^+\pi^-\psi(2S)$.}
\end{figure}

A new era in the study of QCD exotica began in 2003 with the
accidental discovery of the $X(3872)$.  While studying the process
$B\to K\psi(2S)$ with $\psi(2S)\to\pi^+\pi^-J/\psi$, the Belle
Collaboration noticed a narrow peak in the invariant mass spectrum of
the $\pi^+\pi^-J/\psi$ system higher than the $\psi(2S)$
mass~(Fig.~\ref{fig:discoveries}a)~\cite{Choi:2003ue}.  The peak was
surprisingly narrow and did not correspond to any of the expected
charmonium states from potential models. We remark that, in a not
atypical happenstance, the $X$ had previously been sighted by the E705
Collaboration at Fermilab; however, the significance of the novel peak
was not appreciated at the time~\cite{Antoniazzi:1993jz}.

Immediate efforts to clarify the nature of the $X(3872)$ focused on
searching for other decay modes and other production mechanisms.
However, these searches only led to new experimental discoveries.  It
is this pattern of one unexpected result after another, with the
emergence of desperately few connections, that has characterized the
last 14 years of experimental studies in this field.  A brief
historical sketch of a few of the discoveries between 2003 and 2007
illustrates this rapidly expanding collection of QCD exotica.

(1)~In the initial discovery of the $X(3872)$ with
$X(3872)\to\pi^+\pi^-J/\psi$, it was noticed that the $\pi^+\pi^-$
system appears to originate from a $\rho$.  If so, either the $X(3872)$
is an isovector (which cannot be the case for ordinary charmonium), or
the $X(3872)\to\rho J/\psi$ decay violates isospin.  Assuming it is
the latter, a natural place to search for the $X(3872)$ is in $B\to
K(\omega J/\psi)$, since the $X(3872)\to\omega J/\psi$ decay would
conserve isospin.  This search was quickly performed in 2005 by the
Belle Collaboration, but instead of finding the $X(3872)$, a broader
peak was found at a higher
mass~(Fig.~\ref{fig:discoveries}b)~\cite{Abe:2004zs}.  This peak
became known as the $Y(3940)$.

(2)~Since the quantum numbers of the $X(3872)$ were still unknown
after its discovery, it became important to search for it using
several different production mechanisms.  If the $X(3872)$ had $J^{PC}
= 1^{--}$, it should be produced in $e^+e^-$ annihilation.  The BaBar
Collaboration searched for $e^+e^- \to X(3872) \to \pi^+\pi^-J/\psi$
using {\it Initial State Radiation\/} (ISR), where the initial
$e^+e^-$ beams had center-of-mass energy around 10~GeV, but radiated
photons before colliding, thus allowing the search to cover a wide
range of collision energies.  Rather than finding evidence for the
$X(3872)$~\cite{Aubert:2005eg}, a different peak was discovered in
2005, referred to as the
$Y(4260)$~(Fig.~\ref{fig:discoveries}c)~\cite{Aubert:2005rm}.

(3)~After the discovery of the $Y(4260)$ in $e^+e^-\to Y(4260)\to
\pi^+\pi^-J/\psi$, it was natural to search for the $Y(4260)$ in other
decay modes.  Since the $Y(4260)$ decayed to $\pi^+\pi^-J/\psi$, it is
expected also to decay to $\pi^+\pi^-\psi(2S)$.  However, a 2007
search for the $Y(4260)$ in $e^+e^-\to \pi^+\pi^-\psi(2S)$ by the
BaBar Collaboration, using the same ISR technique used in the
discovery of the $Y(4260)$, did not find the $Y(4260)$, but instead
found a peak at an even higher
mass~(Fig.~\ref{fig:discoveries}d)~\cite{Aubert:2007zz}.  This peak
became known as the $Y(4360)$.

\begin{figure}
\begin{center}
\vspace{-0.5cm}
\includegraphics[width=19.8cm]{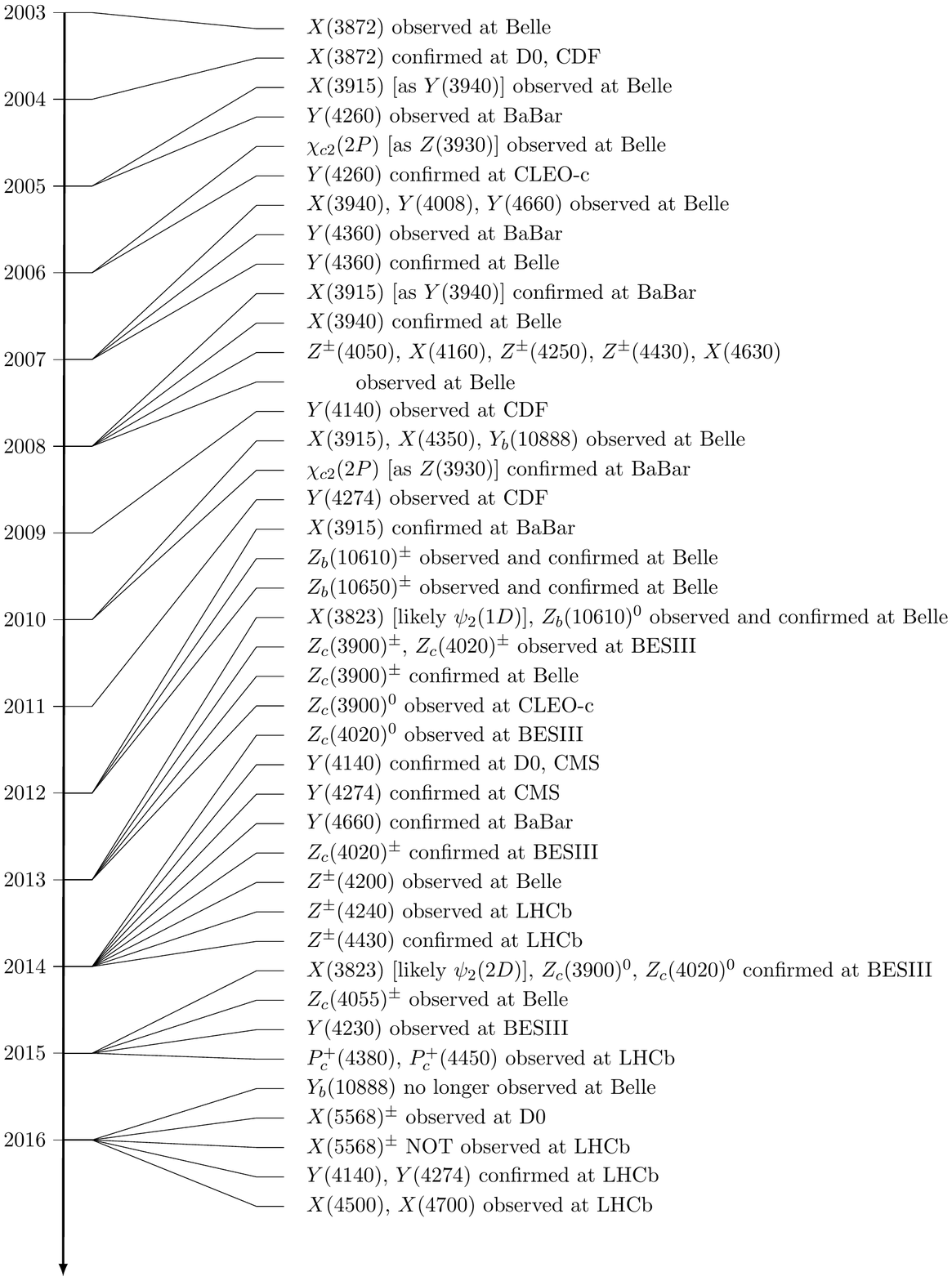}
\end{center}
\vspace{-4.6cm}
\caption{Timeline of discoveries of heavy-quark exotic
  candidates.\label{fig:TIMELINE}}
\end{figure}

Thus, by 2007, the collection of QCD exotica had already grown to a
half-dozen or so.  Attempts to understand the existing peaks had only
led to further peaks and puzzles.  This pattern of discovery, attempts
to clarify, and then new discovery, has largely continued to the
present.  The timeline in Fig.~\ref{fig:TIMELINE} shows a steady
stream of new discoveries.  While a few patterns have emerged, such as
between the $Z_c$ states (in the charmonium region) and the $Z_b$
states (in the bottomonium region) both being observed in $e^+e^-\to
\pi Z_{c,b}$, there are still many states that appear in only one
production mechanism.  For example, it is still not clear why the
$Z_c$ states observed in $e^+e^-\to\pi Z_c$ [such as $Z_c(3900)$]
 and the $Z_c$ states
observed in $B\to KZ_c$ [such as $Z_c(4430)$] are apparently
mutually exclusive.

Because there have been so few connections made between different
production mechanisms, this section is organized by production
mechanism.  Table~\ref{tab:XYZbyProduction} sorts the $XY \! Z$
states according to production mechanism and serves as a loose outline
for the following discussions.  For reference,
Table~\ref{tab:XYZbyMass} also lists the $XY \! Z$ states organized
(roughly) by mass.  A glossary of all observed exotic
states~(Appendix~\ref{app:states}) also serves as a reference.

As a final note, all of the results covered in the following are
experimentally robust, unless otherwise stated.

\begin{table}[p]
  \caption{\label{tab:XYZbyProduction} Exotica organized by the way
    they are produced.  References are given in the decay column.}
\centering
\begin{tabular}{|c|c|c|c|}
\hline
    Process
        & Production
            & Decay
                & Particle \\
\hline
    \multirow{20}{*}{$B$ and $\Lambda_b$ Decays}
        & \multirow{11}{*}{$B \to K + X$}
            & $X \to \pi^+\pi^-J/\psi$~\cite{Choi:2003ue,Aaij:2015eva,Choi:2011fc,Aaij:2013zoa,Aubert:2008gu,Aubert:2004ns,Aubert:2005zh}
                & \multirow{4}{*}{$X(3872)$}
                    \\
        &
            & $X \to D^{*0}\bar{D}^0$~\cite{Gokhroo:2006bt,Aubert:2007rva,Adachi:2008sua}
                &
                    \\
        &
            & $X \to \gamma J/\psi$~\cite{Aaij:2014ala,Aubert:2006aj,Aubert:2008ae,Bhardwaj:2011dj}
                &
                    \\
        &
            & $X \to \gamma \psi(2S)$~\cite{Aaij:2014ala,Aubert:2008ae}
                &
                    \\ \cline{3-4}
        &   
            & \multirow{2}{*}{$X \to \omega J/\psi$~\cite{Abe:2004zs,delAmoSanchez:2010jr,Aubert:2007vj}}
                & $X(3872)$
                    \\
        &   
            &
                & $Y(3940)$
                    \\ \cline{3-4}
        &   
            & $X \to \gamma \chi_{c1}$~\cite{Bhardwaj:2013rmw}
                & $X(3823)$
                    \\ \cline{3-4}
        &   
            & \multirow{4}{*}{$X \to \phi J/\psi$~\cite{Aaij:2016iza,Aaltonen:2009tz,Aaij:2012pz,Lees:2014lra,Abazov:2013xda,Chatrchyan:2013dma,Aaij:2016nsc,Aaltonen:2011at}}
                & $Y(4140)$
                    \\
        &   
            &
                & $Y(4274)$
                    \\ 
        &   
            &
                & $X(4500)$
                    \\ 
        &   
            &
                & $X(4700)$
                    \\ \cline{2-4}
        & \multirow{6}{*}{$B \to K + Z$}
            & \multirow{2}{*}{$Z \to \pi^\pm \chi_{c1}$~\cite{Mizuk:2008me,Lees:2011ik}}
                & $Z_1(4050)$
                    \\
        &   
            &
                & $Z_2(4250)$
                    \\ \cline{3-4}
        &
            & \multirow{2}{*}{$Z \to \pi^\pm J/\psi$~\cite{Chilikin:2014bkk,Aubert:2008aa}}
                & $Z_c(4200)$
                    \\
        &   
            &
                & $Z_c(4430)$
                    \\ \cline{3-4}
        &
            & \multirow{2}{*}{$Z \to \pi^\pm \psi(2S)$~\cite{Choi:2007wga,Aubert:2008aa,Mizuk:2009da,Chilikin:2013tch,Aaij:2014jqa,Aaij:2015zxa}}
                & $Z_c(4240)$
                    \\
        &   
            &
                & $Z_c(4430)$
                    \\ \cline{2-4}
        & \multirow{1}{*}{$B \to K\pi + X$}
            & $X \to \pi^+\pi^-J/\psi$~\cite{Bala:2015wep}
                & $X(3872)$
                    \\ \cline{2-4}
        & \multirow{2}{*}{$\Lambda_b \to K + P_c$}
            & \multirow{2}{*}{$P_c \to p J/\psi$~\cite{Aaij:2015tga}}
                & $P_c(4380)$
                    \\
        &
            &
                & $P_c(4450)$
                    \\ \hline
    \multirow{21}{*}{$e^+e^-$ Annihilation}
        & \multirow{8}{*}{$e^+e^- \to Y$}
            & \multirow{2}{*}{$Y \to \pi \pi J/\psi$~\cite{Liu:2013dau,Aubert:2005rm,He:2006kg,Coan:2006rv,Yuan:2007sj,Lees:2012cn,Ablikim:2016qzw}}
                & $Y(4008)$
                    \\
        &
            &
                & $Y(4260)$
                    \\ \cline{3-4}
        &
            & \multirow{2}{*}{$Y \to \pi \pi \psi(2S)$~\cite{Aubert:2007zz,Wang:2007ea,Lees:2012pv,Wang:2014hta}}
                & $Y(4360)$
                    \\
        &
            &
                & $Y(4660)$
                    \\ \cline{3-4}
        &
            & $Y \to \omega \chi_{c0}$~\cite{Ablikim:2014qwy}
                & $Y(4230)$
                    \\ \cline{3-4} \rule{0pt}{2.5ex}
        &
            &  $Y \to \Lambda_c \bar{\Lambda}_c$~\cite{Pakhlova:2008vn}
                & $X(4630)$
                    \\ \cline{3-4}
        &
            & $Y \to \pi \pi \Upsilon(1S,2S,3S)$~\cite{Chen:2008xia,Santel:2015qga}
                & \multirow{2}{*}{$Y_b(10888)$}
                    \\
        &
            & $Y \to \pi \pi h_b(1P,2P)$~\cite{Abdesselam:2015zza}
                &
                    \\ \cline{2-4}
        & \multirow{9}{*}{$e^+e^- \to \pi + Z$}
            & $Z \to \pi J/\psi$~\cite{Ablikim:2013mio,Liu:2013dau,Xiao:2013iha,Ablikim:2015tbp}
                & \multirow{2}{*}{$Z_c(3900)$}
                    \\
        &
            & $Z \to D^* \bar D$~\cite{Ablikim:2015gda,Ablikim:2013xfr,Ablikim:2015swa}
                &
                    \\ \cline{3-4}
        &
            & $Z \to \pi h_c$~\cite{Ablikim:2013wzq,Ablikim:2014dxl}
                & \multirow{2}{*}{$Z_c(4020)$}
                    \\
        &
            & $Z \to D^* {\bar D}^*$~\cite{Ablikim:2013emm,Ablikim:2015vvn}
                &
                    \\ \cline{3-4}
        &
            & $Z \to \pi^\pm \psi(2S)$~\cite{Wang:2014hta}
                & $Z_c(4055)$
                    \\ \cline{3-4}
        &
            & $Z \to \pi \Upsilon(1S,2S,3S)$~\cite{Belle:2011aa,Krokovny:2013mgx,Garmash:2014dhx}
                & $Z_b(10610)$
                    \\
        &
            & $Z \to \pi h_b(1P,2P)$~\cite{Belle:2011aa}
                & $Z_b(10650)$
                    \\ \cline{3-4} \rule{0pt}{2.5ex}
        &
            & $Z \to B {\bar B}^*$~\cite{Garmash:2015rfd}
                & $Z_b(10610)$
                    \\ \cline{3-4} \rule{0pt}{2.5ex}
        &
            & $Z \to B^* {\bar B}^*$~\cite{Garmash:2015rfd}
                & $Z_b(10650)$
                    \\ \cline{2-4}
        & \multirow{1}{*}{$e^+e^- \to \gamma + X$}
            & $X \to \pi^+\pi^-J/\psi$~\cite{Ablikim:2013dyn}
                & $X(3872)$
                    \\ \cline{2-4}
        & \multirow{1}{*}{$e^+e^- \to \pi^+\pi^- + X$}
            & $X \to \gamma \chi_{c1}$~\cite{Ablikim:2015dlj}
                & $X(3823)$
                    \\ \cline{2-4} \rule{0pt}{2.5ex}
        & \multirow{2}{*}{$e^+e^- \to J/\psi + X$}
            & $X \to D {\bar D}^*$~\cite{Abe:2007sya,Abe:2007jna}
                & $X(3940)$
                    \\ \cline{3-4} \rule{0pt}{2.5ex}
        &
            & $X \to D^* {\bar D}^*$~\cite{Abe:2007sya}
                & $X(4160)$
                    \\ \hline
    \multirow{3}{*}{$\gamma\gamma$ Collisions}
        & \multirow{3}{*}{$\gamma\gamma \to X$}
            & $X \to \omega J/\psi$~\cite{Uehara:2009tx,Lees:2012xs}
                & $X(3915)$
                    \\ \cline{3-4} \rule{0pt}{2.5ex}
        &
            & $X \to D{\bar D}$~\cite{Uehara:2005qd,Aubert:2010ab}
                & $Z(3930)$
                    \\ \cline{3-4}
        &
            & $X \to \phi J/\psi$~\cite{Shen:2009vs}
                & $X(4350)$
                    \\ \hline
    \multirow{3}{*}{Hadron Collisions}
        & \multirow{3}{*}{$pp$ or $p\bar{p}\to X + $ anything}
            & $X \to \pi^+\pi^-J/\psi$~\cite{Acosta:2003zx,Abazov:2004kp,Aaltonen:2009vj,Aaij:2011sn}
                & $X(3872)$
                    \\ \cline{3-4}
        &
            & $X \to \phi J/\psi$~\cite{Abazov:2015sxa}
                & $Y(4140)$
                    \\ \cline{3-4}
        &
            & $X \to B_s \pi^\pm$~\cite{D0:2016mwd}
                & $X(5568)$
                    \\ \hline
\end{tabular}
\end{table}

\begin{table}[p]
  \caption{\label{tab:XYZbyMass} Candidates for QCD exotica roughly 
    organized by mass.  Quantum numbers that have not been measured,
    but are assumed, are listed in parentheses.  Unknown quantum
    numbers are left blank or are indicated with a question mark.
    References for mass and width values are given in the mass column.
    When only a single value has been measured or there is one
    dominant measurement, the value from the original reference is
    used.  Otherwise, we quote the PDG average.  References for the
    production processes and decay modes are given in
    Table~\ref{tab:XYZbyProduction}.}
\centering
\begin{tabular}{|c|c|c|c|c|}
\hline
    Particle
        & $I^{G}J^{PC}$
            & Mass [MeV]
                & Width [MeV]
                    & Production and Decay \\ \hline
    \multirow{2}{*}{$X(3823)$ ($\psi_2(1D)$)}
        & \multirow{2}{*}{$(0^-2^{--})$}
            & \multirow{2}{*}{$3822.2 \pm 1.2$~\cite{Olive:2016xmw}}
                & \multirow{2}{*}{$<16$}
                    & $B \to K X$; $X \to \gamma \chi_{c1}$ \\
        &
            &
                &
                    & $e^+e^- \to \pi^+ \pi^- X$; $X \to \gamma \chi_{c1}$  \\ \hline
    \multirow{7}{*}{$X(3872)$}
        & \multirow{7}{*}{$0^+1^{++}$}
            & \multirow{7}{*}{$3871.69 \pm 0.17$~\cite{Olive:2016xmw}}
                & \multirow{7}{*}{$<1.2$}
                    & $B \to K X$; $X \to \pi^+ \pi^- J/\psi$ \\
        &
            &
                &
                    & $B \to K X$; $X \to D^{*0}\bar{D}^0$ \\
        &
            &
                &
                    & $B \to K X$; $X \to \gamma J/\psi, \gamma \psi(2S)$ \\
       &
           &
               &
                   & $B \to K X$; $X \to \omega J/\psi$ \\ 
       &
           &
               &
                   & $B \to K \pi X$; $X \to \pi^+ \pi^- J/\psi$ \\
       &
           &
               &
                   & $e^+e^- \to \gamma X$; $X \to \pi^+ \pi^- J/\psi$ \\
       &
           &
               &
                   & $pp$ or $p\bar{p} \to X$ + any.; $X \to \pi^+ \pi^- J/\psi$ \\ \hline
    \multirow{2}{*}{$Z_c(3900)$}
        & \multirow{2}{*}{$1^+1^{+-}$}
            & \multirow{2}{*}{$3886.6 \pm 2.4$~\cite{Olive:2016xmw}}
                & \multirow{2}{*}{$28.1 \pm 2.6$}
                    & $e^+e^- \to \pi Z$; $Z \to \pi J/\psi$\\
       &
           &
               &
                   & $e^+e^- \to \pi Z$; $Z \to D^* \bar D$ \\ \hline
    $X(3915)$
        & \multirow{2}{*}{$0^+0^{++}$}
            & \multirow{2}{*}{$3918.4 \pm 1.9$~\cite{Olive:2016xmw}}
                & \multirow{2}{*}{$20 \pm 5$}
                    & $\gamma\gamma \to X$; $X\to \omega J/\psi$\\ \cline{1-1} \cline{5-5}
    $Y(3940)$
        &
            &
                &
                    & $B \to KX$; $X\to \omega J/\psi$ \\ \hline
    $Z(3930)$ ($\chi_{c2}(2P)$)
        & $0^+2^{++}$
            & $3927.2 \pm 2.6$~\cite{Olive:2016xmw}
                & $24 \pm 6$
                    & $\gamma\gamma \to Z$; $Z\to D\bar{D}$\\ \hline
    $X(3940)$
        &
            & $3942^{+7}_{-6}\pm 6$~\cite{Abe:2007sya}
                & $37^{+26}_{-15}\pm 8$
                    & $e^+e^- \to J/\psi + X$; $X \to D {\bar D}^*$ \\ \hline
    $Y(4008)$
        & $1^{--}$
            & $3891 \pm 41 \pm 12$~\cite{Liu:2013dau}
                & $255 \pm 40 \pm 14$
                    & $e^+e^- \to Y$; $Y \to \pi^+ \pi^- J/\psi$ \\ \hline
    \multirow{2}{*}{$Z_c(4020)$}
        & \multirow{2}{*}{$1^+?^{?-}$}
            & \multirow{2}{*}{$4024.1 \pm 1.9$~\cite{Olive:2016xmw}}
                & \multirow{2}{*}{$13 \pm 5$}
                    & $e^+e^- \to \pi Z$; $Z \to \pi h_c$\\
       &
           &
               &
                   & $e^+e^- \to \pi Z$; $Z \to D^* \bar {D}^*$ \\ \hline
    $Z_1(4050)$
        & $1^-?^{?+}$
            & $4051\pm 14^{+20}_{-41}$~\cite{Mizuk:2008me}
                & $82^{+21+47}_{-17-22}$
                    & $B \to KZ$; $Z \to \pi^\pm \chi_{c1}$\\ \hline
    $Z_c(4055)$
        & $1^+?^{?-}$
            & $4054 \pm 3 \pm 1$~\cite{Wang:2014hta}
                & $ 45 \pm 11 \pm 6$
                    & $e^+e^- \to \pi^\mp Z$; $Z \to \pi^\pm \psi(2S)$ \\ \hline
    \multirow{2}{*}{$Y(4140)$}
        & \multirow{2}{*}{$0^+1^{++}$}
            & \multirow{2}{*}{$4146.5 \pm 4.5^{+4.6}_{-2.8}$~\cite{Aaij:2016iza}}
                & \multirow{2}{*}{$83 \pm 21^{+21}_{-14}$}
                    & $B \to K Y$; $Y \to \phi J/\psi$ \\ 
        &
            &
                &
                    & $pp$ or $p\bar{p} \to Y$ + any.; $Y \to \phi J/\psi$ \\ \hline
    $X(4160)$
        &
            & $4156^{+25}_{-20}\pm 15$~\cite{Abe:2007sya}
                & $139^{+111}_{-61}\pm 21$
                    & $e^+e^- \to J/\psi + X$; $X \to D^* {\bar D}^*$ \\ \hline
    $Z_c(4200)$
        & $1^+1^{+-}$
            & $4196^{+31+17}_{-29-13}$~\cite{Chilikin:2014bkk}
                & $370^{+70+70}_{-70-132}$
                    & $B \to KZ$; $Z \to \pi^\pm J/\psi$ \\ \hline
    $Y(4230)$
        & $0^-1^{--}$
            & $4230 \pm 8 \pm 6$~\cite{Ablikim:2014qwy}
                & $38 \pm 12 \pm 2$
                    & $e^+e^- \to Y$; $Y \to \omega \chi_{c0}$ \\ \hline
    $Z_c(4240)$
        & $1^+0^{--}$
            & $4239 \pm 18^{+45}_{-10}$~\cite{Aaij:2014jqa}
                & $220 \pm 47^{+108}_{-74}$
                    & $B\to KZ$; $Z \to \pi^\pm \psi(2S)$ \\ \hline
    $Z_2(4250)$
        & $1^-?^{?+}$
            & $4248^{+44+180}_{-29-35}$~\cite{Mizuk:2008me}
                & $177^{+54+316}_{-39-61}$
                    & $B \to KZ$; $Z \to \pi^\pm \chi_{c1}$ \\ \hline
    $Y(4260)$
        & $0^-1^{--}$
            & $4251 \pm 9$~\cite{Olive:2016xmw}
                & $120 \pm 12$
                    & $e^+e^- \to Y$; $Y \to \pi \pi J/\psi$ \\ \hline
    $Y(4274)$
        & $0^+1^{++}$
            & $4273.3 \pm 8.3^{+17.2}_{-3.6}$~\cite{Aaij:2016iza}
                & $52 \pm 11^{+8}_{-11}$
                    & $B \to K Y$; $Y \to \phi J/\psi$ \\ \hline
    $X(4350)$
        & $0^+?^{?+}$
            & $4350.6 ^{+4.6}_{-5.1} \pm 0.7$~\cite{Shen:2009vs}
                & $ 13^{+18}_{-9} \pm 4$
                    & $\gamma\gamma \to X$; $X\to \phi J/\psi$ \\ \hline
    $Y(4360)$
        & $1^{--}$
            & $4346 \pm 6$~\cite{Olive:2016xmw}
                & $102 \pm 10$
                    & $e^+e^- \to Y$; $Y \to \pi^+ \pi^- \psi(2S)$ \\ \hline
    \multirow{2}{*}{$Z_c(4430)$}
        & \multirow{2}{*}{$1^+1^{+-}$}
            & \multirow{2}{*}{$4478^{+15}_{-18}$~\cite{Olive:2016xmw}}
                & \multirow{2}{*}{$181 \pm 31$}
                    & $B\to KZ$; $Z \to \pi^\pm J/\psi$ \\ 
        &
            &
                &
                    & $B\to KZ$; $Z \to \pi^\pm \psi(2S)$ \\ \hline
    $X(4500)$
        & $0^+0^{++}$
            & $4506 \pm 11^{+12}_{-15}$~\cite{Aaij:2016iza}
                & $92 \pm 21^{+21}_{-20}$
                    & $B \to K X$; $X \to \phi J/\psi$ \\ \hline
    $X(4630)$
        & $1^{--}$
            & $4634^{+8+5}_{-7-8}$~\cite{Pakhlova:2008vn}
                & $92^{+40+10}_{-24-21}$
                    & $e^+e^- \to X$; $X \to \Lambda_c \bar{\Lambda}_c$ \\ \hline
    $Y(4660)$
        & $1^{--}$
            & $4643 \pm 9$~\cite{Olive:2016xmw}
                & $72 \pm 11$
                    & $e^+e^- \to Y$; $Y \to \pi^+ \pi^- \psi(2S)$ \\ \hline
    $X(4700)$
        & $0^+0^{++}$
            & $4704 \pm 10^{+14}_{-24}$~\cite{Aaij:2016iza}
                & $120 \pm 31^{+42}_{-33}$
                    & $B \to K X$; $X \to \phi J/\psi$ \\ \hline
    $P_c(4380)$
        &
            & $4380 \pm 8 \pm 29$~\cite{Aaij:2015tga}
                & $205 \pm 18 \pm 86$
                    & $\Lambda_b \to K P_c$; $P_c \to p J/\psi$ \\ \hline
    $P_c(4450)$
        &
            & $4449.8 \pm 1.7 \pm 2.5$~\cite{Aaij:2015tga}
                & $39 \pm 5 \pm 19$
                    & $\Lambda_b \to K P_c$; $P_c \to p J/\psi$ \\ \hline
    $X(5568)$
        &
            & $5567.8 \pm 2.9^{+0.9}_{-1.9}$~\cite{D0:2016mwd}
                & $21.9 \pm 6.4^{+5.0}_{-2.5}$
                    & $p\bar{p} \to X$ + anything; $X \to B_s \pi^\pm$ \\ \hline
    \multirow{3}{*}{$Z_b(10610)$}
        & \multirow{3}{*}{$1^+1^{+-}$}
            & \multirow{3}{*}{$10607.2 \pm 2.0$~\cite{Olive:2016xmw}}
                & \multirow{3}{*}{$18.4 \pm 2.4$}
                    & $e^+e^- \to \pi Z$; $Z \to \pi \Upsilon(1S,2S,3S)$ \\
        &
            &
                &
                    & $e^+e^- \to \pi Z$; $Z \to \pi h_b(1P,2P)$ \\
        &
            &
                &
                    & $e^+e^- \to \pi Z$; $Z \to B {\bar B}^*$ \\ \hline
    \multirow{3}{*}{$Z_b(10650)$}
        & \multirow{3}{*}{$1^+1^{+-}$}
            & \multirow{3}{*}{$10652.2\pm 1.5$~\cite{Olive:2016xmw}}
                & \multirow{3}{*}{$11.5\pm 2.2$}
                    & $e^+e^- \to \pi Z$; $Z \to \pi \Upsilon(1S,2S,3S)$ \\
        &
            &
                &
                    & $e^+e^- \to \pi Z$; $Z \to \pi h_b(1P,2P)$ \\
        &
            &
                &
                    & $e^+e^- \to \pi Z$; $Z \to B^* {\bar B}^*$ \\ \hline
    \multirow{2}{*}{$Y_b(10888)$}
        & \multirow{2}{*}{$0^-1^{--}$}
            & \multirow{2}{*}{$10891\pm 4$~\cite{Olive:2016xmw}}
                & \multirow{2}{*}{$54\pm 7$}
                    & $e^+e^- \to Y$; $Y \to \pi\pi \Upsilon(1S,2S,3S)$ \\
        &
            &
                &
                    & $e^+e^- \to Y$; $Y \to \pi\pi h_b(1P,2P)$ \\ \hline
\end{tabular}
\end{table}

\subsection{Experiments and Production Mechanisms}

Before describing the individual candidates for QCD exotica, it is
useful to survey a few of the general features of the experimental
mechanisms used to produce them.  Two of these production mechanisms,
weak decays of the $B$ and $\Lambda_b$ and $e^+e^-$ annihilation, have
proven to be particularly rich in new phenomena.  The experiments
using each technique are listed in each of the following sections, but
more detailed information on the experimental collaborations driving
this field is given in Table~\ref{tab:Experiments}.

\begin{table}[p]
  \caption{Major experiments in the past, present, and future of
    heavy-quark exotics studies.\label{tab:Experiments}}
\centering
\begin{tabular}{|c|c|c|c|c|c|}
\hline
    Experiment
        & Highlights
            & Accelerator
                & Years
                    & Institute
                        & Production  \\ \hline
    \multirow{4}{*}{BaBar}
        & \multirow{3}{*}{$Y(4260)$~\cite{Aubert:2005rm}} 
            & \multirow{4}{*}{PEP-II}
                & \multirow{3}{*}{1999--}
                    & \multirow{1}{*}{SLAC}
                        & \multirow{5}{*}{$e^+e^-$ annihilation} \\
        & \multirow{3}{*}{$Y(4360)$~\cite{Aubert:2007zz}}
            &
                & \multirow{3}{*}{2008 }
                    & \multirow{1}{*}{(Menlo Park,}
                        & \multirow{5}{*}{($E_{\rm CM}\approx$ 10~GeV):} \\
        &
            &
                &
                    & \multirow{1}{*}{California,}
                        &  \\
        &
            &
                &
                    & \multirow{1}{*}{USA)}
                        & \multirow{5}{*}{$e^+e^- \to B\bar{B}$; $B\to K X$} \\  \cline{1-5}
    \multirow{7}{*}{Belle}
        & \multirow{1}{*}{$X(3872)$~\cite{Choi:2003ue}}
            & \multirow{7}{*}{KEKB}
                & \multirow{6}{*}{1998--}
                    & \multirow{8}{*}{KEK}
                        & \multirow{5}{*}{$e^+e^- \to Y_b$} \\
        & \multirow{1}{*}{$Y(3940)$~\cite{Abe:2004zs}}
            &
                & \multirow{6}{*}{2010 }
                    & \multirow{8}{*}{(Tsukuba,}
                        & \multirow{5}{*}{$e^+e^- \to \pi Z_b$} \\
        & \multirow{1}{*}{$X(3915)$~\cite{Uehara:2009tx}} 
            &
                &
                    & \multirow{8}{*}{Japan)}
                        & \multirow{5}{*}{$e^+e^-(\gamma_{\rm ISR}) \to Y$} \\ 
        & \multirow{1}{*}{$Z_c(4430)$~\cite{Choi:2007wga,Mizuk:2009da,Chilikin:2013tch}} 
            & 
                &
                    & 
                        & \multirow{5}{*}{$e^+e^-(\gamma_{\rm ISR}) \to \pi Z_c$} \\
        & \multirow{1}{*}{$Z_b(10610)$,} 
            & 
                &
                    & 
                        & \multirow{5}{*}{$e^+e^- \to J/\psi + X$} \\
        & \multirow{1}{*}{$Z_b(10650)$~\cite{Belle:2011aa,Garmash:2014dhx,Garmash:2015rfd}} 
            & 
                &
                    & 
                        & \multirow{5}{*}{$\gamma \gamma \to X$} \\
        & \multirow{1}{*}{$Y_b(10888)$~\cite{Chen:2008xia,Santel:2015qga}} 
            & 
                &
                    & 
                        &  \\ \cline{1-4}
    \multirow{3}{*}{Belle II}
        & \multirow{1}{*}{Upcoming}
            & \multirow{3}{*}{SuperKEKB}
                & \multirow{3}{*}{2018--}
                    &
                        & \\
        & \multirow{1}{*}{continuation of}
            &
                &
                    &
                        & \\
        & \multirow{1}{*}{Belle}
            &
                &
                    &
                        & \\ \hline
    \multirow{4}{*}{CLEO-c}
        & \multirow{3}{*}{$Y(4260)$~\cite{Coan:2006rv}}
            & \multirow{4}{*}{CESR-c}
                & \multirow{3}{*}{2003--}
                    & \multirow{1}{*}{Cornell U.}
                        & \multirow{3}{*}{$e^+e^-$ annihilation} \\
        & \multirow{3}{*}{$\pi^+\pi^-h_c$~\cite{CLEO:2011aa}}
            &
                & \multirow{3}{*}{2008 }
                    & \multirow{1}{*}{(Ithaca,}
                        & \multirow{3}{*}{($E_{\rm CM}\approx$ 4~GeV):} \\
        &
            &
                &
                    & \multirow{1}{*}{New York,}
                        & \\
        &
            &
                &
                    & \multirow{1}{*}{USA)}
                        & \multirow{3}{*}{$e^+e^- \to Y$} \\ \cline{1-5}
    \multirow{4}{*}{BESIII}
        & \multirow{1}{*}{$Z_c(3900)$~\cite{Ablikim:2013mio,Ablikim:2013xfr}}
            & \multirow{4}{*}{BEPCII}
                & \multirow{4}{*}{2008--}
                    & \multirow{2}{*}{IHEP}
                        & \multirow{3}{*}{$e^+e^- \to \pi Z$} \\
        & \multirow{1}{*}{$Z_c(4020)$~\cite{Ablikim:2013wzq,Ablikim:2013emm}}
            &
                &
                    & \multirow{2}{*}{(Beijing,}
                        & \multirow{3}{*}{$e^+e^- \to \gamma X$} \\
        & \multirow{1}{*}{$Y(4230)$~\cite{Ablikim:2014qwy}}
            &
                &
                    & \multirow{2}{*}{China)}
                        & \\
        & \multirow{1}{*}{$X(3872)$~\cite{Ablikim:2013dyn}}
            &
                &
                    &
                        & \\ \hline
    \multirow{3}{*}{CDF}
        & \multirow{1}{*}{$Y(4140)$~\cite{Aaltonen:2009tz}}
            & \multirow{6}{*}{Tevatron}
                & \multirow{5}{*}{1985--}
                    & \multirow{3}{*}{Fermilab}
                        & \multirow{2}{*}{$p\bar{p}$ collisions} \\
        & \multirow{1}{*}{$Y(4274)$~\cite{Aaltonen:2011at}}
            &
                & \multirow{5}{*}{2011 }
                    & \multirow{3}{*}{(Batavia,}
                        & \multirow{2}{*}{($E_{\rm CM}\approx$ 2~TeV):} \\
        & \multirow{1}{*}{$X(3872)$~\cite{Abulencia:2005zc,Abulencia:2006ma,Aaltonen:2009vj}}
            &
                &
                    & \multirow{3}{*}{Illinois,}
                        & \\ \cline{1-2}
    \multirow{3}{*}{D0}
        & \multirow{1}{*}{$X(3872)$~\cite{Abazov:2004kp}}
            &
                & 
                    & \multirow{3}{*}{USA)}
                        & \multirow{2}{*}{$p\bar{p}\to X$ + any} \\
        & \multirow{1}{*}{$Y(4140)$~\cite{Abazov:2015sxa}}
            &
                & 
                    &
                        & \multirow{2}{*}{$p\bar{p}\to B$ + any; $B\to KX$} \\
        & \multirow{1}{*}{$X(5568)$~\cite{D0:2016mwd}}
            &
                & 
                    &
                        & \\  \hline
    \multirow{3}{*}{ATLAS}
        & \multirow{3}{*}{$\chi_b(3P)$~\cite{Aad:2011ih}} 
            & \multirow{12}{*}{LHC}
                & \multirow{12}{*}{2010--}
                    & \multirow{15}{*}{CERN}
                        & \multirow{7}{*}{$pp$ collisions} \\
        &
            &
                &
                    & \multirow{15}{*}{(Geneva,}
                        & \multirow{7}{*}{($E_{\rm CM}=$ 7, 8, 13~TeV):} \\
        &
            &
                &
                    & \multirow{15}{*}{Switzerland)}
                        & \\ \cline{1-2}
    \multirow{3}{*}{CMS}
        & \multirow{1}{*}{$X(3872)$~\cite{Chatrchyan:2013cld}} 
            &
                & 
                    &
                        & \multirow{7}{*}{$pp\to X$ + any} \\
        & \multirow{1}{*}{$Y(4140)$,} 
            &
                & 
                    &
                        & \multirow{7}{*}{$pp\to B$ + any; $B\to KX$} \\
        & \multirow{1}{*}{$Y(4274)$~\cite{Chatrchyan:2013dma}} 
            &
                & 
                    &
                        & \multirow{7}{*}{$pp\to \Lambda_b$ + any; $\Lambda_b\to KP_c$} \\  \cline{1-2}
    \multirow{6}{*}{LHCb}
        & \multirow{1}{*}{$Z_c(4430)$~\cite{Aaij:2014jqa,Aaij:2015zxa}} 
            &
                & 
                    &
                        & \\
        & \multirow{1}{*}{$X(3872)$~\cite{Aaij:2015eva}} 
            &
                & 
                    &
                        & \\
        & \multirow{1}{*}{$P_c(4380)$,} 
            &
                & 
                    &
                        & \\
        & \multirow{1}{*}{$P_c(4450)$~\cite{Aaij:2015tga}} 
            &
                & 
                    &
                        & \\
        & \multirow{1}{*}{$Y(4140)$,} 
            &
                & 
                    &
                        & \\
        & \multirow{1}{*}{$Y(4274)$~\cite{Aaij:2016iza,Aaij:2016nsc}} 
            &
                & 
                    &
                        & \\ \cline{1-4} \cline{6-6}
    \multirow{5}{*}{COMPASS}
        & \multirow{4}{*}{photoproduction~\cite{Adolph:2014hba}} 
            & \multirow{5}{*}{SPS}
                & \multirow{5}{*}{2002-2011}
                    &
                        & \multirow{1}{*}{$\mu$/$\pi$ beam on $N$ target} \\
        & \multirow{4}{*}{$a_1(1420)$~\cite{Adolph:2015pws}} 
            &
                &
                    &
                        & \multirow{1}{*}{($p_{beam}\approx$ 160, 200~GeV)} \\
        &
            &
                &
                    &
                        & \multirow{1}{*}{} \\
        &
            &
                &
                    &
                        & \multirow{1}{*}{$\pi N \to X N$} \\
        &
            &
                &
                    &
                        & \multirow{1}{*}{$\gamma N \to X N$} \\ \hline
    \multirow{5}{*}{${\bar{\rm P}}$ANDA}
        & \multirow{5}{*}{Upcoming}
            & \multirow{5}{*}{HESR}
                &
                    & \multirow{3}{*}{GSI}
                        & \multirow{1}{*}{$\bar{p}$ beam on $p$ target} \\
        &
            &
                & 
                    & \multirow{3}{*}{(Darmstadt,}
                        & \multirow{1}{*}{($p_{beam}\approx$ 1.5--15~GeV):} \\
        &
            &
                & 
                    & \multirow{3}{*}{Germany)}
                        & \\
        &
            &
                & 
                    &
                        & \multirow{1}{*}{$p\bar{p}\to X$} \\
        &
            &
                & 
                    &
                        &  \multirow{1}{*}{$p\bar{p}\to X$ + any} \\  \hline
    \multirow{2}{*}{GlueX}
        & \multirow{2}{*}{Beginning}
            & \multirow{4}{*}{CEBAF}
                & \multirow{4}{*}{2016--}
                    & \multirow{1}{*}{Jefferson Lab}
                        & \multirow{1}{*}{$\gamma$ beam on $p$ target} \\
        & \multirow{2}{*}{(searches for light}
            &
                & 
                    & \multirow{1}{*}{(Newport News,}
                        &  \multirow{1}{*}{($E_{beam} \le$ 11~GeV):} \\ \cline{1-1}
    \multirow{2}{*}{CLAS12}
        & \multirow{2}{*}{quark hybrid mesons)}
            &
                & 
                    & \multirow{1}{*}{Virginia,}
                        & \\
        &
            &
                & 
                    & \multirow{1}{*}{USA)}
                        & \multirow{1}{*}{$\gamma p\to X p$} \\  \hline
\end{tabular}
\end{table}

\subsubsection{$B$ and $\Lambda_b$ Decays}

Mesons and baryons containing a single bottom~($b$) quark, such as a
$B$ meson or the $\Lambda_b$ baryon, provide a good source of
charmonium through the weak decay $b\to W^-c$ followed by $W^- \to
s\bar{c}$.  This decay of the $b$ quark generates the decays $B\to
K\psi$ and $\Lambda_b \to K p \, \psi$, where $\psi$ stands for any
state containing a $c\bar{c}$ pair.
With $e^+e^-$ center-of-mass energies near the
$\Upsilon(4S)$ mass (which dominantly decays to $B\bar{B}$), the BaBar
and Belle experiments have traditionally led these studies.  The LHCb
experiment, however, using $B$ mesons and $\Lambda_b$ baryons produced
in $pp$ collisions, has recently exceeded the statistics of Belle and
BaBar.  Belle~II, an upgrade of the Belle experiment, will also start
collecting data soon.

For the decay $B\to K \psi$, the ``$\psi$'' can either be electrically
neutral or charged.  In the case it is neutral, and if it does not
correspond to a traditional state of charmonium, it is generally
referred to as an ``$X$''.  This is the case for the $X(3872)$ seen to
decay to $\pi^+\pi^-J/\psi$~\cite{Choi:2003ue}.  For historical
reasons, it is also sometimes called a ``$Y$'', such as for the
$Y(3940)$, found decaying to $\omega J/\psi$~\cite{Abe:2004zs}.

By rearranging the light quarks in the decay $B \to K \psi$, the
``$\psi$'' can also be electrically charged.  In this case it is
usually referred to as a ``$Z$''.  These electrically charged $Z$
states are especially interesting since, if they are truly states,
they must contain quarks in addition to the neutral $c\bar{c}$ pair.
Prominent examples are the $Z_c(4430)$ decaying to $\pi^\pm
\psi(2S)$~\cite{Choi:2007wga} and the $Z_1(4050)$ and $Z_2(4250)$
decaying to $\pi \chi_{c1}$~\cite{Mizuk:2008me}.

Similarly, in the decay $\Lambda_b \to K p J/\psi$, there appears to
be non-trivial structure in the $p J/\psi$ system, which cannot
originate from a traditional three-quark baryon~\cite{Aaij:2015tga}
(See the discussion of the $P_c(4380)$ and the $P_c(4450)$ in
Sec.~\ref{sec:Pc} for more detail).  Other decays of the form
$\Lambda_b \to K p \, \psi$ are yet to be thoroughly explored.

\subsubsection{$e^+e^-$ Annihilation}

Both the charmonium and bottomonium systems can be conveniently
accessed through $e^+e^-$ annihilation in a number of ways. The
simplest is direct production through a virtual photon.  In this way,
the $J^{PC} = 1^{--}$ states (the $\psi$ states in charmonium and the
$\Upsilon$ states in bottomonium) can be produced.  Using this method,
BESIII and CLEO-c can produce charmonium states and BaBar, Belle, and
Belle~II can produce bottomonium states.  Resonances typically appear
as peaks in the cross section as a function of $e^+e^-$ center-of-mass
energy.

As a powerful extension of the above technique, $e^+e^-$ annihilation
experiments can also use Initial State Radiation~(ISR) to probe
$e^+e^-$ collisions below the nominal center-of-mass energy.  In this
process, a photon is radiated by the initial $e^+$ or $e^-$,
effectively lowering the center-of-mass energy of the collision.  One
advantage of this method is that it provides access to a whole range
of $e^+e^-$ center-of-mass energies.  This improvement has allowed
BaBar and Belle to survey a number of cross sections in the charmonium
region, despite having nominal center-of-mass energies in the
bottomonium region.  In addition to the expected $\psi$ states, a
number of unexpected ones have been found as well, such as the
$Y(4260)$ in $e^+e^- \to Y(4260)
\to \pi^+\pi^- J/\psi$~\cite{Aubert:2005rm}.  A disadvantage of the
ISR method is that the rate is severely suppressed with respect to
direct production by the extra power of $\alpha_{\rm EM}$.

In $e^+e^-$ annihilation, one can also analyze the decay products of
the directly produced $\psi$, $\Upsilon$, or $Y$.  This approach has
led to, for example, the discovery of the electrically charged $Z_c$
and $Z_b$ states in the process $e^+e^- \to \pi^\mp Z_{b,c}^\pm$.
Using $e^+e^-$ collisions in the bottomonium region, the states
$Z_b(10610)$ and $Z_b(10650)$ were found~\cite{Belle:2011aa}, while
$e^+e^-$ collisions in the charmonium region led to the discovery of
the $Z_c(3900)$~\cite{Ablikim:2013mio,Liu:2013dau} and the
$Z_c(4020)$~\cite{Ablikim:2013wzq}.  It is still unclear if the
$e^+e^-$ annihilation in these processes proceeds through traditional
$\psi$ or $\Upsilon$ states, or through exotic $Y$ states, or neither.
Similarly, one can look for radiative transitions, such as the process
$e^+e^- \to \gamma X(3872)$~\cite{Ablikim:2013dyn}, or for dipion
transitions, such as the process $e^+e^- \to \pi^+\pi^-
X(3823)$~\cite{Ablikim:2015dlj}.

Another method used in $e^+e^-$ annihilation is the double-charmonium
production process $e^+e^-\to J/\psi X$, where $X$ also contains
charm, and the initial $e^+e^-$ collision energy is in the bottomonium
region.  Using this technique, Belle has been able to observe
traditional charmonium states, such as the $\eta_c(1S,2S)$, recoiling
against the $J/\psi$, but has also seen the possibly exotic $X(3940)$
and $X(4160)$~\cite{Abe:2007jna,Abe:2007sya}.  This technique
remains relatively unexplored.

\subsubsection{$\gamma\gamma$ Collisions}

The $e^+e^-$ experiments with center-of-mass energies in the
bottomonium region (BaBar, Belle, and Belle~II) can explore
$\gamma\gamma$ collisions in the charmonium region through the process
$e^+e^-\to e^+e^-X$.  This technique has proven to be a powerful way
to produce conventional charmonium states.  For example, the BaBar
Collaboration has been able to make precision measurements of the mass
and width of the $\eta_c(2S)$, as well as measure new decay modes of
the $\eta_c(2S)$, using $\gamma\gamma$
collisions~\cite{delAmoSanchez:2011bt,Lees:2014iua}.  But there are
several more observations that are yet to be fully understood.  The
$X(3915)$~(decaying to $\omega
J/\psi$~\cite{Uehara:2009tx,Lees:2012xs}) and the $Z(3930)$~(decaying
to $D\bar{D}$~\cite{Uehara:2005qd,Aubert:2010ab}) are both seen
clearly and are often identified with the $\chi_{c0}(2P)$ and
$\chi_{c2}(2P)$ states of charmonium, respectively (although the
former assignment is more controversial).  The $X(4350)$~(decaying to
$\phi J/\psi$~\cite{Shen:2009vs}) needs further experimental
confirmation.

\subsubsection{Hadron Collisions}

The CDF and D0 experiments at Fermilab and the CMS, ATLAS, and LHCb
experiments at CERN have had success producing QCD exotica in very
high-energy $p\bar{p}$~(Fermilab) and $pp$~(CERN) collisions.  Direct
production of particles from the initial collision~({\it prompt\/}
production) is generally separated from production from subsequent $B$
decays~({\it nonprompt\/} production) using the position of the decay
vertex.  The $X(3872)$ appears to have a significant prompt cross
section when compared to prompt production of the
$\psi(2S)$~\cite{Chatrchyan:2013cld}.  Other states seen in hadron
collisions include the $Y(4140)$~(decaying to $\phi
J/\psi$~\cite{Abazov:2015sxa}) and the recently reported
$X(5568)$~(decaying to $B_s\pi$~\cite{D0:2016mwd}).

\subsection{The $X(3872)$ as the First of the $XY \! Z$}
\label{sec:X3872}

\begin{figure}[htb]
\includegraphics*[width= 1.0\columnwidth]{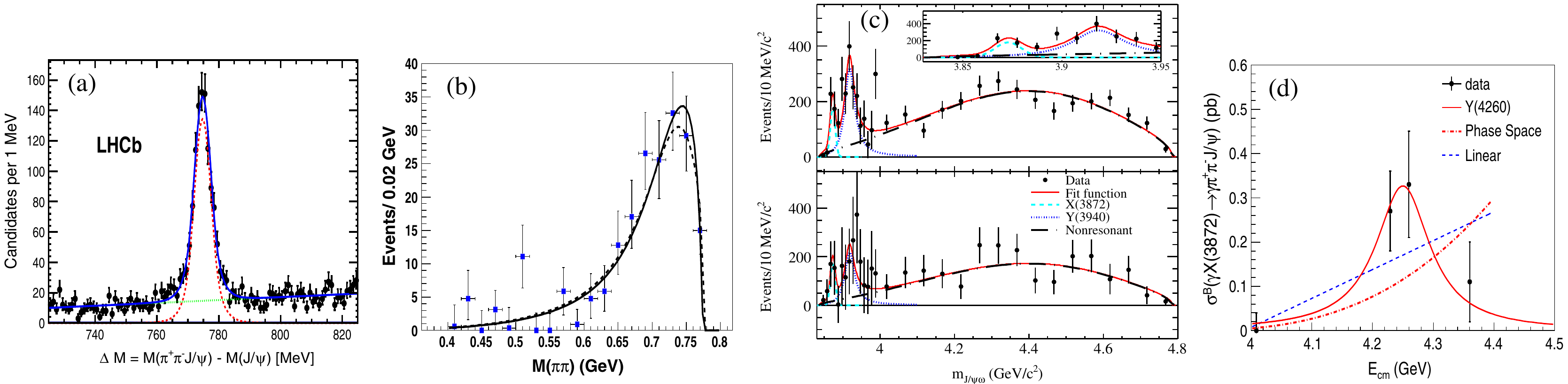}
\caption{\label{fig:X3872} Properties of the $X(3872)$. 
(a)~The latest observation of the $X(3872)$ in $B\to K(\pi^+\pi^-
J/\psi)$ from LHCb~\cite{Aaij:2015eva}.  Compare the size of the data
sample to the earliest observation of the
$X(3872)$~(Fig.~\ref{fig:discoveries}a).  This sample was used in the
determination of the $J^{PC}$ of the $X(3872)$.  (b)~The $\pi^+\pi^-$
mass spectrum from the decay $X(3872)\to \pi^+\pi^-J/\psi$ from
Belle~\cite{Choi:2011fc}, showing the $\pi^+\pi^-$ system originates
from a $\rho$.  The two lines are for different assumptions about the
orbital angular momentum in the decay to $\rho J/\psi$.
(c)~Observation of the decay $X(3872)\to \omega J/\psi$ from
BaBar~\cite{delAmoSanchez:2010jr}.  The top plot is for $B^+\to K^+
(\omega J/\psi)$ and the bottom is for $B^0\to K^0 (\omega J/\psi)$.
The $X(3872)$ appears just below the $Y(3940)$.  (d)~The cross section
as a function of center-of-mass energy for $e^+e^-\to \gamma X(3872)$
from BESIII~\cite{Ablikim:2013dyn}.  The $Y(4260)$ assumption~(solid
line) is more consistent with the data than phase space or
linear~(dashed lines) assumptions.  With only four data points, more
data is required.
}
\end{figure}

As the first of the $XY \! Z$ states to be discovered, the $X(3872)$
is also the most ubiquitous and thoroughly studied.  But even in 2003,
after its initial discovery by Belle in the process $B\to KX(3872)$
with $X(3872)\to \pi^+\pi^-J/\psi$~\cite{Choi:2003ue}, it was already
known that the $X(3872)$ was out of the ordinary.  It was narrow and
had a mass suspiciously close to the $D^{*0}\bar{D}^0$ threshold.
Even while its quantum numbers were not yet known, it was difficult to
fit the $X(3872)$ into any of the unoccupied places in the charmonium
spectrum.  For example, the $^3D_2$ ($J^{PC}=2^{--}$) state of
charmonium could be ruled out because the upper limit on the ratio of
branching fractions
$B(X(3872)\to\gamma\chi_{c1})/B(X(3872)\to\pi^+\pi^-J/\psi)$ was too
restrictive.  And it was also too light to be the $\chi_{c1}(2P)$.
Furthermore, the $\pi^+\pi^-$ system in the decay
$X(3872)\to\pi^+\pi^-J/\psi$ appeared to come from a $\rho$, making
the $X(3872)$ either isospin~1, or meaning that the $X(3872)$ has
significant isospin violation in its decay.  Rather than trace the
historical development of facts, below we list a number of results
that we currently know about the $X(3872)$ and how we know them.

\begin{enumerate}

\item
{\bf The $X(3872)$ exists.}  The initial observation of the
$X(3872)$~[discovered in $B\to KX(3872)$ with
$X(3872)\to\pi^+\pi^-J/\psi$] already had a statistical significance
of $10.3\sigma$~(Fig.~\ref{fig:discoveries}a)~\cite{Choi:2003ue}.
Later observations, including using the same process used in its
discovery, but with a massive increase in the size of the data
sample~(compare Figs.~\ref{fig:discoveries}a
and~\ref{fig:X3872}a)~\cite{Aaij:2015eva}, have put the existence of
the $X(3872)$ beyond any doubt.

\item
{\bf The mass of the $X(3872)$ is close to the $D^{*0}\bar{D}^0$
threshold.}  The average value of all measurements of the $X(3872)$
mass is currently $3871.69\pm0.17$~MeV~\cite{Olive:2016xmw}.  Using
the current value for the $D^{*0}$ mass, $2006.85\pm0.05$~MeV, and the
$D^0$ mass, $1864.83\pm0.05$~MeV, the $D^{*0}\bar{D}^0$ threshold is
$3871.68\pm0.07$~MeV~\cite{Olive:2016xmw}.  The difference between
the $X(3872)$ mass and the $D^{*0}\bar{D}^0$ threshold is therefore
remarkably small, $0.01\pm0.18$~MeV.  Notice that the error is
dominated by the error on the $X(3872)$ mass.

\item
{\bf The $X(3872)$ is narrow.}  The upper limit on the width of the
$X(3872)$ is currently 1.2~MeV.  This value was set by the Belle
experiment in an analysis of $B\to KX(3872)$ with
$X(3872)\to\pi^+\pi^-J/\psi$~\cite{Choi:2011fc}.  Using a simultaneous
fit to the $B$ mass, the $B$ energy, and the $\pi^+\pi^-J/\psi$ mass
spectrum, they were able to overconstrain the area of the $X(3872)$
peak in the $\pi^+\pi^-J/\psi$ mass spectrum.  This technique improved
sensitivity to the width of the $X(3872)$, allowing for such a tight
upper limit, even though the detector resolution for the mass of the
$\pi^+\pi^-J/\psi$ system was around 4~MeV.

\item
{\bf The $X(3872)$ has no isospin partners.}  The electrically neutral
$X(3872)$ has been well-established in both of the processes, $B^+\to
K^+X(3872)$ and $\bar{B}^0 \to \bar{K}^0 X(3872)$, with
$X(3872)\to\pi^+\pi^-J/\psi$.  If the $X(3872)$ had an electrically
charged isospin partner, it would be evident in the related processes
$B^+ \to K^0 X^+$ and $\bar{B}^0 \to K^- X^+$ with $X^+ \to \pi^+
\pi^0 J/\psi$, according to predictable isospin ratios.  However, only
upper limits have been determined for these related processes,
inconsistent with the predicted isospin
ratios~\cite{Aubert:2004zr,Choi:2011fc}.

\item {\bf The $X(3872)$ radiatively decays to both $\gamma J/\psi$
    and $\gamma \psi(2S)$.}  The LHCb experiment has made the most
  precise measurements of both radiative decays $X(3872)\to\gamma
  J/\psi$ and $X(3872)\to\gamma \psi(2S)$~\cite{Aaij:2014ala}.  The
  current average value for the ratio of branching fractions is
  $B(X(3872)\to\gamma \psi(2S))/B(X(3872)\to\gamma J/\psi) =
  2.6\pm0.6$~\cite{Olive:2016xmw}.

\item {\bf The $X(3872)$ decays to $\rho J/\psi$.}  Once the radiative
  decays $X(3872)\to\gamma J/\psi$, $\gamma\psi(2S)$ are established,
  it follows that the $X(3872)$ has $C=+$.  In the decay $X(3872)\to
  \pi^+\pi^-J/\psi$, the $\pi^+\pi^-$ system must then have $C=-$.
  Since the $\pi^+\pi^-$ system must have $C = P = (-1)^L = (-1)^J$,
  the only $J^{PC}$ possibilities are $1^{--}$, $3^{--}$, {\it etc.},
  of which the only plausible combination, considering the low
  $\pi^+\pi^-$ mass, is $1^{--}$.  This result is consistent with
  analyses of the $\pi^+\pi^-$ mass distribution, showing
  $X(3872)\to\rho^0J/\psi$
  (Fig.~\ref{fig:X3872}b)~\cite{Chatrchyan:2013cld, Choi:2011fc,
    Abulencia:2005zc}.  Note that this decay violates isospin if the
  $X(3872)$ has isospin~0.

\item
{\bf The $X(3872)$ has $J^{PC}=1^{++}$.}  The LHCb experiment
conclusively determined the $J^{PC}$ of the $X(3872)$ to be $1^{++}$
using a five-dimensional angular analysis of the process $B^+\to K^+
X(3872)$ with $X(3872)\to \rho^0 J/\psi$ and $\rho^0\to
\pi^+\pi^-$~\cite{Aaij:2015eva}.  The analysis was based on a large
sample of $X(3872)$ decays~(Fig.~\ref{fig:X3872}a), and built upon
earlier $J^{PC}$ analyses~\cite{Abulencia:2006ma,Aaij:2013zoa}.

\item
{\bf The $X(3872)$ decays to $\omega J/\psi$.}  The BaBar experiment
found evidence for the decay $B\to K X(3872)$ with $X(3872)\to\omega
J/\psi$~\cite{delAmoSanchez:2010jr}.  The mass spectrum of the $\omega
J/\psi$ system is dominated by the $Y(3940)$; the $X(3872)$ appears
just below it~(Fig.~\ref{fig:X3872}c).  Comparing $X(3872)$ decays to
$\omega J/\psi$ with its decays to $\pi^+ \pi^- J/\psi$, where the
$X(3872)$ is produced in $B\to K X(3872)$ in both cases, one can
determine the ratio of branching fractions, ${\cal B}(X(3872)\to
\omega J/\psi)/{\cal B}(X(3872)\to \pi^+\pi^-
J/\psi)=0.8\pm0.3$~\cite{Aaij:2014ala}.  Note that the presence of
both of these decays implies that there is isospin violation.

\item
{\bf The $X(3872)$ decays to $D^{*0}\bar{D}^0+c.c$.}  The $X(3872)$
appears as a peak just above $D^{*0}\bar{D}^0$ threshold in the
process $B\to K X(3872)$ with $X(3872) \to
D^{*0}\bar{D}^0+c.c.$~\cite{Gokhroo:2006bt,Aubert:2007rva,
Adachi:2008sua}. Because of the limited available phase space, it is
difficult to determine if there is a continuum $X(3872)\to
D^0\bar{D}^0\pi^0$ decay in addition to the $X(3872) \to
D^{*0}\bar{D}^0$ decay.  Using the latest value of ${\cal B}(B^+\to
X(3872)K^+)\times{\cal B}(X(3872)\to\pi^+\pi^-J/\psi) = (0.84 \pm 0.15
\pm 0.07)\times 10^{-5}$~\cite{Aubert:2008gu} and the latest value of
${\cal B}(B^+\to X(3872)K^+)\times{\cal B}(X(3872)\to
D^{*0}\bar{D}^0+c.c.) = (7.7 \pm 1.6 \pm 1.0)\times
10^{-5}$~\cite{Adachi:2008sua}, one obtains the ratio ${\cal
B}(X(3872)\to D^{*0}\bar{D}^0+c.c.)/{\cal
B}(X(3872)\to\pi^+\pi^-J/\psi) = 9.2 \pm 2.9$, where statistical and
systematic errors have been added in quadrature.

\item
{\bf There are lower limits on $X(3872)$ branching fractions.}  The
BaBar experiment set an upper limit ${\cal B}(B^+\to K^+X(3872)) <
3.2\times10^{-4}$ in a search for inclusive decays of the
$X(3872)$~\cite{Aubert:2005vi}.  This upper limit, combined with
measured product branching fractions, such as ${\cal B}(B^+\to
K^+X(3872))\times{\cal B}(X(3872) \to \pi^+\pi^-J/\psi)$, allows lower
limits to be calculated for $X(3872)$ branching fractions.  In this
way, we know ${\cal B}(X(3872)\to \pi^+\pi^-J/\psi) > 2.6\%$ and
${\cal B}(X(3872)\to D^{*0}\bar{D}^0+c.c.) > 24\%$.

\item
{\bf The $X(3872)$ is produced in hadron collisions.}  The $X(3872)$
has been seen in $p\bar{p}$ collisions with $\sqrt{s}=1.96$~TeV at the
Tevatron~\cite{Acosta:2003zx,Abulencia:2005zc,Abazov:2004kp,
Aaltonen:2009vj} and in $pp$ collisions with $\sqrt{s}=7$~TeV at the
LHC~\cite{Chatrchyan:2013cld,Aaij:2011sn}. The CMS experiment studied
the production of the $X(3872)$ in relation to the production of the
$\psi(2S)$, and found the ratio
\begin{equation}
R = \frac {\sigma(pp\to
X(3872)+\mathrm{anything})\times{\cal B}(X(3872)\to \pi^+\pi^-J/\psi)}
{\sigma(pp\to \psi(2S)+\mathrm{anything})\times{\cal B}(\psi(2S)\to
\pi^+\pi^-J/\psi)} = 0.0656 \pm 0.0029 \pm 0.0065 \, ,
\end{equation}
in a region of rapidity~($|y|<1.2$) and transverse
momentum~($10<p_T<50$~GeV)~\cite{Chatrchyan:2013cld}.  They also
determined the fraction of these $X(3872)$ produced in $B$~decays to
be $0.263\pm0.023\pm0.016$, the remainder being the so-called
``prompt'' production.  In the kinematic region studied, the ratio $R$
appears to have no dependence on $p_T$.

\item
{\bf The $X(3872)$ is possibly produced in radiative decays of the
$Y(4260)$.}  The BESIII experiment found clear evidence for $e^+e^-\to
\gamma X(3872)$ with $X(3872)\to \pi^+\pi^- J/\psi$, where the
$e^+e^-$ center-of-mass energy was in the region of the
$Y(4260)$~\cite{Ablikim:2013dyn}.  The cross section of this process
as a function of center-of-mass energy is suggestive that it proceeds
through a $Y(4260)$~(Fig.~\ref{fig:X3872}d), which would imply the
existence of the radiative decay $Y(4260)\to \gamma X(3872)$, but more
data is needed before this prospect can be determined definitively.

\end{enumerate}

\subsection{Structure in $B$ and $\Lambda_b$ Decays}
\label{sec:WeakDecays}

Besides the $X(3872)$, 
a series of other structures have been
observed in $B$ decays through $B\to KX$, where the ``$X$'' decays to
charmonium and can be either electrically charged or neutral.  Two of
these additional structures, the $X(3823)$ decaying to $\gamma
\chi_{c1}$ and the $Y(3940)$ decaying to $\omega J/\psi$, are
relatively narrow and can likely be accommodated in the traditional
spectrum of $c\bar{c}$ states.  These will be discussed in
Section~\ref{sec:3800}.  In this section we discuss the more exotic
remaining structures.

Recall that if the ``$X$'' is charged [as is the case with, for
example, the $Z_c(4430)$], and if the peak is not generated by a
dynamical effect, then that state must be composed of at least four
quarks, since additional quarks are needed beyond the neutral
$c\bar{c}$ pair to give a unit of electric charge.  It is the presence
of this signature for an exotic state that has brought so much
attention to many of these processes.  But even the neutral ``$X$''
[such as the $Y(4140)$] do not fit in the traditional spectrum of
$c\bar{c}$ states.

These additional ``$X$'' structures appearing in $B\to KX$ are broad,
unlike the $X(3872)$, with widths ranging from roughly 100 to
400~MeV\@.  And, with the possible exception of the $Z_c(4430)$, each
has been seen in only one decay channel.  It is also interesting, but
possibly only a coincidence, that these structures seem to appear in
pairs.  The $Z_c(4430)$ and $Z_c(4240)$ are observed decaying to
$\pi^\pm\psi(2S)$; the $Y(4140)$, $Y(4274)$, $X(4500)$, and $X(4700)$
are observed in $\phi J/\psi$; the $Z_1(4050)$ and $Z_2(4250)$ are
reported in $\pi^\pm \chi_{c1}$; and the $Z_c(4200)$ and
$Z_c(4430)$~[perhaps the same $Z_c(4430)$ as seen in $\pi^\pm
\psi(2S)$] are reported in $\pi^\pm J/\psi$.

This section also includes a discussion of the decay $\Lambda_b\to
K(pJ/\psi)$, since there are many similarities between this process
and $B\to KX$.  The physical process is similar (both including a weak
decay of the bottom quark, $b\to sc\bar{c}$), and the methods used to
analyze them are similar.  Because anything decaying to $p J/\psi$ is
electrically charged, contains a $c\bar{c}$ pair, and is a baryon, it
must contain at least five quarks.  Here again, a pair of broad states
is seen decaying to $p J/\psi$, the $P_c(4380)$ and $P_c(4450)$.

In the case of $B \to K X(3872)$ with $X(3872)\to\pi^+\pi^-J/\psi$
discussed above, the $X(3872)$ was sufficiently narrow to allow the
neglect of interference with any possible structure in, for example,
the $K\pi$ system.  To determine the properties of the $X(3872)$, a
one-dimensional fit to the $\pi^+\pi^-J/\psi$ mass spectrum was
therefore reliable.  This is not the case for the wider ``$X$''
structures produced in $B\to KX$, which require more complex methods.
For example, in the decay $B\to K[\pi^{\pm}\psi(2S)]$, in addition to
any exotic structure in the $\pi^\pm \psi(2S)$ system, one has to also
contend with the $K^*$ resonances in the $K\pi^{\pm}$ system.  In
fact, these resonances are generally larger than the interesting
structure in, for example, the $\pi^\pm \psi(2S)$ system.  And, due to
their non-trivial angular momenta, the $K^*$ decays can populate the
$K\pi^\pm \psi(2S)$ Dalitz plot in a way that affects the projections
onto the $\pi^\pm \psi(2S)$ mass, leading to pollution by {\it
kinematical reflections}.  A number of methods have been employed to
handle this problem, such as full-amplitude analyses or methods that
attempt to parameterize all reasonable angular structure in the
$K\pi^\pm$ system.  Since many methods have been used to analyze the
$Z_c(4430)$ [decaying to $\pi^\pm \psi(2S)$], they will be discussed
in the following section.

\subsubsection{$B\to K\pi\psi(2S)$ and the $Z_c(4430)$ Tetraquark Candidate}
\label{sec:Z4430}

The $Z_c(4430)$ was first reported by the Belle experiment in the
process $B\to K Z_c(4430)$ with $Z_c(4430)\to \pi^\pm
\psi(2S)$~\cite{Choi:2007wga}.  It was the first claim of an
electrically charged state in the charmonium region and therefore
received a lot of attention.  Rather than presenting a full analysis
of the Dalitz plot and angular distributions, this initial observation
dealt with $K^*$ contributions by vetoing $K\pi^\pm$ combinations with
a mass within 100~MeV of the $K^*(890)$ or $K_2^*(1430)$.  After
applying this $K^*$ veto, the $\pi^\pm \psi(2S)$ mass distribution was
fit with a smooth background function and a Breit-Wigner distribution.
The resulting $Z_c(4430)$ had a significance of $6.5\sigma$.

The BaBar experiment objected to this method, arguing that there are
many other $K^*$ resonances besides the $K^*(890)$ and $K_2^*(1430)$
that could have an influence on the $\pi^\pm \psi(2S)$ mass spectrum.
To explore the influence of these other $K\pi^\pm$ resonances, BaBar
analyzed its own sample of data, which was of a comparable size to the
Belle sample, using a model-independent approach~\cite{Aubert:2008aa}.
They first described the angular distributions of the $K\pi^\pm$
system in bins of $K\pi^\pm$ mass, using a series of Legendre
polynomials $P_l(\cos\theta)$, where $\theta$ is the angle in the
$K\pi^\pm$ rest frame between the $K$ momentum and the boost direction
that takes the $K\pi^\pm$ to its lab-frame momentum.  They used
polynomials up to $l=6$, which allowed for $K\pi^\pm$ resonances with
spin $\le 3$.  This series of Legendre polynomials has the feature
that the coefficients, called moments, can be determined without
needing to fit the data.  BaBar then generated a Dalitz plot based
upon these moments and projected it onto the $\pi^\pm \psi(2S)$ mass.
They found that this projection described the $\pi^\pm
\psi(2S)$ mass spectrum well, and therefore found no need for the
$Z_c(4430)$ state, despite the Belle and BaBar data sets being
statistically consistent.

The Belle experiment quickly improved upon the one-dimensional
analysis of the $\pi^\pm \psi(2S)$ system by performing a
two-dimensional Dalitz plot analysis, simultaneously analyzing both
the $\pi^\pm \psi(2S)$ and $K\pi^\pm$ systems~\cite{Mizuk:2009da}.
They later also performed a full amplitude analysis~(four-dimensional)
of the $B\to K\pi^\pm \psi(2S)$ decay, also taking into account
angular distributions~\cite{Chilikin:2013tch}.  Both analyses, using
data sets that were almost the same as the original data set,
confirmed the existence of the $Z_c(4430)$.  The latter found evidence
that the $Z_c(4430)$ had a $J^P$ of $1^+$.

\begin{figure}[htb]
\begin{center}
\includegraphics*[width= 1.0\columnwidth]{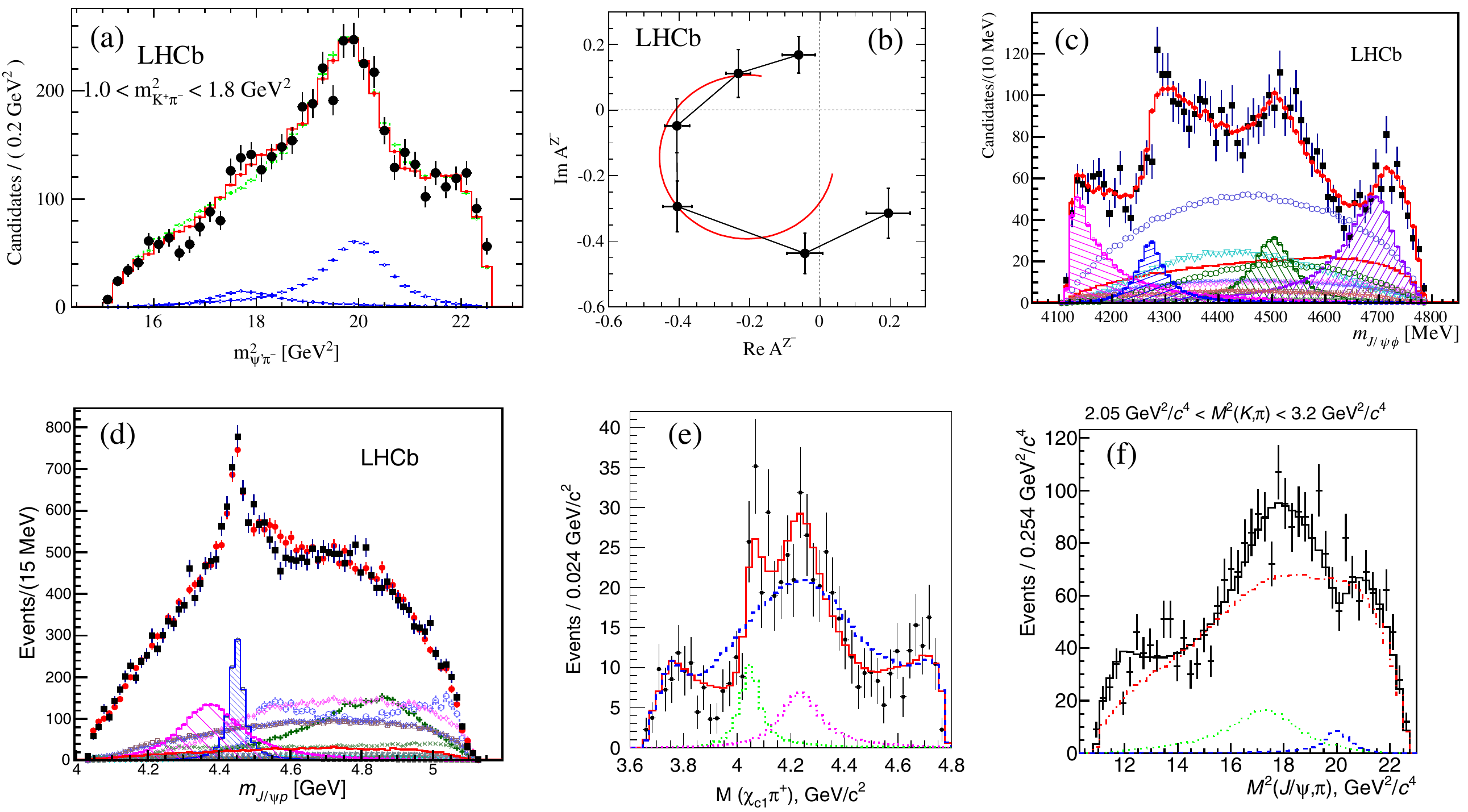}
\end{center}
\caption{\label{fig:Bdecays} 
QCD exotica found in $B$ and $\Lambda_b$ decays.
(a)~Observation of the $Z_c(4200)$ and $Z_c(4430)$ at LHCb in $B\to
K(\pi^\pm \psi(2S))$~\cite{Aaij:2014jqa}.
(b)~Argand diagram for the $Z_c(4430)$~\cite{Aaij:2014jqa}.
(c)~Observation of the $Y(4140)$, $Y(4274)$, $X(4500)$, and
$X(4700)$ by LHCb in $B\to K(\phi J/\psi)$~\cite{Aaij:2016iza}.
(d)~Observation of the $P_c(4380)$ and $P_c(4450)$ by LHCb in
$\Lambda_b \to K(pJ/\psi)$~\cite{Aaij:2015tga}.
(e)~Observation of the $Z_1(4050)$ and $Z_2(4250)$ by Belle in
$B\to K(\pi^\pm \chi_{c1})$~\cite{Mizuk:2008me}.
(f)~Observation of the $Z_c(4200)$ and evidence for the $Z_c(4430)$ by
Belle in $B\to K(\pi^\pm J/\psi)$~\cite{Chilikin:2014bkk}.  }
\end{figure}

While Belle and BaBar collected samples of a few thousand $B\to
K\pi^\pm \psi(2S)$ events, the LHCb experiment was able to analyze a
sample roughly an order of magnitude larger.  With this increase in
statistics, the LHCb experiment performed a full four-dimensional
amplitude analysis, confirmed the existence of the $Z_c(4430)$, and
conclusively showed its $J^P$ to be
$1^+$~(Fig.~\ref{fig:Bdecays}a)~\cite{Aaij:2014jqa}.  They also
observed a lighter and wider structure in the $\pi^\pm \psi(2S)$
amplitude, the $Z_c(4240)$, with a significance of $6\sigma$ and a
preferred $J^P$ of $0^-$.  In addition, LHCb was able to analyze the
phase motion of the $Z_c(4430)$ by replacing the $Z_c(4430)$
Breit-Wigner amplitude with a piece-wise complex constant as a
function of $\pi^\pm \psi(2S)$ mass.  The motion in the complex
plane~(the Argand diagram) is consistent with what one would expect
for a resonance~(Fig.~\ref{fig:Bdecays}b).  As a final test, LHCb also
repeated the moments method used by the BaBar experiment and found
that the $\pi^\pm \psi(2S)$ mass spectrum could not be described by
using reflections from the $K\pi^\pm$ system; the $Z_c(4430)$ was
still needed~\cite{Aaij:2014jqa,Aaij:2015zxa}.  In this moments study,
the existence of the broader $Z_c(4240)$ was not addressed.

\subsubsection{$B\to K\phi J/\psi$ and the $Y(4140)$ and More}
\label{sec:Y4140}

Like the $Z_c(4430)$, produced in $B\to KZ_c(4430)$ with $Z_c(4430)\to
\pi^\pm \psi(2S)$, the $Y(4140)$, produced in $B\to KY(4140)$ with
$Y(4140)\to \phi J/\psi$, had a controversial beginning.  It was first
reported by the CDF experiment~\cite{Aaltonen:2009tz}, but with a
significance of only 3.8$\sigma$ from a sample of fewer than 100 $B^+$
decays.  It was not confirmed by the LHCb~\cite{Aaij:2012pz} and
BaBar~\cite{Lees:2014lra} experiments, but it was confirmed by the D0
experiment~\cite{Abazov:2013xda}, each using samples of a few hundred
$B$ decays.  The CMS experiment~\cite{Chatrchyan:2013dma}, using a
sample of around 2000 $B$ decays, found a $5\sigma$-significance
signal for the $Y(4140)$.  Complicating the situation, both the
original CDF analysis~\cite{Aaltonen:2009tz} and the higher-statistics
CMS analysis~\cite{Chatrchyan:2013dma} also reported the existence of
a higher-mass state, the $Y(4274)$, although the masses reported for
the state were significantly different.  All of these initial analyses
were performed by fitting only the one-dimensional $\phi J/\psi$ mass
spectrum, neglecting any influence from the $K \phi$ system.

Similar to the story of the $Z_c(4430)$, the status of the $Y(4140)$
remained in limbo until a higher-statistics analysis from the LHCb
experiment was performed~\cite{Aaij:2016iza,Aaij:2016nsc}.  Using more
than 4000 $B^+$ decays with relatively small backgrounds, the LHCb
experiment in fact not only confirmed the existence of the $Y(4140)$
and the $Y(4274)$, with significances of $8.4\sigma$ and $6.0\sigma$,
respectively, but also reported another pair of peaks, the $X(4500)$
and $X(4700)$, with significances greater than
$5\sigma$~(Fig.~\ref{fig:Bdecays}c).  Using a full six-dimensional
amplitude analysis, including $K^*$ resonances in the $K\phi$ system
and descriptions of all decay angular distributions, the $J^{PC}$ of
the $Y(4140)$ and the $Y(4274)$ were both determined to be $1^{++}$.
The $J^{PC}$ values of the higher-mass $X(4500)$ and $X(4700)$ were
both found to be $0^{++}$.

\subsubsection{$\Lambda_b \to K p J/\psi$ and the $P_c$ Pentaquark Candidates}
\label{sec:Pc}

The experimental analysis of the decay $\Lambda_b \to K (p J/\psi)$ is
very similar to that of $B\to K [\pi^\pm \psi(2S)]$ and $B\to K(\phi
J/\psi)$ discussed above.  Using a sample of around 26,000 $\Lambda_b$
decays, LHCb performed a full amplitude analysis of the process
$\Lambda_b \to K (p J/\psi)$, which included all known $\Lambda$
states decaying to $Kp$~\cite{Aaij:2015tga}.  Two additional
amplitudes in the $pJ/\psi$ system were needed to describe the data,
both found with more than $9\sigma$
significance~(Fig.~\ref{fig:Bdecays}d).  The lighter one, the
$P_c(4380)$, was wide, with a width around 200~MeV; the heavier one,
the $P_c(4450)$, was narrow, with a width around 40~MeV.  The favored
$J^P$ of the $P_c(4380)$ and $P_c(4450)$ were found to be
${\frac{3}{2}}^-$ and ${\frac{5}{2}}^+$, respectively, although the
combinations $({\frac{3}{2}}^+, {\frac{5}{2}}^-)$ and
$({\frac{5}{2}}^+, {\frac{3}{2}}^-)$ could not be ruled out.  The
Argand diagram for the narrower $P_c(4450)$ was found to be consistent
with a resonance; the Argand diagram for the wider $P_c(4380)$ was
more uncertain and depends more upon the details of the $pK$
amplitudes, which are not precisely known.

\subsubsection{Other $B$ Decays}
\label{sec:otherB}

Like the decays $B\to K\pi^{\pm}\psi(2S)$, $B\to K \phi J/\psi$, and
$\Lambda_b \to K p J/\psi$ discussed above, the decays $B \to K
\pi^{\pm} \chi_{c1}$ and $B \to K \pi^{\pm} J/\psi$ also possibly show
evidence for pairs of exotic structures decaying to charmonium.  The
$Z_1(4050)$ and $Z_2(4250)$, decaying to $\pi^\pm \chi_{c1}$, were
reported by the Belle experiment in a Dalitz plot analysis of the
decay $B\to K\pi^\pm
\chi_{c1}$~(Fig.~\ref{fig:Bdecays}e)~\cite{Mizuk:2008me}, while the
$Z_c(4200)$ and $Z_c(4430)$, decaying to $\pi^\pm J/\psi$, were
reported by Belle in an amplitude analysis of the decay $B\to K\pi^\pm
J/\psi$~(Fig.~\ref{fig:Bdecays}f)~\cite{Chilikin:2014bkk}.  The
$Z_c(4430)$ decaying to $\pi^\pm J/\psi$ is consistent with the
$Z_c(4430)$ decaying to $\pi^\pm \psi(2S)$ and is perhaps the only one
of the family of $Z$ structures to be seen in multiple decays.  The
three new $Z$ structures reported by Belle were each found to have
significances of greater than $5\sigma$, while the $Z_c(4430)$ decay
to $\pi J/\psi$ was found with a significance of $4.0\sigma$.

The BaBar experiment has also analyzed both of these channels using
the same moments method discussed
above~\cite{Aubert:2008aa,Lees:2011ik}.  No evidence for the $Z$
structures was found in either case.  An investigation of these two
channels with higher statistics, perhaps by the LHCb experiment, is
therefore needed.

\subsection{Structure in $e^+e^-$ Annihilation}
\label{sec:ee}

When proceeding through a single virtual photon, $e^+e^-$ annihilation
should in principle be a relatively straightforward way to produce
vector mesons and study their decays.  The lowest-lying $\psi$ states
of charmonium, the $J/\psi$, $\psi(2S)$, and $\psi(3770)$, and the
lowest-lying states of bottomonium, the $\Upsilon(1S)$,
$\Upsilon(2S)$, $\Upsilon(3S)$, and $\Upsilon(4S)$, have been produced
and studied using $e^+e^-$ annihilation for over 35 years.  However,
raising the center-of-mass energies of the $e^+e^-$ collisions
significantly above the threshold to produce open-charm or open-bottom
states [the $\psi(3770)$ lies just above $D\bar{D}$ threshold and the
$\Upsilon(4S)$ lies just above $B\bar{B}$ threshold] has led to a
number of surprises that are yet to be understood.  Before presenting
more detail about the structures seen in $e^+e^-$ annihilation, we
first provide a short chronology of how these discoveries have
unfolded.  This narrative serves to illustrate the parallels between
charmonium and bottomonium, and how developments in one have led to
new studies and discoveries in the other.

(1) The surprises in $e^+e^-$ annihilation began in 2005 with the
discovery of the $Y(4260)$ by the BaBar
experiment~\cite{Aubert:2005rm}.  BaBar used initial-state radiation
(ISR) to study the energy dependence of the cross section for
$e^+e^-\to \pi^+\pi^- J/\psi$; the $Y(4260)$ appeared as an unexpected
peak at 4.26~GeV\@.  This result was soon followed by the 2007
discovery of the $Y(4360)$ by BaBar in the cross section for
$e^+e^-\to\pi^+\pi^-\psi(2S)$ using the same
procedure~\cite{Aubert:2007zz}.  (Also see Sec.~\ref{sec:expintro} and
Figs.~\ref{fig:discoveries}c and~\ref{fig:discoveries}d.)

(2) In 2008, the Belle experiment, looking for a bottomonium analogue
of the $Y(4260)$ or $Y(4360)$, studied the cross sections for $e^+e^-
\to \pi^+\pi^-\Upsilon(1S,2S)$ at a center-of-mass energy
corresponding to the $\Upsilon(5S)$ mass~\cite{Abe:2007tk}.  The cross
sections were found to be anomalously large, indicating either the
presence of an underlying exotic state, or
$\Upsilon(5S)\to\pi^+\pi^-\Upsilon(1S,2S)$ partial widths several
orders of magnitude larger than the measured
$\Upsilon(4S)\to\pi^+\pi^-\Upsilon(1S,2S)$ partial widths.  In 2010,
Belle extended this study by analyzing several center-of-mass energies
in the region surrounding the $\Upsilon(5S)$~\cite{Chen:2008xia}.  The
peak in the $e^+e^-\to\pi^+\pi^-\Upsilon(1S,2S,3S)$ cross sections
appeared to be shifted from the $\Upsilon(5S)$ mass, leading to the
postulation of the $Y_b(10888)$.

(3) In 2011, the CLEO-c experiment found that the $e^+e^- \to
\pi^+\pi^- h_c(1P)$ cross section in the region of the $Y(4260)$ was
of a comparable size to the $e^+e^-\to \pi^+\pi^- J/\psi$ cross
section~\cite{CLEO:2011aa}.  This result was a surprise, because if
$e^+e^-$ proceeds through the production of a conventional $s_c = 1$
charmonium state, as expected, the process $e^+e^-\to
\pi^+\pi^-h_c(1P)$ would involve a spin flip, and therefore ought to
be strongly suppressed with respect to the process $e^+e^-\to
\pi^+\pi^-J/\psi$, which would not involve a spin flip.

(4) In 2012, motivated by the observation of $e^+e^-\to
\pi^+\pi^-h_c(1P)$, the Belle experiment performed a search for
$e^+e^-\to \pi^+\pi^- h_b(1P,2P)$~\cite{Adachi:2011ji}.  Neither the
$h_b(1P)$ nor the $h_b(2P)$ had yet been discovered.  Belle not only
discovered both states, but also found that the $e^+e^- \to
\pi^+\pi^-h_b(1P,2P)$ cross sections in the region of the
$\Upsilon(5S)$ were of comparable size to the cross sections for
$e^+e^-\to \pi^+\pi^-\Upsilon(1S,2S,3S)$, which parallels the
situation in charmonium.

(5) Also in 2012, as a follow-up to their discovery of $e^+e^-\to
\pi^+\pi^- h_b(1P,2P)$, the Belle experiment analyzed the substructure
in the five processes $e^+e^-\to \pi^+\pi^- h_b(1P,2P)$ and $e^+e^-\to
\pi^+\pi^- \Upsilon(1S,2S,3S)$, where the $e^+e^-$ collisions were
again in the region of the $\Upsilon(5S)$~\cite{Belle:2011aa}.  They
found two electrically charged $Z_b$ states, the $Z_b(10610)$ and the
$Z_b(10650)$, in the process $e^+e^-\to \pi^\pm Z_b$, where both $Z_b$
states decayed to all of $\pi^\pm h_b(1P,2P)$ and $\pi^\pm
\Upsilon(1S,2S,3S)$.  The $Z_b(10610)$ has a mass near the
$B\bar{B}^*$ threshold; the $Z_b(10650)$ has a mass near the
$B^*\bar{B}^*$ threshold.

(6) In 2013, the BESIII experiment used $e^+e^-$ collisions with
center-of-mass energies at the $Y(4260)$ mass to study substructure in
the process $e^+e^-\to\pi^+\pi^- J/\psi$~\cite{Ablikim:2013mio}.  They
observed the electrically charged $Z_c(3900)$ in the process $e^+e^-
\to \pi^\pm Z_c(3900)$, with $Z_c(3900) \to \pi^\mp J/\psi$.  This
process was simultaneously discovered by the Belle experiment, except
using ISR instead of direct production of the
$Y(4260)$~\cite{Liu:2013dau}.  The $Z_c(4020)$ was also discovered in
2013 by the BESIII experiment in the process $e^+e^- \to \pi^\pm
Z_c(4020)$, with $Z_c(4020) \to \pi^\mp
h_c(1P)$~\cite{Ablikim:2013wzq}.  Similar to the case of bottomonium,
the charmoniumlike $Z_c(3900)$ and $Z_c(4020)$ are near the
$D\bar{D}^*$ and $D^*\bar{D}^*$ thresholds, respectively.

We have therefore uncovered a number of parallels between charmonium
and bottomonium.  In the charmonium system, there is a series of
unexplained ``$Y$'' states decaying to charmonium, such as the
$Y(4260)$ and the $Y(4360)$; in bottomonium, there may be an exotic
state with mass similar to the $\Upsilon(5S)$ [or at least
unexpectedly large decays of the $\Upsilon(5S)$ to other bottomonium
states].  In the charmonium system, the $Z_c(3900)$ and $Z_c(4020)$
lie near the $D\bar{D}^*$ and $D^*\bar{D}^*$ thresholds, respectively;
in bottomonium, the $Z_b(10610)$ and $Z_b(10650)$ lie near the
$B\bar{B}^*$ and $B^*\bar{B}^*$ thresholds, respectively.  In the
following, we discuss the bottomonium and charmonium regions
separately.

\subsubsection{Cross Sections in the Bottomonium Region}
\label{sec:eeBottomonium}

\begin{figure}[htb]
\begin{center}
\includegraphics*[width= 1.0\columnwidth]{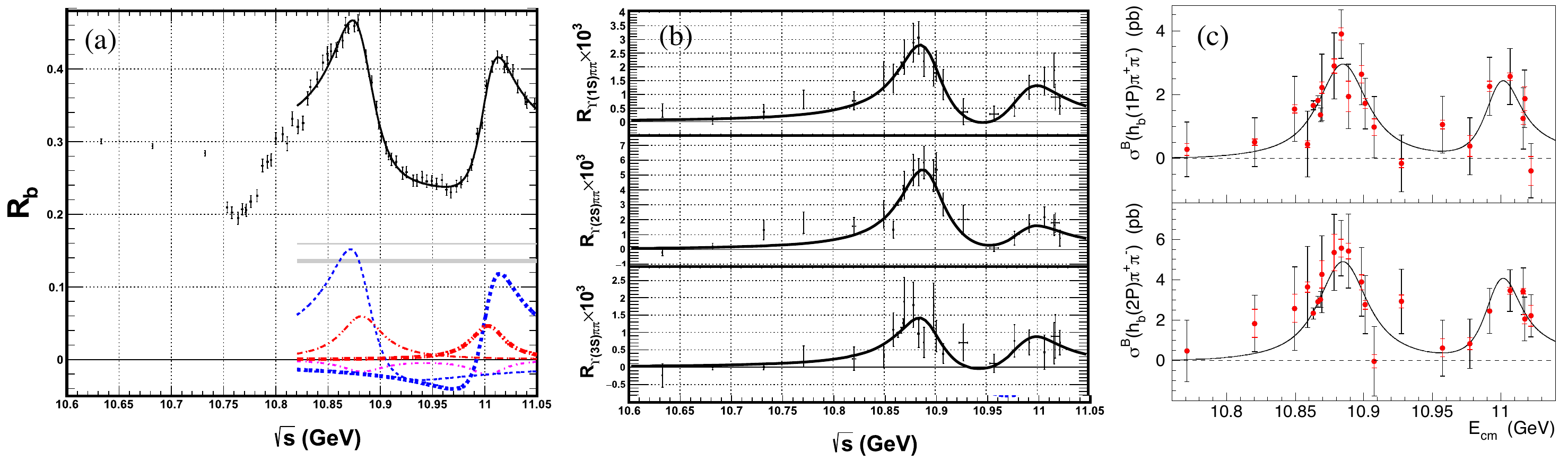}
\end{center}
\caption{\label{fig:eeBottomonium} 
Inclusive and exclusive $e^+e^-$ cross sections in the bottomonium
region as a function of center-of-mass energy~($\sqrt{s}$ or
$E_{CM}$).  The $\Upsilon(5S)$ and $\Upsilon(6S)$ are present in each
reaction.
(a)~The inclusive $e^+e^-$ cross section~(shown as $R_b \equiv
\sigma(b\overline{b})/\sigma_{\mu\mu}^0$).  The solid lines are for a
fit that includes interfering $\Upsilon(5S)$ and $\Upsilon(6S)$ states
as well as coherent and incoherent backgrounds~\cite{Santel:2015qga}.
(b)~The exclusive $e^+e^- \to \pi^+\pi^-\Upsilon(1S,2S,3S)$ cross
sections~\cite{Santel:2015qga}.
(c)~The exclusive $e^+e^- \to \pi^+\pi^-h_b(1P,2P)$ cross
sections~\cite{Abdesselam:2015zza}. Note that all five of the
exclusive cross sections are dominated by the $\Upsilon(5S)$ and
$\Upsilon(6S)$. All figures are from Belle.
}
\end{figure}

Above the $\Upsilon(4S)$, which has a mass just above the threshold to
produce $B\bar{B}$ pairs, the inclusive $e^+e^- \to b\overline{b}$
cross section $\sigma(b\overline{b})$ clearly shows two additional
peaks~\cite{Chen:2008xia,Aubert:2008ab,Santel:2015qga}.  These peaks
are illustrated in Fig.~\ref{fig:eeBottomonium}a, where
$\sigma(b\overline{b})$ is normalized by the Born cross section
$\sigma_{\mu\mu}^0$ (for $e^+e^- \to \mu^+\mu^-$) to form the variable
$R_b \equiv \sigma(b\overline{b})/\sigma_{\mu\mu}^0$.  These two peaks
are the $\Upsilon(10860)$ and the $\Upsilon(11020)$, often abbreviated
as the $\Upsilon(5S)$ and the $\Upsilon(6S)$, respectively, even
though the $5S$ and $6S$ quark-model assignments are not certain.
Determining the parameters of the $\Upsilon(5S)$ and $\Upsilon(6S)$
from the $R_b$ spectrum is complicated by large interference effects
between the resonant $\Upsilon(5S)$ and $\Upsilon(6S)$ amplitudes and
the nonresonant $b\overline{b}$ amplitude (which itself is not
expected to be a simple function in this region of multiple
thresholds)~\cite{Santel:2015qga}.  For the same reason, it is
difficult to precisely determine the electronic widths of the
$\Upsilon(5S)$ and $\Upsilon(6S)$.

The exclusive cross sections for $e^+e^-\to\pi^+\pi^- \Upsilon(nS)$
(where $n=1,2,3$) also show two peaks, but apparently without any of
the nonresonant
backgrounds~(Fig.~\ref{fig:eeBottomonium}b)~\cite{Santel:2015qga}.
The Belle experiment has performed three separate analyses of these
cross sections, with each analysis including progressively more data
and more sophistication.

In the first analysis~\cite{Abe:2007tk}, completed in 2008, data at a
single center-of-mass energy near the peak of the $\Upsilon(5S)$ was
taken, and the cross sections for $e^+e^-\to\pi^+\pi^- \Upsilon(nS)$
were measured at this point.  Assuming the entire inclusive $b\bar{b}$
cross section at the same point was from the $\Upsilon(5S)$, the
$\Upsilon(5S)$ partial widths to $\pi^+\pi^-\Upsilon(nS)$ could be
computed from the ratio of exclusive to inclusive cross sections.
These partial widths were found to be much larger than those for the
lower-lying $\Upsilon$ states.  For example, the
$\Upsilon(5S)\to\pi^+\pi^-\Upsilon(1S)$ partial width was found to be
0.59~MeV, compared to the $\Upsilon(4S)\to\pi^+\pi^-\Upsilon(1S)$
partial width of 0.0019~MeV\@.  In fact, due to the assumption about
the inclusive $b\bar{b}$ cross section, it is now thought that these
$\Upsilon(5S)$ partial widths were underestimated, making the
discrepancy even larger.

In the second analysis~\cite{Chen:2008xia}, completed in 2010, Belle
used seven center-of-mass energies around the $\Upsilon(5S)$ to
roughly map out the shape of the $e^+e^-\to\pi^+\pi^- \Upsilon(nS)$
cross sections.  The peak in these exclusive cross sections was found
to be at a higher mass than the $\Upsilon(5S)$ mass as it appears in
the inclusive cross section.  The discrepancy was $9\pm4$~MeV.  This
result led to the postulation of the $Y_b(10888)$ as a separate state
from the $\Upsilon(5S)$.

Finally, in the third analysis~\cite{Santel:2015qga}, completed in
2016, Belle used a much larger number of center-of-mass energy points
to map out the region of the $\Upsilon(5S)$ and $\Upsilon(6S)$.  Here
the argument shifted.  Since two peaks could be seen clearly in the
exclusive $e^+e^-\to\pi^+\pi^- \Upsilon(nS)$ cross sections, and with
negligible backgrounds, these peaks were now used to define the
parameters of the $\Upsilon(5S)$ and $\Upsilon(6S)$.  The fit to the
$R_b$ spectrum yielded consistent parameters, but the interference
with the nonresonant $b\bar{b}$ continuum makes the fits to the $R_b$
spectrum unreliable.  Hence the status of an exotic $Y_b(10888)$
remains unsettled, and so does the reason for the anomalously large
$\pi^+\pi^- \Upsilon(nS)$ partial widths of the ``$\Upsilon(5S)$''.

The same two peaks are also apparent in the exclusive cross sections
for $e^+e^-\to\pi^+\pi^- h_b(nP)$ with $n=1,2$
(Fig.~\ref{fig:eeBottomonium}c)~\cite{Abdesselam:2015zza}.  Again, the
$\Upsilon(5S)$ and $\Upsilon(6S)$ appear with little nonresonant
background.  The sizes of the cross sections are similar to those for
$e^+e^-\to\pi^+\pi^- \Upsilon(nS)$.

\subsubsection{Cross Sections in the Charmonium Region}
\label{sec:eeCharmonium}

\begin{figure}[htb]
\begin{center}
\includegraphics*[width= 1.0\columnwidth]{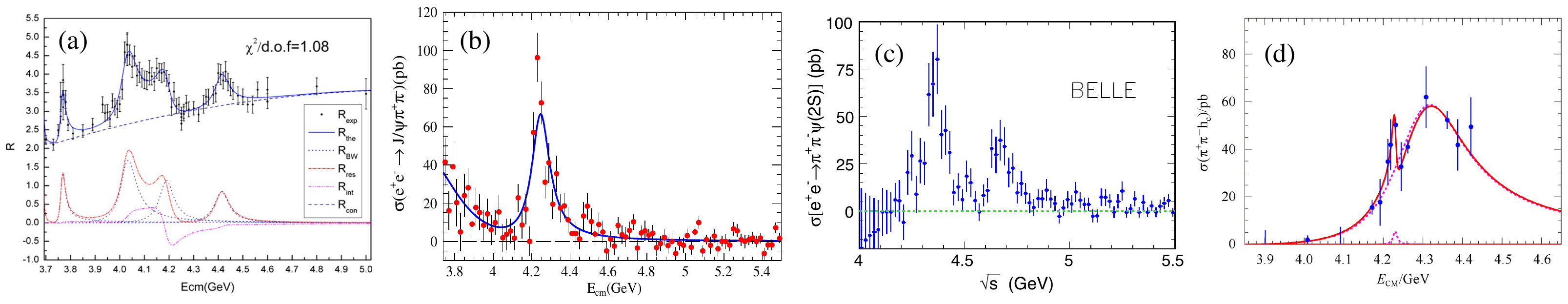}
\end{center}
\caption{\label{fig:eeCharmonium} 
Inclusive and exclusive $e^+e^-$ cross sections in the charmonium
region as a function of center-of-mass energy~($\sqrt{s}$ or
$E_{CM}$).
(a)~The inclusive $e^+e^-$ cross section~(shown as $R \equiv
\sigma(q\overline{q})/\sigma_{\mu\mu}^0$) from
BESII~\cite{Ablikim:2007gd}.  The solid lines are for a fit that
includes interfering $\psi(3770)$, $\psi(4040)$, $\psi(4160)$, and
$\psi(4415)$ states as well as a non-interfering continuum background.
(b)~The exclusive $e^+e^- \to \pi^+\pi^-J/\psi$ cross section from
BaBar showing the $Y(4260)$~\cite{Lees:2012cn}.
(c)~The exclusive $e^+e^- \to \pi^+\pi^-\psi(2S)$ cross section from
Belle showing the $Y(4360)$ and $Y(4660)$~\cite{Wang:2014hta}.
(d)~The exclusive $e^+e^- \to \pi^+\pi^-h_c(1P)$ cross section from
BESIII with a fit to a narrow peak with a mass near 4.23~GeV and a
wider peak at higher mass~\cite{Chang-Zheng:2014haa}.
}
\end{figure}

While the inclusive $e^+e^-$ cross section at center-of-mass energies
in the bottomonium region above the $\Upsilon(4S)$ shows two peaks,
the $\Upsilon(5S)$ and $\Upsilon(6S)$, the inclusive $e^+e^-$ cross
section in the charmonium region above the $\psi(3770)$ shows three,
the $\psi(4040)$, $\psi(4160)$, and
$\psi(4415)$~(Fig.~\ref{fig:eeCharmonium}a)~\cite{Ablikim:2007gd}.
These peaks match well with potential model expectations for the
$n^{2S+1}L_J = 3^3S_1$, $2^3D_1$, and $4^3S_1$ states of charmonium,
respectively~\cite{Barnes:2005pb}.  However, many complications arise
when exclusive $e^+e^-$ cross sections are considered.

The first of the puzzling exclusive $e^+e^-$ cross sections to be
measured was $e^+e^- \to \pi^+\pi^- J/\psi$, where the $Y(4260)$
appeared as a peak in the cross section around 4.26~GeV
(Fig.~\ref{fig:eeCharmonium}b)~\cite{Liu:2013dau,Aubert:2005rm,
  He:2006kg,Coan:2006rv,Yuan:2007sj,Lees:2012cn}, and with a cross
section around two orders of magnitude smaller than the inclusive
cross section.  The mass of the $Y(4260)$ lies between the masses of
the $\psi(4140)$ and $\psi(4415)$.  In fact, in the inclusive $e^+e^-$
cross section, the region of the $Y(4260)$ has an apparently
featureless depletion of events.  Setting an upper limit on the
inclusive decays of the $Y(4260)$ has allowed a lower limit to be
calculated for the branching fraction of the decay
$Y(4260)\to\pi^+\pi^-J/\psi$ of 0.6\%~\cite{Mo:2006ss}, although this
calculation involves a relatively difficult fit to the inclusive cross
section.  Besides corresponding to a dip in the inclusive cross
section, the shape of the $e^+e^-\to Y(4260) \to \pi^+\pi^-J/\psi$
cross section also appears strange: It rises rapidly below the peak
and falls more slowly above the peak.  The Belle experiment attributed
this asymmetry to interference with a lower-mass
$Y(4008)$~\cite{Liu:2013dau,Yuan:2007sj}, although the BaBar
experiment could not confirm this hypothesis~\cite{Lees:2012cn}.
 The BESIII experiment has reported that the $Y(4260)$ may in 
 fact consist of two peaks, a narrow peak around 4.22~GeV and a wider
 peak around 4.31~GeV, accounting for the asymmetry~\cite{Ablikim:2016qzw}.
The shape of the $e^+e^- \to \pi^0\pi^0 J/\psi$ cross section is
consistent with that of the $e^+e^- \to \pi^+\pi^- J/\psi$ cross
section and is suppressed by a factor of two, consistent with
expectations for an isosinglet $Y(4260)$~\cite{Ablikim:2015tbp}.

Finding new decay modes of the $Y(4260)$ has proven to be difficult.
It is not sufficient to measure a single exclusive $e^+e^-$ cross
section at 4.26~GeV, but one must instead measure the cross section at
a range of energies in order to determine whether the energy
dependence of the cross section corresponds to the $Y(4260)$.  For
example, the CLEO-c experiment measured a non-zero cross section for
$e^+e^- \to K^+K^-J/\psi$ at 4.26~GeV~\cite{Coan:2006rv}, but this
single point is not sufficient to establish the existence of the decay
$Y(4260) \to K^+K^-J/\psi$.  Attempts to establish this decay by the
Belle experiment have lacked the required
statistics~\cite{Yuan:2007bt,Shen:2014gdm}.

When the data has been sufficient to map exclusive cross sections as a
function of center-of-mass energy, the $Y(4260)$ has not been found.
In $e^+e^-\to\pi^+\pi^-\psi(2S)$, there are two clear peaks, the
$Y(4360)$ and the $Y(4660)$
(Fig.~\ref{fig:eeCharmonium}c)~\cite{Wang:2007ea,Lees:2012pv,
Wang:2014hta}.  In $e^+e^-\to\pi^+\pi^-h_c(1P)$, the data is also
clearly inconsistent with a $Y(4260)$; there is some evidence for a
narrow peak around 4.23~GeV and a much wider peak at higher mass
(Fig.~\ref{fig:eeCharmonium}d)~\cite{Ablikim:2013wzq,
Chang-Zheng:2014haa}.  The $\omega\chi_{c0}$ cross section also shows
evidence for peaking at a mass lower than that of the $Y(4260)$, a
feature that has been named the $Y(4230)$~\cite{Ablikim:2014qwy}.
Other cross sections, such as $\eta J/\psi$~\cite{Ablikim:2012ht,
Wang:2012bgc,Ablikim:2015xhk}, $\omega
\chi_{c1,2}$~\cite{Ablikim:2015uix}, and
$\Lambda_c\bar{\Lambda}_c$~\cite{Pakhlova:2008vn} [where the $X(4630)$
has been reported], have also proved to be remarkably complex.

Understanding the open-charm cross sections, which are typically an
order of magnitude larger than the closed-charm cross sections listed
above, is likely a prerequisite for sorting out all of the structure
seen in exclusive $e^+e^-$ cross sections in the charmonium region.
Many open-charm cross sections have been measured by the CLEO-c
experiment~\cite{CroninHennessy:2008yi}, BaBar~\cite{Aubert:2006mi,
Aubert:2009aq,delAmoSanchez:2010aa}, and Belle~\cite{Abe:2006fj,
Pakhlova:2008zza,Pakhlova:2007fq,Pakhlova:2009jv,Pakhlova:2010ek}, but
higher-statistics measurements should be soon provided by the BESIII
experiment.

\subsubsection{Substructure in the Bottomonium Region}
\label{sec:eeSubBottomonium}

\begin{figure}[htb]
\begin{center}
\includegraphics*[width= 1.0\columnwidth]{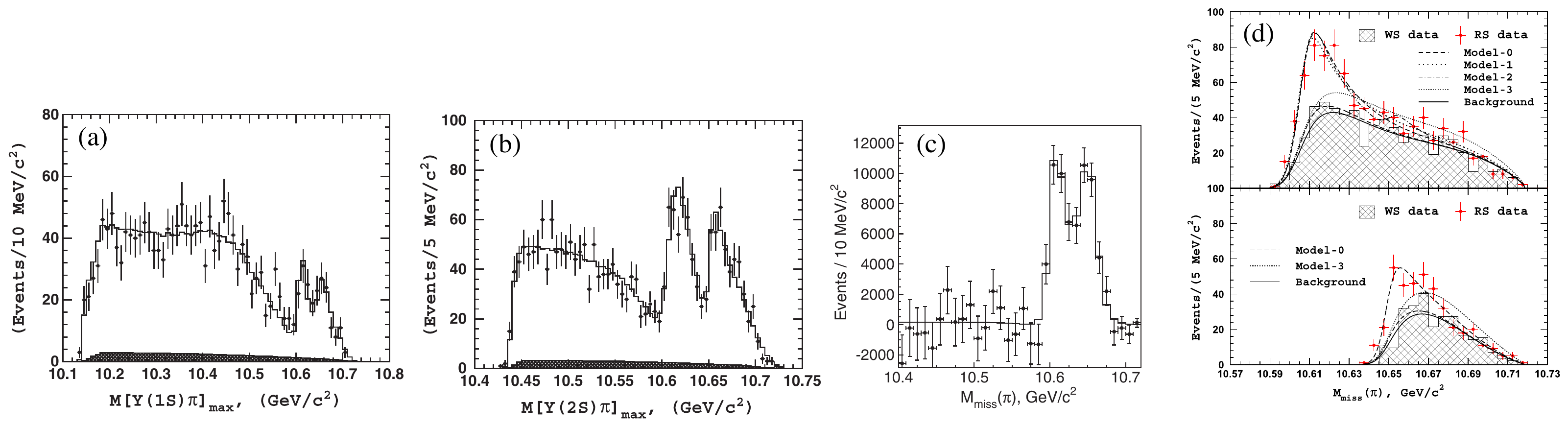}
\end{center}
\caption{\label{fig:eeZb} 
The $Z_b$ states observed in $e^+e^-$ annihilation in the bottomonium
region.   (a,b,c)~Observation of the $Z_b(10610)$ and $Z_b(10650)$ in
$e^+e^- \to \pi^\mp Z_b$ with the $Z_b$ decaying to $\pi^\pm
\Upsilon(1S)$~(a), $\pi^\pm \Upsilon(2S)$~(b), and $\pi^\pm
h_b(1P)$~(c)~\cite{Belle:2011aa}.  (d)~Observation of the $Z_b(10610)$
decaying to $(B\bar{B}^*)^\pm$~(top) and the $Z_b(10650)$ decaying to
$(B^*\bar{B}^*)^\pm$~(bottom)~\cite{Garmash:2015rfd}.  All figures are
from the Belle experiment.  }
\end{figure}

In the bottomonium region, we have already seen that there are
surprisingly large cross sections for $e^+e^- \to
\pi^+\pi^-\Upsilon(1S,2S,3S)$ and $e^+e^- \to \pi^+\pi^-h_b(1P,2P)$ at
center-of-mass energies near the $\Upsilon(5S)$
mass~(Sec.~\ref{sec:eeBottomonium}).  Perhaps more interesting is the
fact that all five of these reactions proceed, either entirely or
partially, through the intermediate processes $e^+e^- \to \pi^\pm
Z_b(10610)$ and $e^+e^- \to \pi^\pm Z_b(10650)$, where the
$Z_b(10610)$ and $Z_b(10650)$ are electrically charged, have widths on
the order of 20~MeV, and decay to $\pi^\mp \Upsilon(1S,2S,3S)$ and
$\pi^\mp h_b(1P,2P)$.  These results were discovered by the Belle
experiment in 2012 in an analysis of all five
reactions~\cite{Belle:2011aa}.  In the study of $e^+e^- \to \pi^\pm
Z_b$ with $Z_b\to\pi^\mp\Upsilon(nS)$, separate two-dimensional
Dalitz-plot fits for $n=1,2,3$ were
performed~(Figs.~\ref{fig:eeZb}a,b).  The $Z_b \to \pi^\mp h_b(nP)$
(with $n=1,2$) processes were studied using one-dimensional fits to
the $\pi^\mp h_b(nP)$ mass distributions, which were obtained by
fitting for the $h_b(nP)$ yield in bins of $\pi^\mp h_b(nP)$
mass~(Fig.~\ref{fig:eeZb}c).  The masses and widths of the
$Z_b(10610)$ and $Z_b(10650)$ were consistent in all five reactions,
and the combined significance of both $Z_b$ states was over $10\sigma$
in each reaction.  In 2015, the study of the $e^+e^- \to
\pi^+\pi^-\Upsilon(1S,2S,3S)$ processes was extended to include a
six-dimensional amplitude analysis~\cite{Garmash:2014dhx}.  The $J^P =
1^+$ hypothesis was favored for both the $Z_b(10610)$ and the
$Z_b(10650)$.

A neutral version of the $Z_b(10610)$ was seen by Belle in 2013 in the
related processes $e^+e^- \to \pi^0 Z_b(10610)$ with
$Z_b(10610)\to\pi^0\Upsilon(2S,3S)$, with a combined significance of
$6.5\sigma$~\cite{Krokovny:2013mgx}.  The ratio of cross sections for
the charged and neutral processes was consistent with expectations for
an isovector $Z_b(10610)$.  The statistics were not sufficient to
observe the $Z_b(10610)\to\pi^0\Upsilon(1S)$ decay or the
$Z_b(10650)\to\pi^0\Upsilon(1S,2S,3S)$ decays, but the upper limits
were consistent with isospin expectations.

One of the most striking features of the $Z_b(10610)$ and $Z_b(10650)$
is that their masses are just above the thresholds needed to produce
$B\bar{B}^*$ and $B^*\bar{B}^*$, respectively.  This fact prompted a
study of the processes $e^+e^-\to B^{(*)}\bar{B}^{(*)}\pi$ with
center-of-mass energy near the $\Upsilon(5S)$ mass by the Belle
experiment~\cite{Garmash:2015rfd}.  By fully reconstructing one $B$
meson and the pion, Belle was able to observe the decays
$Z_b(10610)\to B\bar{B}^*$ (where $B\bar{B}^*$ is shorthand for
$B^+\bar{B}^{*0}$ and $\bar{B}^0 B^{*+}$ and their charge conjugates)
and $Z_b(10650)\to B^*\bar{B}^*$ (where $B^*\bar{B}^*$ is shorthand
for $B^{*+}\bar{B}^{*0}$ and its charge conjugate), shown in
Fig.~\ref{fig:eeZb}d.  No evidence was found for the kinematically
allowed $Z_b(10650)\to B\bar{B}^*$ decay, and no evidence was found
for the process $e^+e^-\to B\bar{B}\pi$.  Assuming the charged
$Z_b(10610)$ and $Z_b(10650)$ decay only to $\pi^\pm
\Upsilon(1S,2S,3S)$, $\pi^\pm h_b(1P,2P)$, and $B\bar{B}^{(*)}$ (which
is supported by the study of the inclusive $\Upsilon(5S)$ cross
section~\cite{Santel:2015qga}), branching fractions could be
calculated.  It was found that the open-bottom decays are roughly an
order of magnitude larger than the closed-bottom decays.

\subsubsection{Substructure in the Charmonium Region}
\label{sec:eeSubCharmonium}

While there are two $Z_b$ states in the bottomonium region, one with
mass near the $B\bar{B}^*$ threshold and one with mass near the
$B^*\bar{B}^*$ threshold, there are analogous $Z_c$ states in the
charmonium region near the $D\bar{D}^*$ and $D^*\bar{D}^*$ thresholds,
although with a few additional complications.  Note that these $Z_c$
states produced in $e^+e^-$ annihilation are distinct from those
produced in $B$ decays~(Sec.~\ref{sec:WeakDecays}).

\begin{figure}[htb]
\begin{center}
\includegraphics*[width= 1.0\columnwidth]{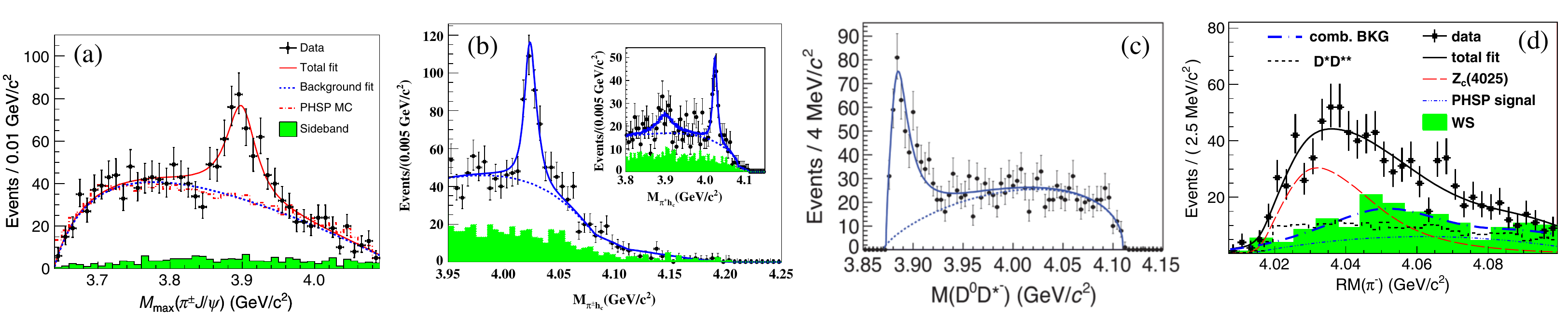}
\end{center}
\caption{\label{fig:eeZc} 
The $Z_c$ states observed in $e^+e^-$ annihilation in the charmonium
region.
(a)~Observation of the $Z_c(3900)$ in $e^+e^- \to \pi^\mp Z_c$ with
the $Z_c$ decaying to $\pi^\pm J/\psi$~\cite{Ablikim:2013mio}.
(b)~Observation of the $Z_c(4020)$ in $e^+e^- \to \pi^\mp Z_c$ with
the $Z_c$ decaying to $\pi^\pm h_c(1P)$~\cite{Ablikim:2013wzq}.
(c)~Observation of the $Z_c(3900)$ decaying to
$(D\bar{D}^*)^\pm$~\cite{Ablikim:2013xfr}.
(d)~Observation of the $Z_c(4020)$ decaying to
$(D^*\bar{D}^*)^\pm$~\cite{Ablikim:2013emm}.
All figures are from the BESIII experiment.
}
\end{figure}

The first of the $Z_c$ states discovered in $e^+e^-$ annihilation was
the $Z_c(3900)$.  The $Z_c(3900)$ was simultaneously discovered by
BESIII and Belle in 2013 in analyses of the process $e^+e^- \to
\pi^+\pi^- J/\psi$ with center-of-mass energies near the $Y(4260)$
mass.  The BESIII experiment used a single center-of-mass energy at
4.26~GeV~\cite{Ablikim:2013mio}; the Belle experiment covered a wider
range by using the initial-state radiation
technique~\cite{Liu:2013dau}.  For both experiments, the $Z_c(3900)$
appeared as a peak in the mass spectrum of the $\pi^\pm J/\psi$
system, with a width of around 50~MeV~(Fig.~\ref{fig:eeZc}a).  Only
one-dimensional fits were performed, but studies of the $\pi^+\pi^-$
system were carried out to demonstrate that the $Z_c(3900)$ peak did
not originate from kinematic reflections.  Analogous to the
$Z_b(10610)$ of bottomonium, the $Z_c(3900)$ is near the $D\bar{D}^*$
threshold.  But unlike bottomonium, the $\pi^\pm J/\psi$ system showed
no sign of a second state near the $D^*\bar{D}^*$ threshold (which is
just above 4~GeV).

Shortly after the discovery of the $Z_c(3900)$, the BESIII experiment
did observe a second state near the $D^*\bar{D}^*$ threshold,
analogous to the $Z_b(10650)$ of bottomonium~\cite{Ablikim:2013wzq}.
It was discovered in the process $e^+e^- \to \pi^+\pi^- h_c(1P)$,
where three center-of-mass energies~(4.23, 4.26, and 4.36~GeV) were
analyzed near the $Y(4260)$ mass.  The $Z_c(4020)$ was observed as a
narrow peak~(with a width of roughly 8~MeV) in the $\pi^\pm h_c(1P)$
mass spectrum~(Fig.~\ref{fig:eeZc}b).  No evidence for the
$Z_c(3900)\to\pi^\pm h_c(1P)$ could be found and only an upper limit
could be set.

The BESIII experiment also studied the $Z_c(3900)$ and $Z_c(4020)$ in
open-charm decays.  Like the $Z_b(10610)$, the $Z_c(3900)$ was found
to decay to $D\bar{D}^*$ in the process $e^+e^- \to
D\bar{D}^*\pi$~(where $D\bar{D}^*$ stands for both $D^+\bar{D}^{*0}$
and $\bar{D}^0 D^{*+}$ and their charge conjugates), both by
reconstructing a single $D$
meson~(Fig.~\ref{fig:eeZc}c)~\cite{Ablikim:2013xfr} and by
reconstructing both $D$ mesons~\cite{Ablikim:2015swa}.  The first of
these analyses also demonstrated the $J^P$ of the $Z_c(3900)$ to be
$1^+$.  And, like the $Z_b(10650)$, the $Z_c(4020)$ was found in the
process $e^+e^- \to D^*\bar{D}^*\pi$ decaying to $D^*\bar{D}^*$, where
$D^*\bar{D}^*$ stands for $D^{*+}\bar{D}^{*0}$ and its charge
conjugate~(Fig.~\ref{fig:eeZc}c)~\cite{Ablikim:2013emm}.  Also similar
to bottomonium, the decays of the $Z_c(3900)$ and $Z_c(4020)$ to open
charm are roughly an order of magnitude larger than their decays to
closed charm.  The masses and widths of the $Z_c$ states as observed
in their closed- and open-charm decays are not entirely
consistent---the $Z_c(3900)$ is lighter and narrower in its open-charm
decay, while the $Z_c(4020)$ is heavier and wider in its open-charm
decay---but it is highly probable that the closed- and open-charm
channels are related.

Neutral partners to the $Z_c(3900)$ and $Z_c(4020)$ were subsequently
discovered in the neutral versions of all four reactions listed above.
The $Z_c(3900)$ was found to decay to $\pi^0
J/\psi$~\cite{Xiao:2013iha,Ablikim:2015tbp} and
$(D\bar{D}^*)^0$~\cite{Ablikim:2015gda}; the $Z_c(4020)$ was found to
decay to $\pi^0 h_c(1P)$~\cite{Ablikim:2014dxl} and
$(D^*\bar{D}^*)^0$~\cite{Ablikim:2015vvn}.  In the analysis of the
$Z_c(3900)\to \pi^0 J/\psi$ decay~\cite{Ablikim:2015tbp}, the ratio of
the cross section for $e^+e^-\to\pi^0Z_c(3900)$ followed by
$Z_c(3900)\to\pi^0 J/\psi$ to the cross section for
$e^+e^-\to\pi^0\pi^0 J/\psi$ was measured at a number of different
center-of-mass energies.  The sizes of the data samples, however, were
not sufficient to determine whether or not the
$e^+e^-\to\pi^0Z_c(3900)$ process proceeds through a $Y(4260)$.

A third $Z_c$ state, the $Z_c(4055)$, was reported by the Belle
Collaboration in the process $e^+e^- \to \pi^\pm Z_c$ with $Z_c \to
\pi^\mp \psi(2S)$ for center-of-mass energies near the
$Y(4360)$~\cite{Wang:2014hta}.  Its mass and width are clearly
inconsistent with both the $Z_c(3900)$ and the $Z_c(4020)$.  These
results also present a striking dissimilarity with the bottomonium
system, where the parameters of the $Z_b(10610)$ and the $Z_b(10650)$
are consistent in all three reactions $e^+e^-\to
\pi^+\pi^-\Upsilon(1S,2S,3S)$.  The $Z_c(4055)$ requires further
study.

It is interesting to note that neither the $Z_c(3900)$ nor the
$Z_c(4055)$ has been seen in $B$ decays.  The $Z_c(3900)$ could have
been seen in the decay $B\to K\pi J/\psi$, but instead the $Z_c(4200)$
and $Z_c(4430)$ were found~(Sec.~\ref{sec:otherB}).  Similarly, the
$Z_c(4055)$ could have been seen in the decay $B\to K\pi \psi(2S)$,
but instead the $Z_c(4240)$ and $Z_c(4430)$ were
found~(Sec.~\ref{sec:Z4430}).
The fact that the $Z_c(4055)$ and $Z_1(4050)$ (the latter produced in
$B\to K Z_1$ and decaying to $\pi^\pm \chi_{c1}$,
Sec.~\ref{sec:otherB}) have a similar mass and width must be
coincidence.  If the $Z_c(4055)$ were produced in $B$ decays, like the
$Z_1(4050)$, it would be seen in $B\to K\pi \psi(2S)$.  And if the
$Z_1(4050)$ were produced in $e^+e^-$ annihilation like the
$Z_c(4055)$, then in $e^+e^- \to \pi^\pm Z_1(4050)$ with $Z_1(4050)
\to \pi^\mp \chi_{c1}$ and $\chi_{c1}\to \gamma J/\psi$ would produce
a prominent $Z_1(4050)$ signal in $e^+e^- \to \gamma \pi^+\pi^-
J/\psi$, which is not seen~\cite{Ablikim:2013dyn}.  A search for the
$Z_c(4020)$ in $B$ decays has not yet been performed.

\subsection{The Region Between 3.8 and 4.0 GeV}
\label{sec:3800}
\label{sec:X3823}
\label{sec:XYZ3940}

The majority of the exotic states discussed above exhibit properties
that have clearly identified them as exotic: the $Z_c(4430)$ contains
a $c\bar{c}$ pair and has an electric charge; the $Y(4260)$ has a mass
that is incompatible with the predicted, and already discovered,
$J^{PC}=1^{--}$ quark-model states; the $X(3872)$ is extremely narrow
and has a mass remarkably close to the $D^0\bar{D}^{*0}$ threshold.
Other candidates for QCD exotica, especially in the region between 3.8
and 4.0~GeV, cannot be identified so obviously as exotic.  The
challenge in this region is to try to separate exotic candidates from
quark-model states, many of which are yet to be identified.  A few
assignments appear to be straightforward: the $X(3823)$ is likely the
$\psi_2(1D)$ ($n^{2s+1}L_J = 1^3D_2$) state of charmonium; and the
$Z(3930)$ is likely the $\chi_{c2}(2P)$ state.  But other assignments
are not settled: the $X(3915)$ [which is likely the same as the
$Y(3940)$] was previously identified as the $\chi_{c0}(2P)$ state, but
this assignment is problematic; and the interpretation of the
$X(3940)$ remains an outstanding issue.  Here we provide a few notes
on quark-model assignments.

(1)~{\bf The $X(3823)$ is the $\psi_2(1D)$.}
The $X(3823)$ was seen by Belle in the process $B\to
KX(3823)$~\cite{Bhardwaj:2013rmw} and by BESIII in the process $e^+e^-
\to \pi^+\pi^-X(3823)$~\cite{Ablikim:2015dlj}, where in both cases the
$X(3823)$ decayed to $\gamma \chi_{c1}$.  While the sizes of the data
samples in these two measurements were not sufficient to determine the
quantum numbers, the $J^{PC} = 2^{--}$ assignment is highly likely,
based upon its close match to the quark-model predictions for the
$\psi_2(1D)$ state of charmonium.  First, the mass of the $X(3823)$
closely matches the quark-model predictions for the mass of the
$\psi_2(1D)$ state, which is well constrained, given the
identification of the $\psi(3770)$ with the related $\psi(1D)$ state.
Second, the $X(3823)$ decays to $\gamma \chi_{c1}$, and the
$\psi_2(1D)$ state is expected to have a large partial width to
$\gamma \chi_{c1}$.  Upper limits on the $X(3823)$ decay to $\gamma
\chi_{c2}$ are also consistent with expectations for the $\psi_2(1D)$.
Finally, the $X(3823)$ is narrow, as expected for a $2^{--}$ state,
since the $D\bar{D}$ decay is forbidden by quantum numbers, and the
$X(3823)$ has a mass below the $D\bar{D}^*$ threshold.

\begin{figure}[htb]
\begin{center}
\includegraphics*[width= 1.0\columnwidth]{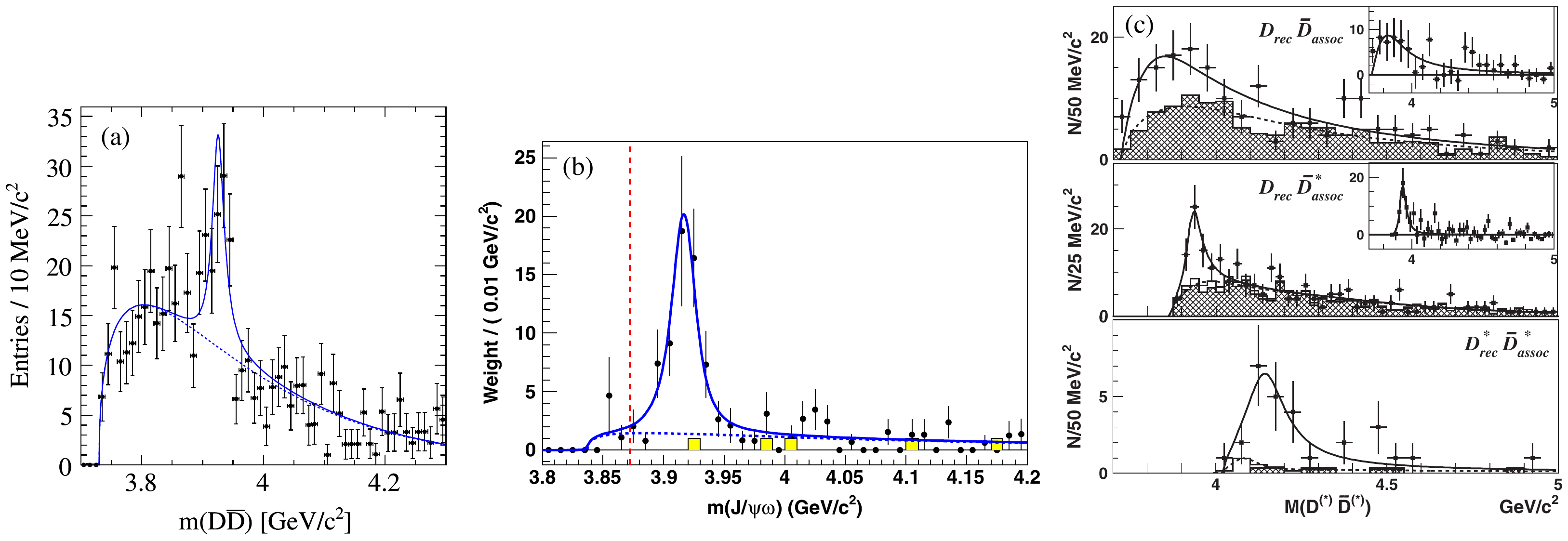}
\end{center}
\caption{\label{fig:XYZ3940} 
The $XY \! Z$ around 3.9~GeV\@.  (a)~Observation of the $Z(3930)$ by
BaBar in $\gamma\gamma \to Z$ with $Z \to
D\bar{D}$~\cite{Aubert:2010ab}.  (b)~Observation of the $X(3915)$ by
BaBar in $\gamma\gamma \to X$ with $X \to \omega
J/\psi$~\cite{Lees:2012xs}.  (c)~A study of $e^+e^- \to J/\psi +
D\bar{D},D\bar{D}^*,D^*\bar{D}^*$ at Belle~\cite{Abe:2007sya}.  The
$D\bar{D}$ system~(top) shows a broad excess of events; the
$D\bar{D}^*$ system~(middle) shows the $X(3940)$; the $D^*\bar{D}^*$
system~(bottom) shows the $X(4160)$.  }
\end{figure}

(2)~{\bf The $Z(3930)$ is the $\chi_{c2}(2P)$.}
The $Z(3930)$ was seen by both Belle~\cite{Uehara:2005qd} and
BaBar~\cite{Aubert:2010ab} in the process $\gamma \gamma \to Z(3930)$
with $Z(3930) \to D\bar{D}$~(Fig.~\ref{fig:XYZ3940}a).  Both
measurements could conclusively determine the $J^{PC}$ to be $2^{++}$.
Since the mass of the $Z(3930)$ is near the quark-model prediction for
the $\chi_{c2}(2P)$, and since it decays to $D\bar{D}$ as is expected
for the $\chi_{c2}(2P)$, the $\chi_{c2}(2P)$ assignment appears
reasonable.

(3)~{\bf Is the $X(3915)$ [identified with the $Y(3940)$] the
$\chi_{c0}(2P)$?}
The $Y(3940)$ was seen by both Belle~\cite{Abe:2004zs} and
BaBar~\cite{delAmoSanchez:2010jr,Aubert:2007vj} in the process $B\to K
Y(3940)$ with $Y(3940)\to \omega J/\psi$~(Fig.~\ref{fig:X3872}c).  The
initial mass measurement was near 3940~MeV~(hence the
name)~\cite{Abe:2004zs}, but subsequent measurements were near
3915~MeV~\cite{delAmoSanchez:2010jr,Aubert:2007vj}.  The $X(3915)$ was
seen by Belle~\cite{Uehara:2009tx} and BaBar~\cite{Lees:2012xs} in the
process $\gamma \gamma \to X(3915)$ with $X(3915)\to\omega
J/\psi$~(Fig.~\ref{fig:XYZ3940}b).  BaBar was also able to show that
the $J^{PC}$ is likely $0^{++}$~\cite{Lees:2012xs}.  Since their
masses and widths are consistent, and since they both decay to $\omega
J/\psi$, the $Y(3940)$ and $X(3915)$ are usually considered to be the
same state [referred to as the $X(3915)$].  The $X(3915)$ was
originally identified with the $\chi_{c0}(2P)$ state of charmonium,
based on its mass and likely $J^{PC}$, but this assignment has a
number of problems~\cite{Olsen:2014maa}.  First, the mass difference
between the $X(3915)$ and the $\chi_{c2}(2P)$ [or $Z(3930)$],
$8.8\pm3.2$~MeV, is far smaller than the expected
$\chi_{c0}(2P)$-$\chi_{c2}(2P)$ mass difference.  Second, if the
$X(3915)$ were the $\chi_{c0}(2P)$, then it should be seen in decays
to $D\bar{D}$, which is expected to be the dominant mode.  These
$D\bar{D}$ decays would have been evident in the analysis of $B\to
KD^0\bar{D}^0$ by Belle~\cite{Brodzicka:2007aa} if ${\cal
B}[X(3915)\to D^0\bar{D}^0] > 1.2\times {\cal B}[X(3915)\to\omega
J/\psi]$~\cite{Olsen:2014maa}.  The $X(3915)\to D\bar{D}$ decay should
also have been evident in the process $\gamma\gamma\to X(3915)$ with
$X(3915)\to D\bar{D}$, which is not seen in Fig.~\ref{fig:XYZ3940}a.

(4)~{\bf What is the $X(3940)$ [and the $X(4160)$]?}
The $X(3940)$ was first reported by Belle in the process $e^+e^- \to
J/\psi \, X(3940)$ with the $X(3940)$ decaying to
anything~\cite{Abe:2007jna}.  A later analysis examined the processes
$e^+e^- \to J/\psi \, D^{(*)}\bar{D}^{(*)}$, and the $X(3940)$ was
seen only in the $D\bar{D}^*$ decay~\cite{Abe:2007sya}.  In addition,
a peak named the $X(4160)$ was seen in $D^*\bar{D}^*$, and a broad
excess of events was seen in $D\bar{D}$~(Fig.~\ref{fig:XYZ3940}c).
None of these peaks currently have clear interpretations.

\subsection{Results Waiting for Confirmation}
\label{sec:confirmation}

The majority of the candidates for QCD exotica discussed above are
experimentally on solid ground.  Even many of the states that were
controversial initially, such as the $Z_c(4430)$ and the $Y(4140)$,
have become firmly established over the last several years.  In this
section we single out a few states, though, that remain unsettled and
require confirmation.

LHCb, with its larger samples of $B$ decays than those collected by
the $B$ factories, has confirmed the existence of a number of the
states seen in $B$ decays.  There are a few more channels, however,
that need to be revisited.  The existence of the $Z_1(4050)$ and the
$Z_2(4250)$ in $B\to K(\pi^\pm \chi_{c1})$, and the existence of the
$Z_c(4200)$ and the $Z_c(4430)$ in $B\to K (\pi^\pm J/\psi)$, both
reported by the Belle experiment but not seen by BaBar, remain
somewhat controversial.

In the $e^+e^-$ sector, the large number of $Y$ states in the
charmonium region needs to be investigated.  While many features of
the data are statistically significant, there is apparently little
order from channel to channel.  A more global analysis of the data is
required to understand the effects of cross-channel scattering.  Such
an analysis could settle the existence or non-existence of a few of
the $Y$ states, such as the $Y(4230)$, and may help clarify the
properties of the $Y(4260)$.

In $\gamma\gamma$ collisions, the $X(3915)$ (decaying to $\omega
J/\psi$) is firmly established. The presumably related $X(4350)$,
reported by Belle to decay to $\phi J/\psi$~\cite{Shen:2009vs},
however, requires confirmation.

The issue of the $X(5568)$, recently reported by the D0 experiment in
inclusive $p\overline{p}$ production at a center-of-mass energy of
1.96~TeV~\cite{D0:2016mwd}, also remains unsettled.  Because it decays
to $B_s \pi^\pm$, it could be a tetraquark state that contains four
separate quark flavors, $b$, $s$, $u$, and $d$.  It could be related
to the electrically charged $Z_c$ (containing $c\bar{c}$ and light
quarks) or the $Z_b$ (with $b\bar{b}$ and light quarks), but it
differs in the fact that its mass is significantly below the threshold
to decay to two open-(heavy)-flavor mesons, in this case a $B$ and a
$K$, while the $Z_c$ and $Z_b$ states have masses above the open-charm
and open-bottom thresholds, respectively.  The D0 experiment reported
that a significant fraction (around 10\%) of the $B_s$ produced in the
transverse momentum region between 10 and 30~GeV originated from
$X(5568)$ decays.  The LHCb experiment searched for the same state,
but with $pp$ collisions and with center-of-mass energies at 7 and
8~TeV, but found no evidence for it~\cite{Aaij:2016iev}.  LHCb set an
upper limit of around 2\% for the fraction of $B_s$ originating from
$X(5568)$ decays for transverse momentum of the $B_s$ above 10~GeV.
The $X(5568)$ certainly deserves further study.


\section{Theory Applications} \label{sec:Theory}

\subsection{Molecular Picture} \label{sec:Molecule}

We begin with the first theoretical picture proposed to describe the
structure of hidden-flavor multiquark hadrons, that of hadronic
molecules.  The original proposal of charmed-meson molecules, as
already noted, far predated~\cite{Voloshin:1976ap,DeRujula:1976zlg}
the discovery~\cite{Choi:2003ue} of the first confirmed exotic
candidate, the $X(3872)$.

\subsubsection{General Considerations: Binding Energy and Size}

The only states thus far absolutely known to be hadronic molecules are
the composite nuclei, lending hope that one may attempt to draw some
useful insights from their attributes.  The deuteron stands alone as
the only confirmed two-hadron bound state, making it a suitable
prototype for heavy-hadron molecules~\cite{Tornqvist:2004qy}.  The
essential properties of the deuteron for this purpose are (i) that its
quantum numbers ($Q = 1$, $J^P = 1^+$, $I = 0$) are accessible to a
bound state of a component proton and a neutron, (ii) the proximity of
its mass $m_D \equiv 2m_N - B$ to the threshold for dissociation into
$p + n$ (binding energy $B = 2.2$~MeV), which suggests a large
characteristic size $R$ for the state,
\begin{equation}
R \equiv \frac{\hbar}{\sqrt{2\mu B}} = 4.3 \ {\rm fm} \, ,
\end{equation}
where the reduced mass $\mu \simeq m_N/2$, and (iii) a large $p$-$n$
spin-triplet ($t$) {\it scattering length\/} $a_t = 5.3$~fm in the
corresponding channel, supporting its interpretation as a bound
state~\cite{Sakurai:2011zz}.  Indeed, for sufficiently small $B$, the
only length parameter describing the bound state is the scattering
length, $R \to a$, a phenomenon known as {\it low-energy
universality}~\cite{Braaten:2003he}.

In fact, the scattering length by itself provides only partial
information on the structure of the state.  A more incisive test comes
through considering the next moment in the {\it effective-range
  expansion\/} of the low-momentum ($k$), $s$-wave ($\ell = 0$)
scattering amplitude $f_0$, which is called the {\it effective
  range\/} $r_0$:
\begin{equation}
f_0 = \frac{1}{k \cot \delta_0 (k) - i k} = \frac{1}{\frac 1 a +
  r_0 \frac{k^2}{2} - i k} \, .
\end{equation}
For the deuteron channel, $r_0 = 1.75$~fm.  Weinberg long ago derived
a criterion~\cite{Weinberg:1965zz} for determining in terms of $a$,
$r_0$, and $R$ whether a state is primarily extended (composite) or
compact (elementary).  The parameter connecting the observables is the
wave function renormalization pole residue $Z$, which is 0 for purely
composite particles like molecules and approaches 1 for an elementary
state.  The relations read
\begin{equation}
a   = 2 \left( \frac{1-Z}{2-Z} \right) R + O \left( \frac{1}{m}
\right) \, , \ \ \
r_0 = - \left( \frac{Z}{1-Z} \right) R + O \left( \frac{1}{m} \right)
\, ,
\end{equation}
where $m$ represents corrections due to the momentum scale of the
binding interactions ({\it i.e.}, $m$ is set to $m_\pi$ if one-pion
exchange is the primary binding mechanism).  Noting that the deuteron
satisfies $a_t > R$, and especially that $r_0 > 0$, Weinberg deduced
that $Z$ cannot be too close to 1, and indeed that the deuteron is
dominated by its composite component since $Z$ lies much closer to 0
than 1\@.  In principle, such measurements for the heavy-quark exotics
should become feasible in the future when near-threshold production
experiments become possible and detailed {\it line shapes\/} of the
production amplitudes for the states become available.  At present,
however, not even the sign of $B$ for $X(3872)$ has been uniquely
fixed.

As one further figure of merit for studies of hadronic molecules, the
largest binding energies per nucleon for compound nuclei are $<
9$~MeV\@.  Such numbers are obtained, for example, in nuclear shell
models by starting with a basic attractive nucleon-nucleon potential
of depth $\approx 50$~MeV and then adding various
corrections~\cite{Krane:1987ky}.  One therefore expects all true
heavy-hadron molecules to lie not far below dissociation thresholds
(tens of MeV or less) and to have large spatial extent
[$O$(1--10~fm)].

\subsubsection{Dynamics of Binding}

Of course, bound states must also possess a dynamical mechanism that
can provide a sufficiently attractive binding interaction.  In the
deuteron, the long-distance attraction necessary for this extended
bound state to persist is provided largely, but not exclusively,
through pion exchange.  The detailed mechanism is a variant of the
original Yukawa interaction, via a potential energy function of the
form
\begin{equation}
V(r) = ({\rm couplings} \times \mbox{spin-isospin-orbital structure})
\times \frac{e^{-\mu r}}{r} \times \left[ 1 + O\left( \frac{1}{\mu r}
\right) \right] \, ,
\end{equation}
where $\mu$ is the mass of the exchanged meson.  Contact
[$\delta^{(3)}({\bf r})$] terms are also frequently included as
contributions to $V(r)$.  At large $r$ the pion, being the lightest
meson, dominates.  An intermediate-range attraction is interpreted as
a two-$\pi$ correlation or $J^P = 0^+$ $\sigma$-meson exchange, while
a short-distance hard-core repulsion is interpreted as $J^P = 1^-$
$\rho$ or $\omega$ exchange.  Potentials using this basic type of
interaction are used to great effect in modeling complex nuclei, as in
the Nijmegen~\cite{Stoks:1994wp}, Bonn~\cite{Machleidt:2000ge}, and
Argonne~\cite{Wiringa:1994wb} potentials.

Needless to say, meson-exchange models, even for two-body systems, can
become quite intricate and require a substantial number of parameters.
Moreover, many species of compound nuclei have rather long lifetimes
and well-measured properties (in particular, the deuteron is
completely stable).  In contrast, the heavy-quark exotic candidates
all have very short lifetimes; the longest-lived one appears to be the
$X(3872)$, whose width is only known as the bound $< 1.2$~MeV; a
plausible width of, say, 100~keV corresponds to a lifetime of only
$10^{-20}$~s.  Not enough precision data is yet available to perform
the same level of fitting to interaction potentials for heavy-quark
exotics, even for the well-studied $X(3872)$.  Nevertheless, the
extreme closeness of the $X(3872)$ mass to the $D^0 {\bar D}^{*0}$
threshold makes it extremely compelling to model as a molecule of
these mesons~\cite{Tornqvist:2004qy}.

It is also worth recalling that a $qq\bar q\bar q$ system can form a
pair of color-singlet mesons in two ways, corresponding in the case of
$c\bar c q\bar q$ systems like the $X(3872)$ to an open-charm meson
pair, or to a pair of a charmonium and a light-quark meson.  Indeed,
the $X(3872)$ lies not only very close to the $D^0 {\bar D}^{*0}$
threshold, but also to the thresholds for $J/\psi \, \rho^0_{\rm
peak}$ (3872~MeV) and $J/\psi \, \omega$ (3880~MeV).  However,
molecules of the pure $(c\bar c)(q\bar q)$ type would necessarily be
bound by the exchange of the much heavier $D^{(*)}$ mesons, which
propagate shorter distances than $\pi$'s and would have difficulty
accounting for spatially extended bound states.  It therefore appears
much more natural for such states to have a rather larger open-charm
than hidden-charm meson component, and therefore the open-charm decays
are expected to dominate.  For all exotics candidates for which
open-charm modes have been seen, they do indeed provide the dominant
decay channels, although several of the exotics still lack evidence
for such decays despite dedicated searches and plenty of available
phase space.  One cannot eliminate the possibility of a substantial
($c\bar c$)($q\bar q$) component, or indeed, a pure charmonium $(c\bar
c)$ state if quantum numbers allow, to combine with a primarily
open-charm hadron-pair molecule, and such a {\it coupled-channel
analysis\/} may be essential to understanding the detailed structure
of exotics such as the $X(3872)$.

One can also explore {\it quark exchange\/} as a binding mechanism,
using a quark potential model.  Indeed, one of the first
analyses~\cite{Swanson:2003tb} of the $X(3872)$ contained both
quark-exchange and pion-exchange potentials.  Because of color
confinement, one expects quark exchange to be a significant binding
mechanism only at short distances, where the equivalent description in
terms of meson exchanges (due to quark-hadron duality) might require
the inclusion of multiple meson species.  Moreover, quark exchanges
with net non-singlet (octet) color charge are possible and cannot be
expressed in terms in any number of (color-singlet) mesons, although
the bound ``mesons'' in this case would themselves become colored
objects.

Virtually every exotic candidate has been modeled as a hadronic
molecule.  Relevant thresholds appear in
Figs.~\ref{fig:ccneutral}--\ref{fig:cccharged} as dashed or dotted
lines, from which one can assess the ease or difficulty with which the
molecular hypothesis can be supported.  A few exotics lie remarkably
close to hadron thresholds: the $X(3872)$ of course, and also the
$X(3915)$ and possibly $X(3940)$ below $m_{D_s^+} + m_{D_s^-} =
3937$~MeV, $P_c(4380)$ below $m_{\Sigma_c^{*+}} + m_{D^0} = 4388$~MeV,
and $P_c(4450)$ below $m_{\Sigma_c^+} + m_{D^{*0}} =
4461$~MeV~\cite{Karliner:2015ina}.  Others are quite close to and lie
just {\em above\/} thresholds, such as $Z^+_c(3900)$ above $m_{D^0} +
m_{D^{*+}} = 3875$~MeV, $Z^+_c(4020)$ above $m_{D^{*0}} + m_{D^{*+}} =
4017$~MeV, $X(4630)$ above $\Lambda_c^+ + \bar{\Lambda}_c^- =
4573$~MeV, $Z_b(10610)$ above $m_{B^*} + m_{B} = 10604$~MeV, and
$Z_b(10650)$ very slightly above $2m_{B^*} = 10650$~MeV\@.  In these
latter cases, the molecular hypothesis only works if one posits a
mechanism to prevent the instantaneous fall-apart decay into the
component hadrons, such as an intermediate-range potential barrier
that must be tunneled through in order for decay to occur.
Alternately, such states may be considered molecular resonances rather
than true bound states~\cite{Zhao:2014gqa}; a strong attraction
between the component hadrons can persist above threshold, creating an
enhancement exhibiting a width that nevertheless remains observably
small.  As the distance of the state from threshold increases, such
objects gradually merge into ones better described as the threshold
kinematical effects to be discussed in Sec.~\ref{subsec:kinem}.

Lastly, one should note that the mass of the heavy quark $Q$
influences the ease with which hadronic molecules can be formed.  In
particular, molecules containing $b\bar b$ should be more likely to
form than those containing the lighter pair $c\bar
c$~\cite{Tornqvist:1993ng}, since Fermi motion and other effects
suppressed as $1/m_Q$ are rather larger in the charm case (in
particular, when compared with the typical binding scales provided by
$m_\pi$), and can be more effective in $c\bar c$ states in
counteracting the binding obtained through light-meson exchanges.

\subsubsection{Case Study: $X(3872)$ as a Molecule} \label{sec:X3872molec}

As mentioned, virtually every heavy-quark exotic candidate has been
considered in the molecular picture, which would make a full
examination of the literature rather cumbersome for the purposes of
this pedagogical summary.  Instead, we present here a qualitative
chronological overview of studies of the exotic state most likely to
be a molecule by virtue of its proximity to a hadronic threshold, the
$X(3872)$.

Despite intensive studies since 2003, the exact nature of the $J^{PC}
= 1^{++}$ state $X(3872)$ still remains elusive.  Its most remarkable
feature remains its extreme closeness to the $D^0 {\bar D}^{*0}$
threshold, $m_{X(3872)} - m_{D^{*0}} - m_{D^0} = +0.01 \pm
0.18$~MeV\@.  In fact, the threshold $m_{D^{*+}} + m_{D^+}$ lies about
7~MeV higher, meaning that molecular $X(3872)$ should have a larger
$D^0 {\bar D}^{*0}$ than $D^+ {\bar D}^{*-}$ component, thus
manifestly breaking isospin in the $X(3872)$---a unique situation not
previously encountered in hadronic physics.  Nevertheless, no charged
partner to the $X(3872)$ has turned up in a dedicated
search~\cite{Aubert:2004zr}, suggesting that it should be interpreted
as a (largely) $I=0$ state.  Even so, it decays to both $J/\psi \,
\pi^+ \pi^-$~\cite{Aubert:2008gu}---understood as the $I=1$ state
$J/\psi \, \rho^0$ due to the proximity of $m_{X(3872)}$ to the
combination $m_{J/\psi} + m_{\rho^0, \, {\rm peak}}$---and to the
$I=0$ state $J/\psi \,
\omega$~\cite{delAmoSanchez:2010jr,Lees:2012xs}.  These features alone
are enough to demonstrate that $X(3872)$ cannot simply be the
yet-unobserved conventional charmonium state $\chi_{c1}(2P)$, which
was anticipated on the basis of quark-potential models to lie several
tens of MeV higher than 3872~MeV (note the cluster of [blue] dashed
lines for $J^{PC} = 1^{++}$ in Fig.~\ref{fig:ccneutral}).

The first two analyses~\cite{Swanson:2003tb,Wong:2003xk} of $X(3872)$
as a $D^0 {\bar D}^{*0}$ molecule included explicit quark degrees of
freedom.  The analysis of Ref.~\cite{Swanson:2003tb} also included
$J/\psi \, \{ \omega, \rho \}$ components; however, it was later found
to underpredict the substantial radiative decay branching fractions to
$J/\psi \, \gamma$ and $\psi(2S) \gamma$.  Meanwhile,
Ref.~\cite{Wong:2003xk} predicted both $D^0 {\bar D}^{*0}$ and $D^+
{\bar D}^{*-}$ bound states, which the lack of $I=1$ partners to the
$X(3872)$ seems to preclude.

The first hadronic effective Lagrangian
studies~\cite{AlFiky:2005jd,Fleming:2007rp} of $X(3872)$ as a bound
state appeared in 2006--7, with the first chiral unitary calculations
beginning in 2013~\cite{Wang:2013kva}.  In the direction of purely
hadronic-exchange potential models, the first calculation including
$\sigma$ exchange to represent intermediate-range attraction appeared
in 2008~\cite{Liu:2008fh} and $\rho$ exchange in
2009~\cite{Liu:2008tn}.  While even the earliest calculations ({\it
e.g.},~\cite{Swanson:2003tb}) included both central and tensor
interactions, the indispensability of including both $s$ and $d$ waves
in the binding of the $D^0 {\bar D}^{*0}$ pair via a tensor
interaction---analogous to its necessary presence in the deuteron wave
function in order to explain its nonzero electric quadrupole
moment---was first noted in 2008~\cite{Thomas:2008ja}.

Calculations with separate treatment of the $D^0 {\bar D}^{*0}$ and
$D^+ {\bar D}^{*-}$ and isospin breaking ({\it i.e.}, not just the
charged and neutral $D^{(*)}$ mass differences, but the relative
weight of these states in the $I=0$ and $I=1$ Hamiltonian eigenstates)
began in 2009~\cite{Lee:2009hy}.

State-of-the-art meson-exchange models for
$X(3872)$~\cite{Li:2012cs,Zhao:2015mga} now include coupled-channel
effects, isospin breaking, $s$-$d$ mixing, and now also explicit
$1/m_Q$ effects.

In contrast, in 2009 QCD sum rules calculations~\cite{Matheus:2009vq}
were found to favor a much larger $(c\bar c)$ component (97\%)
compared to the $D^0 {\bar D}^{*0}$ component (3\%) when the tiny
width of the $X(3872)$ is taken into account.  An $X(3872)$ with such
a composition was found to satisfactorily accommodate its radiative
decays~\cite{Nielsen:2010ij}, which can be quite challenging for pure
meson-exchange models.  In fact, an admixture for the $X(3872)$
favoring the $c\bar c$ component had been anticipated already in
2005~\cite{Suzuki:2005ha} on other grounds, as we discuss next.  But
the central message should already be clear: Although the technology
for describing the $X(3872)$ as a primarily $D^0 {\bar D}^{*0}$
molecule is quite mature, solid reasons exist for questioning this
interpretation.

\subsubsection{Prompt Production of the $X(3872)$}

The suggestion that the $\chi_{c1}(2P)$ $(c\bar c)$ component should
dominate the $X(3872)$ wave function compared to the $D^0 {\bar
D}^{*0}$ component, despite its closeness to the $D^0 {\bar D}^{*0}$
threshold, was emphasized in 2005 in Ref.~\cite{Suzuki:2005ha}, in
part due to the (then) newly discovered
fact~\cite{Acosta:2003zx,Abazov:2004kp} that $X(3872)$ was produced in
high-energy colliders with a rate comparable to that of ordinary
charmonium $\psi(2S)$.  This so-called {\it prompt production\/}
(production at the primary collision point, as opposed to production
through the subsequent decay of a $b$-containing hadron originally
produced from the initial collision) of the
$X(3872)$~\cite{Acosta:2003zx,Chatrchyan:2013cld,Abazov:2004kp,
Aaij:2011sn}---providing a cross section of about 30~nb---is
surprisingly large and creates quite a problem for the molecular
picture.

The essential physics is simple to describe, but its correct
implementation remains controversial.  If the $X(3872)$ is primarily a
$D^0 {\bar D}^{*0}$ molecule, then presumably the strongly bound
$D^{(*)}$ hadrons must form first, and then {\it coalesce\/} into the
weakly bound molecule.  The component hadrons must have a sufficiently
small relative momentum less than some $k_{\rm max}$ in order to have
an opportunity to form a bound state, or else they simply fly off as
free particles.  One expects the probability of finding such
correlated pairs to drop drastically for large beam energies such as
those at the Tevatron and especially at the LHC, in particular for
high values of transverse momentum $p_T$ with respect to the beam.

Again drawing on the analogy between the deuteron and the $X(3872)$,
one can ask about the rate of production of anti-deuterons in $pp$ or
$p\bar p$ collisions (whose component $\bar p \bar n$ baryons must
clearly be produced in the collision).  By modeling the coalescence in
conjunction with standard hadronization Monte Carlo algorithms and
limiting to $k_{\rm max} = 50$~MeV, Ref.~\cite{Bignamini:2009sk}
showed the prompt-production cross section to be only about $0.1$~nb,
hundreds of times smaller than the observed value.  That this
coalescence model produces the correct rate for antideuteron
production (with $k_{\rm max} = 80$~MeV) was demonstrated in
Ref.~\cite{Guerrieri:2014gfa}.

Hadronization is, however, a complicated process, and an analysis
based on correlated free particles may not directly translate into
their bound states.  In particular, Ref.~\cite{Artoisenet:2009wk}
argued that strong {\it final-state interactions\/} (FSI) between the
hadrons are sufficient to allow $k_{\rm max}$ to be as high as 500~MeV
and still form a bound state, making the large prompt production rate
not so surprising.  A rebuttal~\cite{Bignamini:2009fn} argued that
such strong FSI would produce unobserved results, like the generation
of a $D_s {\bar D}_s^*$ molecule at the Tevatron, and that strong FSI
did not appear to be needed for deuteron studies.  The same
collaboration also proposed an analysis~\cite{Esposito:2013ada} to
consider the effect of multiple scattering of the $D^0$ and ${\bar
D}^{*0}$ from pions in the interaction region in order to test how
many $D^0 {\bar D}^{*0}$ pairs can thereby be rescattered into a state
of relative momentum $< k_{\rm max}$, and
showed~\cite{Guerrieri:2014gfa} that the prompt production rate of
$X(3872)$ can be brought in this way closer to the experimental
value---but again, to values still far below it, unless particularly
strong FSI are included.

A direct comparison between prompt production of (anti)deuterons and
$X(3872)$ at values of $p_T \approx 15$~GeV, at which the $X(3872)$
has already been seen at CMS~\cite{Chatrchyan:2013cld}, will
illuminate the relative importance of FSI in the two processes and
will provide a more decisive probe of the structure of the $X(3872)$.
Such experiments are well within the capabilities of the LHC, and
these future measurements will provide crucial information for studies
of exotics.

\subsection{Hadrocharmonium Picture}

\subsubsection{Motivation and Origin}

Some of the heavy-quark exotic candidates preferentially decay to
conventional charmonium plus light hadrons, rather than to open-charm
meson pairs ($D \bar D$ or $D {\bar D}^*$).  In particular, the
$J^{PC} = 1^{--}$ candidates $Y(4008)$, $Y(4230)$, $Y(4260)$,
$Y(4360)$, and $Y(4660)$ (See Appendix~A) fit into this category.
Moreover, no open-charm decay of $Z_c^\pm (4430)$ has yet been seen,
and it strongly prefers to decay to $\psi^\prime \, \pi^\pm$ rather
than to $J/\psi \, \pi^\pm$, and the $Y(4008)$ and $Y(4260)$ decay to
$\pi^+ \pi^- J/\psi$, while the $Y(4360)$ and $Y(4660)$ decay to
$\pi^+ \pi^- \psi^\prime$: Some of the exotics clearly have specific
preferred charmonium decay products.

These observations have a natural explanation if the exotic state can
be described as a particular compact charmonium species embedded in a
larger cloud of light-quark hadronic matter, an idea dating back to
the proposal of {\it nuclear-bound quarkonium\/} in
1990~\cite{Brodsky:1989jd}.  In this picture for exotics, the heavy
$c\bar c$ pair can be supposed to act as a sort of nucleus for the
system.  This proposal was first qualitatively mentioned by Voloshin
in the discussion of Ref.~\cite{Voloshin:2007dx} in 2008, and
developed into a model some months later in
Ref.~\cite{Dubynskiy:2008mq}, where it was dubbed {\it
hadrocharmonium}.

\subsubsection{Structure and Binding}

A first observation about the hadrocharmonium picture is that it is
qualitatively distinct from a simple molecular picture of charmonium
plus a light meson, in which the wave functions of the two hadrons
have a somewhat suppressed spatial overlap, as in a diatomic molecule.
In hadrocharmonium, the core is purported to live entirely within the
light-quark cloud.  Such a distinction should be kept in mind when
considering the interpretation of calculations such as in
Ref.~\cite{Guo:2008zg}, in which the $Y(4660)$ is proposed to be a
$f_0(980) \psi^\prime$ bound state.

The binding mechanism for hadrocharmonium~\cite{Dubynskiy:2008mq} is a
color van der Waals attraction between a compact, color-singlet $c\bar
c$ core and a larger $q\bar q$ cloud interacting chiefly through the
chromoelectric dipole ($E1$ multipole) interaction, the QCD analogue
of the atomic van der Waals attraction.  While this interaction is
manifestly attractive, it does not guarantee the existence of bound
states, especially because of the counteracting effect of the Fermi
motion of the light degrees of freedom (mass labeled by $M_X$).  As
found in Ref.~\cite{Dubynskiy:2008mq}, a value of $M_X$ exceeding
1~GeV, perhaps approaching 2~GeV, is necessary for the net effect of
all interactions to give binding for the hadrocharmonium system.
Interestingly, this result shows that hadrocharmonium with more highly
excited light degrees of freedom is more likely to form and be
observed.  In the case of {\it hadrobottomonium}, the effect of Fermi
motion decreases (since $m_b
\simeq 3m_c$), but so does the strength of the chromoelectric dipole
interaction, due to the smaller size of $\Upsilon$ states compared to
$\psi$ states; owing to these competing effects, hadrobottomonium
states may still exist, but likely not as exact siblings to
hadrocharmonium states.  In particular, Ref.~\cite{Dubynskiy:2008mq}
anticipates hadrobottomonium states no lower than 11~GeV, too heavy to
accommodate the known $Z_b$ states (at 10610 and 10650~MeV).

\subsubsection{Hadrocharmonium and Heavy-Quark Spin Symmetry}

Owing to heavy-quark spin symmetry (HQSS) and the attendant hypothesis
that the charmonium wave function is largely decoupled from the light
degrees of freedom, one expects that any particular charmonium
structure existing at the center of the hadrocharmonium state should
leave its imprint on the final state.  This restriction not only
provides a natural explanation for the preference of particular states
to decay to a particular charmonium state ({\it e.g.}, $J/\psi$ for
$Y(4008)$, $Y(4260)$, $Z_c(3900)$ {\it vs.} $\psi^\prime$ for
$Y(4360)$, $Y(4660)$, $Z_c(4430)$~\cite{Dubynskiy:2008mq,
Voloshin:2013dpa} because the $J/\psi = \psi(1S)$ wave function is
much more compact than that of the $\psi^\prime =
\psi(2S)$~\cite{Eichten:1978tg,Eichten:1979ms}, but it also predicts
that the open-charm decay modes should be relatively suppressed
because of the dynamical difficulty of breaking up the compact $(c\bar
c)$ core and rearranging the constituents with $(c\bar c)(q\bar q)$
color structure into $(c\bar q)(\bar c q)$.  For the $Z_c(3900)$ this
interpretation is problematic, as the $D{\bar D}^*$ mode appears to
dominate its decay width~\cite{Ablikim:2013xfr}.

In addition, HQSS predicts that the $c\bar c$ spin in hadrocharmonium
is approximately conserved, so that states with a spin-triplet
(-singlet) core should decay preferentially to $\psi$ or $\chi_c$
($\eta_c$ or $h_c$).  Experimental evidence that $Y(4260)$ and
$Y(4360)$ decay not only to spin-triplet $\psi$ states but
spin-singlet $h_c$ as well~\cite{Ablikim:2013wzq} inspired an
extension of the hadrocharmonium hypothesis~\cite{Li:2013ssa} that
asserts the core can be a mixture of spin-singlet and spin-triplet
$c\bar c$ and still satisfy HQSS.

In contrast, in a truly molecular model, the hadronic components have
well-defined quantum numbers, and the HQSS predictions can be somewhat
different.  Following the aforementioned proposal of a $f_0(980)
\psi^\prime$ state in Ref.~\cite{Guo:2008zg}, the authors then
predicted~\cite{Guo:2009id} the existence of an $f_0(980)
\eta_c^\prime$ bound state, using that $\psi^\prime$ and
$\eta_c^\prime$ are degenerate states in the HQSS limit.  A
side-by-side comparison of the HQSS predictions of molecular,
hadrocharmonium, and diquark pictures is presented in
Ref.~\cite{Cleven:2015era}.

\subsubsection{Can Hadrocharmonium Coexist with Other Pictures?}

The question of whether hadrocharmonium states really occur in nature
comes down to an assessment of the relative strength of valid
competing dynamical effects.  It seems extremely likely that, by
allowing $\Lambda_{\rm QCD}$ and the heavy-quark mass $m_Q$ to assume
a variety of numerical values, one can find regimes in which hadronic
molecular states occur and regimes in which hadroquarkonium states
occur.  These regimes may be distinct, or they may overlap, in which
case the eigenstates of the hadronic QCD Lagrangian for a given set of
parameters may be combinations of the two.  Without being able to
solve QCD for physical values of quark masses, one must rely on hints
from data such as spectroscopy, decay modes and ratios, and in the
future, detailed production line shapes.

One very interesting piece of data in this regard is the suggestion of
a significant measured branching fraction for the radiative decay
$Y(4260) \to \gamma X(3872)$~\cite{Ablikim:2013dyn}.  The $X(3872)$
was touted in Sec.~\ref{sec:Molecule} as the best candidate for a
hadronic molecule (although not without some conceptual difficulties),
while $Y(4260)$ was described in this section as a prime candidate for
a hadrocharmonium state.  If indeed they are connected by a prominent
radiative transition, then one expects a high degree of similarity in
the structure of their wave functions, and hence of what kinds of
state they are.  For example, the possibility that both are
molecular-$c\bar c$ combinations is studied in
Ref.~\cite{Dong:2014zka}.

\subsection{Diquark Picture}

\subsubsection{General Considerations: Nature of Diquarks}

The prediction by Eq.~(\ref{eq:CasimirNums}) of a attractive channel
in which two color-{\bf 3} quarks can combine into a color-$\bar{\bf
  3}$ diquark (or two color-$\bar{\bf 3}$ antiquarks into a color-{\bf
  3} antidiquark) immediately suggests the possibility of composite
but colored subcomponents inside of hadrons.  This fact alone explains
the rich history of diquark phenomenology~\cite{Anselmino:1992vg},
particularly for baryons.  The diquark itself can be considered to be
either a fairly compact object, with a size similar to that of an
ordinary meson (a few tenths of a fm), or it can be considered merely
as a correlated state between two quarks in a hadron.  Since the
quarks have spin $\frac 1 2$, the diquark (orbital) ground state can
be scalar (spin 0) or vector (spin 1), and has positive parity.

While numerous papers dating as far back as the 1970s have examined
the possibility of diquarks as constituents of exotic hadrons
(including exotics containing heavy quarks), the modern studies of
diquark models for heavy-quark exotics were originally inspired by
certain peculiar behaviors of the light-quark scalar mesons $a_0(980)$
($I=1$) and $f_0(980)$ ($I=0$) that cast doubt upon a naive $q\bar q$
interpretation for these states.  For instance, their masses lie
extremely close to the $K\bar K$ thresholds (hence little phase space
is available for these channels), and yet their $K\bar K$ decay
branching fractions are in the tens of percent~\cite{Olive:2016xmw}.
Diquark-antidiquark models provide a natural explanation of this fact
by suggesting that each diquark component in the $a_0(980)$ and
$f_0(980)$ carries a valence $s$ (or $\bar s$)
quark~\cite{Jaffe:1976ig}.  A fresh look at more recent light-quark
scalar meson data using the diquark model~\cite{Maiani:2004uc}
inspired an extension of the approach~\cite{Maiani:2004vq} to the
then-newly discovered heavy-quark exotics; this extension, to be
discussed below, constitutes the basis of modern heavy-quark diquark
models.

\subsubsection{Heavy-Quark Diquark Models}

The presence of heavy quarks in the exotic hadron has a number of
interesting implications for its structure in diquark models.  First,
light-quark diquarks were predicted to be more strongly bound in the
spin-0 than spin-1 channel, and hence the former (``good'') diquarks
are expected to be more successful in forming light hadrons than the
latter (``bad'') diquarks~\cite{Jaffe:2004ph}.  However, for a diquark
that contains a heavy quark, the ``good'' and ``bad'' varieties differ
only by the relative orientation of the heavy-quark spin, and
operators sensitive to this spin are suppressed by powers of $1/m_Q$
according to heavy-quark spin symmetry (HQSS)\@.  Second, the
characteristic size associated with a diquark may be identified with
its Compton wavelength, which is inversely proportional to the reduced
mass $\mu$ of its constituents.  For a given light constituent mass
$m$, $\mu$ can vary from $\frac 1 2 m$ (for an equal-mass light-light
system) up to $m$ (for an infinitely heavy-light system).  One thus
expects a heavy-light diquark to be substantially smaller than a
light-light diquark.

Diquark models tend to predict large numbers of states, particularly
when compared with molecular hadron or hadrocharmonium models.  This
proliferation of states is the result of the nonzero net color charge
of the diquarks, meaning that the overall system is bound by strong
fundamental QCD forces rather than by the much weaker color-singlet
van der Waals forces.  As a result, one expects all quark spin and
isospin combinations to produce a state, because the energy cost for
exchanging up and down spins or exchanging $u$ and $d$ quarks is
relatively small compared to the strong-interaction energy scales
responsible for the overall binding of the state.  In contrast, we
have seen that molecular models are highly sensitive to the proximity
of hadronic dissociation thresholds, as well as the spin and isospin
of mesons assumed responsible for their binding.  In particular, not
every two-meson threshold is expected to produce a hadronic molecule.
In the case of hadrocharmonium, we have seen that the spatial extent
of the core charmonium wave function, which depends upon internal
excitation quantum numbers, is significant in determining whether or
not the state binds.  One should not, however, infer from these
observations that all diquark-antidiquark states are equivalent; the
proximity of hadronic thresholds can have profound effects on the
states, as discussed below.

The most common model for the diquark-antidiquark
system\footnote{Hidden heavy flavor is implied here, but $Q$ and $\bar
  Q$ need not be the same flavor.} $(Qq_1)(\bar Q\bar q_2)$ uses an
effective Hamiltonian that is dominated by spin and orbital
interactions amongst the quarks.  In the original model of
Ref.~\cite{Maiani:2004vq}, the Hamiltonian can be written as
\begin{equation} \label{eq:H_full}
H = m_{(Q q_1)} + m_{(\bar Q \bar q_2)} + H_{SS}^{qq} + H_{SS}^{q
\bar q} + H_{SL} + H_L \, ,
\end{equation}
where $m_{(Q q_1)}$ and $m_{(\bar Q \bar q_2)}$ are the diquark
masses.  $H_{SS}^{qq}$ represents spin-spin couplings between the two
quarks (or the two antiquarks), and therefore refers to spin-spin
couplings {\em within\/} either the diquark or the antidiquark:
\begin{equation} \label{eq:HSSqq}
H_{SS}^{qq} = 2\kappa_{(Q q_1)} \, {\bf s}_Q \! \cdot {\bf s}_{q_1} +
2\kappa_{(\bar Q \bar q_2)} \, {\bf s}_{\bar Q} \! \cdot {\bf s}_{\bar
q_2} \, .
\end{equation}
In contrast, $H_{SS}^{q \bar q}$ couples quarks to antiquarks, thereby
providing interactions {\em between\/} the diquark and the
antidiquark:
\begin{eqnarray}
H_{SS}^{q \bar q} & = &
2 \kappa_{Q \bar q_2}   \, {\bf s}_Q     \! \cdot {\bf s}_{\bar q_2} +
2 \kappa_{Q \bar Q}     \, {\bf s}_Q     \! \cdot {\bf s}_{\bar Q}   
+
2 \kappa_{q_1 \bar Q}   \, {\bf s}_{q_1} \! \cdot {\bf s}_{\bar Q}   +
2 \kappa_{q_1 \bar q_2} \, {\bf s}_{q_1} \! \cdot {\bf s}_{\bar q_2}
\, . \label{eq:HSSqqbar}
\end{eqnarray}
The remaining terms are the spin-orbit ($H_{SL}$) and purely orbital
($H_L$) contributions,
\begin{eqnarray}
H_{SL} & = & -2a [{\bf s}_{(Qq_1)} \! \cdot {\bf L} + {\bf s}_{(\bar Q
  \bar q_2)} \! \cdot {\bf L}] = -2a \, {\bf S} \! \cdot
{\bf L} \, , \nonumber \\
H_L & = & \frac{B_c}{2} {\bf L}^2 \, . \label{eq:H_orb}
\end{eqnarray}
Here, ${\bf s}_{(Qq_1)} \equiv {\bf s}_Q + {\bf s}_{q_1}$ is the total
diquark spin (and similarly for the antidiquark), and ${\bf S}$
represents the the total quark spin for the system.  From this
Hamiltonian, one then computes the mass eigenvalues for a full
spectrum of four-quark states, using standard operator techniques.

The original model of Ref.~\cite{Maiani:2004vq} fit the $J^{PC} =
1^{++}$ $X(3872)$---the only exotic candidate known at the time---to
the symmetric combination of $\{s_{(cq)} = 1, \, s_{(\bar c \bar q)} =
0\}$ and $\{s_{(cq)} = 0, \, s_{(\bar c \bar q)} = 1\}$ states, and
predicted a number of other levels, such as a $1^{+-}$ state [the same
quantum numbers as the $Z_c^0(3900)$ and $Z_c(4020)$] at the much
lower mass of 3750~MeV\@.  In 2014, the model was
improved~\cite{Maiani:2014aja} by the inclusion of a significant
dynamical assumption: Spin couplings between the diquark and the
antidiquark are assumed to be negligible.  In terms of the full
Hamiltonian of Eq.~(\ref{eq:H_full}), the contribution $H_{SS}^{q \bar
  q}$ of Eq.~(\ref{eq:HSSqqbar}) is set to zero.

The model of Ref.~\cite{Maiani:2014aja} has many desirable features;
for example, a number of the $1^{--}$ $Y$ states naturally arise as
relative $L=1$ excitations of the diquark-antidiquark pair, meaning
that the electric dipole radiative transition $Y(4260) \to \gamma
X(3872)$ is natural in this picture~\cite{Chen:2015dig}, and the
$Z_c^0(3900)$ arises as $1^{+-}$ partner to the $X(3872)$, the
$\{s_{cq} = 1, \, s_{\bar c \bar q} = 0\}$ and $\{s_{cq} = 0, \,
s_{\bar c \bar q} = 1\}$ states now appearing in the antisymmetric
combination, while the $Z_c^0(4430)$ is the first radial excitation of
$Z_c^0(3900)$.  Nevertheless, the prediction of numerous
yet-unobserved states is a key feature of this model; for example, a
prominent $2^{++}$ state remains to be found.  The analysis can be
applied to the bottom sector as well, where it has been used, {\it
  e.g.}, to study the $Z_b(10610)$ and $Z_b(10650)$~\cite{Ali:2011ug}.
It has also been applied to the $c\bar c s\bar s$
sector~\cite{Lebed:2016yvr}, where the troublesome $0^{++}$ $X(3915)$
[which, in contrast to the expectation for the pure $c\bar c$ state
$\chi_{c0}(2P)$, lacks $D \bar D$ decays; see Sec.~\ref{sec:XYZ3940}]
is suggested to be the $c\bar c s\bar s$ ground state, and states that
decay to $J/\psi \, \phi$ such as the $Y(4140)$ (Sec.~\ref{sec:Y4140})
are naturally accommodated.

\subsubsection{Dynamical Diquarks}
\label{sect:dyndq}

For all its merits, the diquark picture in the Hamiltonian formalism
does not provide detailed dynamics.  One may, for example, relativize
the light quarks~\cite{Ebert:2008kb}, or incorporate a variant of the
static Cornell potential
[Eq.~(\ref{eq:Cornell})]~\cite{Patel:2014vua}, or model the
interaction using a static color flux tube~\cite{Deng:2015lca}, or
introduce nonlocal ({\it e.g.}, Gaussian) vertex functions between the
quarks~\cite{Goerke:2016hxf}.

However, one important feature not taken into account in these
pictures is that the exotic states exist for only a very short time
($\sim 10^{-20}$~s or less), while the techniques described up to this
point refer directly or indirectly to eigenstates of a Hamiltonian,
which suggests a single time coordinate for the whole system and hence
a single (approximate) common rest frame for the components.  In
reality, the diquark-antidiquark pair may be flying apart from their
production point for the entire lifetime of the exotic hadron.  Put
another way, if one treats the exotic as some sort of molecule, it may
not survive long enough to execute a single orbit.

The {\it dynamical diquark picture\/} introduced in
Ref.~\cite{Brodsky:2014xia} instead suggests that confinement is the
primary binding mechanism for the exotic states.  The
diquark-antidiquark pair forms promptly at the production point, and
rapidly separates due to the kinematics of the production process,
whether via the decay of a heavy $b$-containing hadron or through a
hadron collision process.  Since the diquark and antidiquark are
colored objects, they cannot separate asymptotically far apart; they
create a {\it color flux tube\/} or {\it string\/} between them.  Were
sufficient energy available, the string would break as part of a
conventional {\it fragmentation\/} process, producing an additional
$q\bar q$ pair.  In the case of the $(cq)(\bar c \bar q)$ system, the
first available threshold is $\Lambda_c + {\bar \Lambda}_c$ at
4573~MeV, and indeed the $X(4630)$ just above this threshold
(Sec.~\ref{sec:eeCharmonium}) has only been seen so far in the
baryonic decay mode.

\begin{figure}[ht]
\begin{center}
\includegraphics[width=8cm,angle=0]{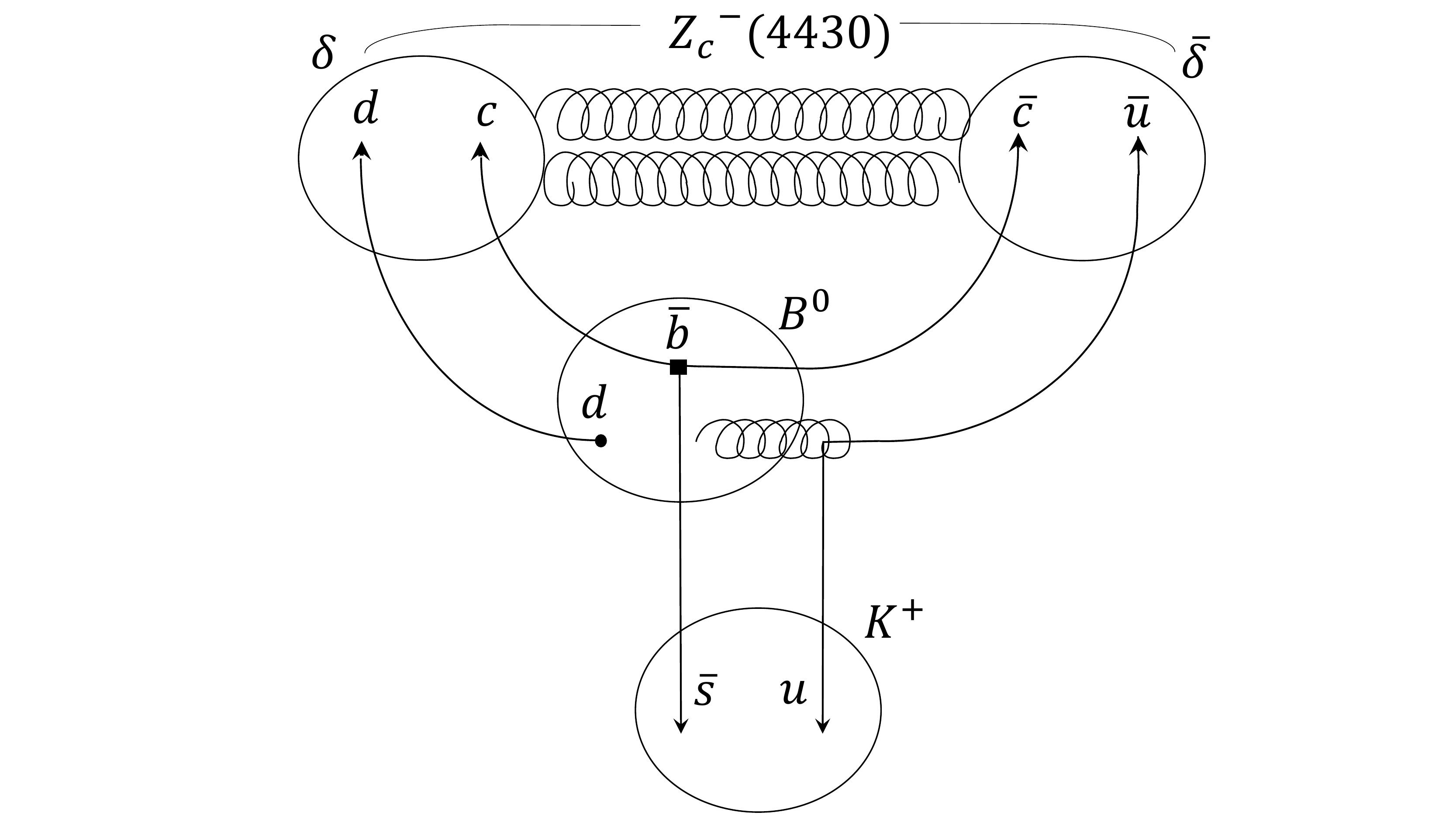}
\end{center}
\caption{Illustration of the dynamical diquark picture mechanism for
  production of the $Z_c^-(4430)$ in the decay $B^0 \to Z_c^-(4430)
  K^+$ (the weak-interaction vertex indicated by a square), adapted
  from Ref.~\cite{Brodsky:2014xia}.  The diquark-antidiquark pair are
  denoted by $\delta$ and $\bar \delta$, and the color flux tube is
  indicated by gluon lines.\label{fig:Z4430}}
\end{figure}

Below the fragmentation threshold, the only available modes for decay
require the quarks (in the diquark) and the antiquarks (in the
antidiquark) to overlap with the wave function of a meson state;
inasmuch as the diquark-antidiquark pair may have achieved a
substantial separation ($r > 1$~fm) during the lifetime of the state,
the overlap is suppressed by the exponentially small meson wave
function tail at large $r$.  The transition rate is therefore also
suppressed, potentially explaining the measurably small exotic widths.
Additionally, the large size of the diquark-antidiquark pair before
coming to rest can explain the preference of more highly excited
exotics like the $Z_c(4430)$ to decay into $\psi(2S)$, which is
spatially much larger than the
$J/\psi$~\cite{Eichten:1978tg,Eichten:1979ms}.  The production
mechanism for $B^0 \to Z_c^-(4430) K^+$ is illustrated in
Fig.~\ref{fig:Z4430}.

Despite these qualitative successes, it should be noted that the
dynamical diquark picture has not yet been developed into a particular
model with uniquely specified interactions.  Necessary ingredients
include modeling of the diquark formation and proper quantization of
the flux tube glue, in order to obtain a specific spectrum and pattern
of decays for the exotic states.

\subsubsection{Pentaquarks from Diquarks}

Both the conventional diquark picture and dynamical diquark picture
can be used to study pentaquark states, including the recently
discovered candidates $P_c(4380)$, $P_c(4450)$.  In the conventional
diquark picture~\cite{Maiani:2015vwa}, the pentaquark may be assembled
as the bound state of three $\bar{\bf 3}$ components, $\bar c
(cq)(qq)$, a composition exploited, {\it e.g.}, in
Refs.~\cite{Anisovich:2015zqa,Li:2015gta,Ghosh:2015ksa}.  Alternately,
pentaquarks can arise in the dynamical diquark
model~\cite{Lebed:2015tna} via the sequential formation of compact
color triplets through the attractive channels ${\bf 3} \times {\bf 3}
\to \bar{\bf 3}$ and $\bar{\bf 3} \times \bar{\bf 3} \to {\bf 3}$ as a
diquark-{\it triquark\/} system, $\bar c_{\bar{\bf 3}} (qq)_{\bar{\bf
    3}} \to [\bar c (qq)]_{\bf 3}$ plus $(c q)_{\bar{\bf 3}}$ , as was
used in Ref.~\cite{Zhu:2015bba}, and applied to hidden-strangeness
system in Refs.~\cite{Lebed:2015fpa,Lebed:2015dca}.  Whether or not
the diquarks in exotics are sufficiently tightly bound to enter as
elementary fields for use in QCD constituent counting rules and alter
the energy scaling behavior of their production amplitudes is
addressed in Ref.~\cite{Brodsky:2015wza}.

Significantly, the observation of opposite parities for the
$P_c(4380)$ and $P_c(4450)$ requires one of the two states to contain
a unit of orbital excitation.  This fact is not difficult to
accommodate in the diquark picture, where the broad $P_c(4380)$ can be
a highly excited $s$-wave resonance ($J^P = {\frac 3 2}^-$), while the
narrower $P_c(4450)$ can be a lower $p$-wave resonance ($J^P = {\frac
  5 2}^+$).

\subsubsection{Other Diquark Approaches}

In addition to the Hamiltonian operator, quark model, and flux tube
approaches, diquarks have also been employed as interpolating fields
in QCD sum rule calculations and in lattice QCD simulations.

A very brief summary of the theory of QCD sum rules has been presented
in Sec.~\ref{sec:QCDsumrules}, while a review of applications to the
charmonium system through 2009 appears in Ref.~\cite{Nielsen:2009uh}.
QCD sum rules can incorporate diquark-antidiquark pair interpolating
operators such as the $J^{PC} = 1^{+-}$ ${\bf 3}$-$\bar{\bf 3}$
current~\cite{Chen:2010ze}, an example of which is ($C$ being the
Dirac matrix representing charge conjugation):
\begin{equation}
J_{2\mu} = q^T_a C c_b
(\bar{q}_a \gamma_{\mu} \gamma_5 C \bar{c}^T_b - \bar{q}_b
\gamma_{\mu} \gamma_5C\bar{c}^T_a) -
q^T_a C \gamma_{\mu} \gamma_5 c_b
(\bar{q}_aC\bar{c}^T_b-\bar{q}_bC\bar{c}^T_a) \, .
\end{equation}
In Ref.~\cite{Chen:2010ze}, to give just one sample result, the
$1^{++}$ $c\bar c q\bar q$ states are found to have masses about
4.0--4.2~GeV, somewhat higher than the $X(3872)$.  Entire spectra may
thus be computed once one has a complete set of interpolating
operators.  Other examples (both tetraquark and pentaquark states)
appear in Refs.~\cite{Wang:2013vex,Wang:2013zra,Wang:2015epa}.  One
must note, however, that QCD sum rules take their interpolating
operators to be local, which in the current context means that the
diquarks are pointlike.

A similar situation arises in lattice QCD simulations (briefly
reviewed in Sec.~\ref{sec:lattice}).  Here, calculations have been
performed that include $D{\bar D}^*$, diquark-antidiquark, and $c\bar
c$ interpolating operators.  In the case of the $X(3872)$, the most
recent simulations~\cite{Prelovsek:2013cra,Padmanath:2015era} include
$J/\psi \, \omega$ and $J/\psi \, \rho$ as well.  Of these
simulations, only Ref.~\cite{Padmanath:2015era} includes diquark
interpolating operators, but finds that the $X(3872)$ appears only if
both $D{\bar D}^*$ and $c\bar c$ interpolators are included, {\it
  i.e.}, the diquark interpolators are unnecessary.  This result is
analogous to the structure for $X(3872)$ suggested in
Sec.~\ref{sec:X3872molec}.  Again, the interpolators in lattice
simulations are nominally pointlike; introducing finite-size effects
is possible, although rather costly in computational time, requiring
the use of nontrivial link variables.  Furthermore, the
state-of-the-art calculations of Ref.~\cite{Padmanath:2015era} use
$m_\pi = 266$~MeV; since the pion with its light mass may very well be
crucial (See Sec.~\ref{sec:Yukawa}) to the successful formation of
exotic states, the results of future simulations with smaller pion
masses are eagerly awaited.

\subsection{Hybrids} \label{sec:hybrids}

Although there is little doubt that hybrid mesons (and baryons) exist,
not much else is known about these states.  The main preliminary
question concerns their observability; in particular, are they
sufficiently long lived to be recognized as resonances?  Assuming no
unexpected experimental impediments to their production and
observation, the main intellectual challenge will be discerning the
degrees of freedom and their dynamics that are relevant to describing
the spectrum, production, and decay of these novel states.

The absence of experimental input has led to a rather broad
evolutionary landscape, with commensurately many ideas concerning the
nature of soft glue.  The chief historical ideas have been that soft
glue forms some sort of string or flux tube, or that it is an
effective constituent confined by a bag or potential.  Alternatively,
nonperturbative glue can be thought of in terms of collective,
nonlocal degrees of freedom, or as a local quasiparticle degree of
freedom.

The steadily improving capabilities of computational lattice gauge
field theory lends hope that this situation will be improved.
Ironically, lattice calculations have so far provided evidence for
both pictures: The adiabatic gluonic surfaces discussed in
Sec.~\ref{sect:bo} can be modeled reasonably well with a bag
picture~\cite{Juge:2002br}, while results from the Lattice Hadron
Collaboration~\cite{Liu:2012ze} provide compelling evidence that
nonperturbative gluons can be thought of as chromomagnetic
quasiparticles of quantum numbers $J^{PC} = 1^{+-}$ with an excitation
energy of approximately 1 GeV\@.  In this way, the lightest hybrid
multiplet contains states with
\be 
J^{PC} = 1^{--} = (1^{+-})_{\textrm{glue}} \times
(0^{-+})_{\textrm{quarks}} \, , \label{eq:hybrid1}
\ee
which corresponds to a vector hybrid with quarks in a spin singlet and
in an $s$ wave, and
\be
J^{PC} = (0,1,2)^{-+} = (1^{+-})_{\textrm{glue}} \times
(1^{--})_{\textrm{quarks}} \, , \label{eq:hybrid2}
\ee
which combines the ``gluon'' with quarks in a spin triplet and in an
$s$ wave.

An early lattice computation of the heavy hybrid-meson spectrum was
made by the CP-PACS collaboration~\cite{Manke:1998qc}.  The authors
worked with the Lagrangian of nonrelativistic QCD and ignored all
spin-dependent operators.  This assumption led to a degenerate
multiplet of states with the quantum numbers given in
Eq.~(\ref{eq:hybrid1}).  The computations yielded a charmonium hybrid
multiplet 1.323(13) GeV above the spin-averaged charmonium ground
state (near 4.39 GeV) and a bottomonium hybrid multiplet near 10.99
GeV\@.


More recently, the Hadron Spectrum Collaboration performed a
large-scale unquenched calculation~\cite{Liu:2012ze} that used a large
variational basis, a fine temporal lattice spacing, two light
dynamical quarks, a dynamical strange quark, and improved lattice
actions to obtain a comprehensive charmonium spectrum.  Despite these
technical advances, the dynamical quarks were still heavy, yielding a
pion mass of 396 MeV, and a $J/\psi$-$\eta_c$ splitting of 80(2)~MeV,
which is too small compared to the experimental value of 113~MeV.

The authors of Ref.~\cite{Liu:2012ze} also probed the internal
structure of their hadrons by measuring state overlaps with various
operators.  Thus, for example, some vectors have significant overlaps
with a quark-antiquark pair in a ${}^3S_1$ state, while others have
larger overlap with ${}^3D_1$ operators.  These overlaps only provide
qualitative indications of state configurations because they are
scale-dependent, and comparison to continuum matrix elements can be
confounded by operator mixing.

\begin{figure}[h]
\centering
\includegraphics[angle=0,width=12cm]{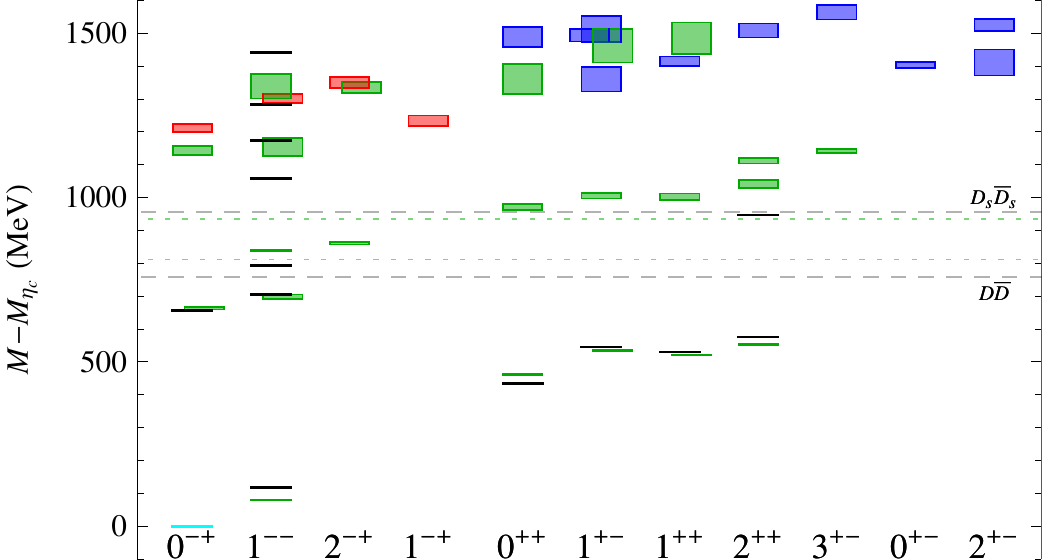}
\caption{A lattice QCD calculation of charmonium states.  (Figure
  reproduced with permission from Ref.~\cite{Liu:2012ze}.)  Solid
  (black) lines are experimental masses, green boxes (the lighter
  grays in a black-and-white representation) refer to predominantly
  conventional charmonia, red boxes (the darker grays in the columns
  up to $J^{PC} = 1^{-+}$) are the lowest-lying hybrid multiplet; blue
  boxes (the darker grays in the columns starting at $J^{PC} =
  0^{++}$) are the first excited hybrid multiplet.  Box heights
  represent statistical uncertainty.}
\label{ccfig}
\end{figure}

This method can be used to determine states having large overlaps with
operators of large gluonic content.  The resulting states are
indicated with red and blue boxes in Fig.~\ref{ccfig} (darker grays,
when the figure is viewed in a black-and-white representation).  As
can be seen, the red boxes form an approximate multiplet with the
expected quantum numbers of Eq.~(\ref{eq:hybrid1}).  The thin (black)
lines in the figure are experimental masses, and (green) boxes are
calculations of predominantly conventional charmonium state masses.
Notice that the agreement with $J^{PC} = 1^{--}$ worsens as one moves
up the spectrum.  In view of this deterioration, one might expect that
an additional 100 MeV of uncertainty should be applied to the
predicted hybrid masses presented in Table~\ref{tab:ccspec}.


\begin{table}[ht]\centering
\begin{tabular}{c|ll}
\hline\hline  
$J^{PC}$ &\multicolumn{2}{c}{Mass (MeV)$^{\vphantom\dagger}$}\\
 \hline
$0^{-+}$ & 4195(13) &  \\
$1^{-+}$ & 4217(16) &  \\
$1^{--}$ & 4285(14) &  \\
$2^{-+}$ & 4334(17) &  \\
\hline
$1^{+-}$ & 4344(38) & 4477(30) \\
$0^{+-}$ & 4386(9) &   \\
$2^{+-}$ & 4395(40) & 4509(18)  \\
$1^{++}$ & 4399(14) &   \\
$0^{++}$ & 4472(30) &   \\
$2^{++}$ & 4492(21) &   \\
$3^{+-}$ & 4548(22) & \\
\hline\hline
\end{tabular}
\caption{Charmonium hybrid mass predictions~\cite{Liu:2012ze}.  Masses
  are (\texttt{lattice mass}) - (\texttt{lattice} $\eta_c$) +
  (\texttt{expt.} $\eta_c$). }
\label{tab:ccspec}
\end{table}

\subsubsection{Transitions}

Models of strong hybrid decays typically find a selection rule that
forbids decay to pairs of identical $s$-wave
mesons~\cite{Page:1996rj,Isgur:1985vy,LeYaouanc:1984gh}.  This
constraint is sometimes extended to forbidding decay to any pairs of
identical mesons~\cite{Page:1998gz}.  Such a rule, leading to the
absence of decay channels, often predicts hybrids to be narrow.

The first lattice calculation of a hadronic transition was made by the
UKQCD collaboration for the case of heavy
hybrids~\cite{McNeile:2002az}.  The static-quark limit imposes
important constraints on the decay process, since the quark-antiquark
configuration must remain invariant.  The authors focused on the
decay of the exotic $1^{-+}$ state and determined that decay into
$s$-wave mesons is forbidden (since production of the light-quark pair
in a spin triplet is forbidden by conservation of gluonic parity and
charge conjugation, while production of a spin singlet is forbidden by
parity reflection in the quark-antiquark axis).

Furthermore, decay to an $s$-wave $(Q\bar q)$ + $p$-wave $(q\bar Q)$
configuration is forbidden because the $p$-wave excitation energy is
typically greater than the hybrid excitation energy.  Thus, the only
allowed transition in the heavy-quark limit is a string de-excitation
process in which a light flavor-singlet meson is produced.

The authors computed two such transitions, using unquenched QCD with
light-quark masses near the strange quark mass.  When the results are
interpreted in terms of bottomonium, the authors obtained
\begin{equation}
\Gamma [ b\bar b g(1^{-+}) \to \eta_b \, \eta(s\bar s) ]
\sim 1\ {\rm  MeV} \, ,
\end{equation}
and
\begin{equation}
\Gamma [ b\bar b g(1^{-+}) \to \chi_b \, \sigma(s\bar s) ]
\sim 60\ {\rm  MeV} \, .
\end{equation}

More recently, charmonium hybrid radiative transitions have been
computed by the Hadron Spectrum Collaboration~\cite{Dudek:2009kk}.
The calculation was made with a large operator basis in the quenched
approximation.  The renormalization constant required to compare the
lattice matrix elements to physical ones was determined
nonperturbatively by conserving charge at zero recoil.  The resulting
widths are presented in Table \ref{tab:cctrans}, where one sees quite
acceptable agreement with experiment.  Notice that the process $c\bar
c g(1^{-+}) \to J/\psi \, \gamma$ is a magnetic dipole transition.
With conventional charmonia, these transitions require a spin flip and
are therefore suppressed for heavy quarks.  In the case of hybrids,
the extra ``gluon'' permits evading the suppression, and the
transition can be large.

\begin{table}[ht]
\centering
\begin{tabular}{lll}
\hline\hline
transition & $\Gamma_{\rm lattice}$ (keV) & $\Gamma_{\rm expt}$ (keV) \\
\hline
$ \chi_{c0} \to J/\psi \gamma$ & 199(6) & 131(14) \\
$ \psi' \to \chi_{c0} \gamma$ & 26(11) & 30(2) \\
$ \psi'' \to \chi_{c0} \gamma$ & 265(66) & 199(26) \\
$ c\bar c g(1^{--}) \to \chi_{c0} \gamma$ & $<$ 20  &  \\
\hline
$J/\psi \to \eta_c \gamma$  & 2.51(8) & 1.85(29) \\
$\psi' \to \eta_c \gamma$  & 0.4(8) &  0.95 -- 1.37 \\
$\psi'' \to \eta_c \gamma$  & 10(11) &   \\
$c \bar c g(1^{--})  \to \eta_c \gamma$  & 42(18) &   \\
\hline
$c \bar c g(1^{-+})  \to J/\psi \gamma$  & 115(16) &   \\
\hline\hline
\end{tabular}
\caption{Quenched lattice charmonia radiative decays~\cite{Dudek:2009kk}.}
\label{tab:cctrans}
\end{table}

\subsubsection{Y(4260) as a Hybrid}

The most popular candidate for a heavy hybrid meson is the $Y(4260)$
(we stress, however, that alternative models for this state exist as
discussed above, such as hadrocharmonium and an $L=1$
diquark-antidiquark state).  As discussed in
Section~\ref{sec:eeCharmonium}, the $Y$ has been observed in the
reaction $e^+e^- \to J/\psi \pi\pi$ by four different experiments and
is evidently a $J^{PC} = 1^{--}$ charmoniumlike meson.  With a mass of
4251(9) MeV, the state lies between quark model predictions for the 2D
vector at 4168(24)~MeV [expt.\ 4191(5)~MeV] and the 4S vector at
4428(22)~MeV [expt.\ 4421(4)~MeV], and is therefore a prime candidate
for an exotic state.

Several groups have noted the following features of the
$Y(4260)$~\cite{Close:2005iz,Kou:2005gt,Zhu:2005hp}:

\begin{itemize}

\item The decay modes $J/\psi \, \sigma$, $J/\psi \, f_0$, $J/\psi \,
  a_0$ appear to dominate.

\item $\Gamma[Y(4260) \to e^+e^-]$ is much smaller than for all other
  vector charmonia.

\item $\Gamma[Y(4260) \to J/\psi \pi^+ \pi^-]$ is much larger than for
  all other vector charmonia.

\item The mass is about 1~GeV greater than the ground state $\eta_c$
  and $J/\psi$, as is expected for a gluonic excitation.
\end{itemize}

Close and Page~\cite{Close:2005iz} also argue that the decay selection
rule implies a preferred decay to $DD^{**}$ states, which lie 40~MeV
above the $Y$ mass.  Rescattering (Sec.~\ref{subsec:kinem}) is then
postulated to yield the $J/\psi \, \pi\pi$ final state.

Subsequent lattice work~\cite{Liu:2012ze} (discussed above) yielded an
estimate of the vector charmonium hybrid mass of 4285~MeV, quite close
to that of the $Y$\@.  Furthermore, as just mentioned, lattice results
strongly indicate that quarks should form a spin singlet in the
low-lying vector hybrid.  Since photons prefer to create quarks in a
spin triplet, a spin flip is required to created the hybrid state.
This observation neatly fits with the idea of an extra gluon being
present in the state, but ``costs'' a factor of approximately $\langle
p\rangle/m_c \sim \Lambda_{\rm QCD}/m_c \sim 0.1$ with respect to the
creation of $s$-wave vector charmonia (Here, $\langle p\rangle$ is a
typical momentum scale in the hybrid).  Close and Page estimated the
electronic width of the $Y$ to be
\be
5 \ \textrm{eV} < \Gamma[Y(4260) \to e^+e^-] < 60 \ \textrm{eV} \, .
\ee
Subsequent measurements have raised the lower limit to 9 eV\@.  This
width is 2--3 orders of magnitude smaller than those of the
well-established conventional charmonium vectors, in rough agreement
with the spin-flip suppression just noted.

Finally, Close and Page noted that the relatively large width,
$\Gamma[Y(4260)] = 120(12)$~MeV, implies that decays to $D_1\bar D$,
$D_1'\bar D$, and $D_0\bar D^*$ should be accessible.  These decay
modes feed $D^*\bar D\pi$ and $D\bar D \pi$ final states, which can be
searched for.  We remark that the current PDG~\cite{Olive:2016xmw}
lists ``not seen'' for these modes, indicating that this expectation
was incorrect.

If the $Y(4260)$ is indeed a hybrid (or predominantly a hybrid), the
remainder of the low-lying multiplet should lie nearby.  In
particular, one expects a $0^{-+}$ hybrid at 4170 MeV, a $2^{-+}$
hybrid at 4310 MeV, and the quantum number-exotic $1^{-+}$ hybrid at
4190 MeV\@.  The discovery of the latter meson would be a watershed
moment in hadron spectroscopy, as it would be the first definitive
sighting of this manifestly exotic, long-expected form of matter.
After discovery, the next task will be a thorough exploration of the
production and decay mechanisms that manifest in the hybrid spectrum.
Finally, developing a robust models of these properties will lend
insight into the behavior of QCD in a new regime.


\subsection{Kinematical Effects} \label{subsec:kinem}

The relevance of ``kinematical effects'' to heavy exotic hadrons is
slowly becoming more appreciated\footnote{Although this nomenclature
  is traditional, it is regrettable, since channel coupling or the
  generation of self-energies is dynamical.  Nevertheless, the term
  persists because it stresses that the relevant phenomena are not
  described solely by $S$-matrix poles.}.  Before discussing the case
of heavy mesons, we remind the reader of a long-standing controversy
in the case of the light scalar $f_0$ and $a_0$ mesons.

\subsubsection{Light Hadrons and Cusps}

In the context of the constituent quark model, scalar mesons are
considered as quark-antiquark states with ${}^{2S+1}L_J = {}^{3}P_0$
quantum numbers.  This assignment is problematic because it leads to
states that are heavier than the lightest $s$-wave states by a typical
orbital excitation energy of several hundred MeV\@.  Thus, the
constituent quark model appears to identify the $f_0(1370)$,
$f_0(1500)$, $a_0(1450)$, and $K_0(1430)$ as the lightest scalar
nonet.  The problem, of course, is that this identification leaves the
$f_0(980)$, $a_0(980)$, and $f_0(500)$ as orphan states.

It should be stressed that this tension only exists in the context of
nonrelativistic constituent quark models.  Relativistic models, for
example, typically have large spin-orbit forces in the light-quark
sector, which can lead to light scalar
mesons~\cite{Ligterink:2003hd,Koll:2000ke}.  Nevertheless, other
issues remain: (i) The width of the $f_0$ is 40--100~MeV, much smaller
than the 500--1000~MeV expected by scaling $\Gamma(b_1\to \omega\pi)$;
(ii) the near-degeneracy of the $f_0$ and $a_0$ suggests that they are
nearly ideally mixed, but then $\Gamma(f_0\to \pi\pi)/\Gamma(a_0\to
\eta\pi)$ should be around 4 ($\pi$ vs.\ $\eta$ light-quark
Clebsch-Gordan coefficients), rather than the observed ratio of
approximately 1; (iii) both states couple strongly to $K\bar K$,
suggesting valence strange-quark content.

A popular alternative interpretation is that four-quark states $qq\bar
q \bar q$ comprise the lowest scalar nonet.  The quarks can either be
in a symmetrical configuration, as in a bag
model~\cite{Jaffe:1976ig,Jaffe:2004ph}, or asymmetrical, as in diquark
models.  In the diquark picture an ``inverted'' light-scalar nonet is
built as follows:

\begin{eqnarray}
  [ud][\bar u\bar d] &&  f_0(500) \, , \nonumber \\
  {}[ud][\bar d \bar s],\ [ud][\bar s \bar d],\ [us][\bar u \bar d],\
  [ds][\bar d \bar u]  &&  \kappa(800) \, , \nonumber \\
  \frac{1}{\sqrt{2}}([su] [\bar s\bar u] + [sd][\bar s\bar d]) &&
  f_0(980) \, , \nonumber \\
  {}[su][\bar s \bar d],\ \frac{1}{\sqrt{2}}([su][\bar s \bar u] -
  [sd][\bar s \bar d]),\ [sd][\bar s \bar u] && a_0(980) \, .
\label{eq:diq}
\end{eqnarray}

Yet another four-quark model assumes that states are predominantly
composed of a meson-meson system.  For example, the possibility that
the $a_0(980)$ and $f_0(980)$ quartet is composed of $K\bar K$ bound
states was examined in Ref.~\cite{Weinstein:1990gu}.  Similar
applications in the light-quark sector abound: $f_1(1420)$ ($K^*\bar
K$)~\cite{Caldwell:1986tg}, $f_2(2010)$
($\phi\phi$)~\cite{Hernandez:1990yc} (p.\ VII.166), and $f_0(1770)$
($K^*\bar K^*$)~\cite{Dooley:1991bg}.

Perhaps the most conservative resolution of the light-scalar issue
relies on the strong coupling of scalar mesons to their decay channels
to generate dynamical poles with masses below
1~GeV~\cite{Tornqvist:1995kr,vanBeveren:1986ea,Kaminski:1993zb}.  The
dynamical states are not arranged according to their $q\bar q$
``seed'' states, but according to the channels that dominate: $\pi\pi$
for the $f_0(500)$, $K\pi$ for the $\kappa(800)$, and $K\bar K$ for
$f_0(980)$ and $a_0(980)$.

\subsubsection{Heavy Hadrons and Cusps}

Not surprisingly, all of these multiquark notions have found
application in descriptions of the new heavy-quark exotic mesons.
Bugg was the first to emphasize the possible importance of cusps for
interpreting several heavy states~\cite{Bugg:2008wu}.  Amongst these
are the $X(3872)$ at the $D^0(1865)$-$D^{*0}(2007)$ threshold,
$Z(4430)$ near the $D^*(2007)$-$D_1(2420)$ threshold, and $Y(4260)$,
which is close to the $D(1865)$-$D_1(2420)$ threshold.  All of these
couplings are $s$ wave.  We remark that the same situation exists with
baryons: the $P_{11}$ $N(1710)$ and $P_{13}$ $N(1720)$ are near the
$N$-$\omega$ threshold, while the $\Lambda_c(2940)$ is near the
$D^*(2007)$-$N$ threshold.

Early applications of loop diagrams in heavy-meson physics
concentrated on their effects on hadronic transitions~\cite{Li:2007au,
  Li:2008xm,Hanhart:2007bd,Lutz:2007sk}.  For example, Guo {\it et
  al.}~\cite{Guo:2010zk,Guo:2010ak} examined the effects of virtual
charmed-meson loops on strong transitions of charmonia and found that
the loops significantly enhanced processes such as $\psi^\prime \to
\pi J/\psi$.  We remark that, although these studies did not address
cusp enhancements, they made the important point that the QCD
multipole expansion does not incorporate the long-distance or
long-time scales that can occur when intermediate hadrons are created
and propagate.  While this observation appears self-evident, it
contradicts several decades of lore surrounding hadronic interactions
that do not involve flavor exchange.

\begin{figure}[ht]
\begin{center}
\includegraphics[width=12cm,angle=0]{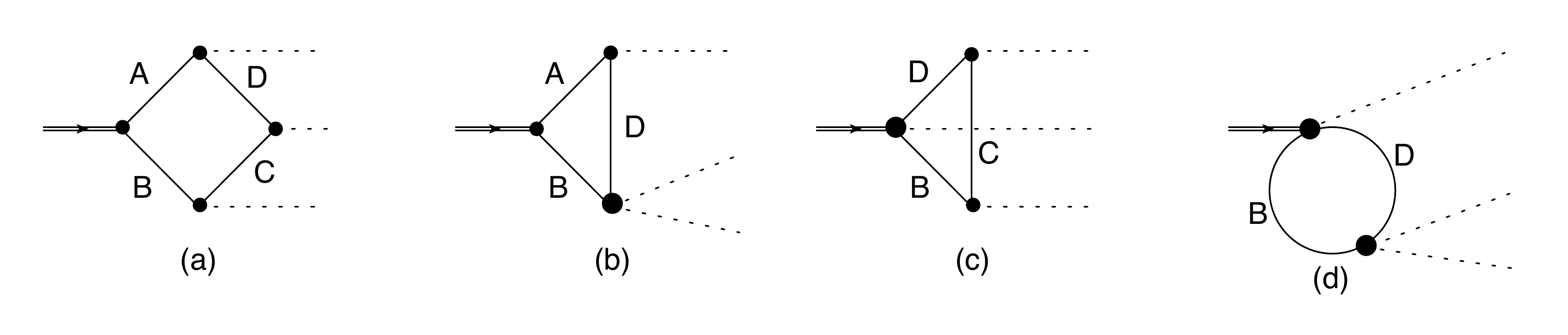}
\end{center}
\caption{Various hadronic loop diagrams.  Diagrams (b), (c), and (d)
  can be obtained from (a) by taking the limits $m_C \to \infty$, $m_A
  \to \infty$, and both, respectively.}
\label{fig:loops2}
\end{figure}

The earliest explicit model that applied cusp effects to heavy exotics
(to our knowledge) is due to Chen and Liu~\cite{Chen:2011pv}, who
developed the ``Initial Single Pion Emission'' (ISPE) model to explain
the $Z_b(10560)$ and $Z_b(10610)$ as cusp effects (although this
terminology was not used).  Ensuing work invoked the same mechanism to
predict charmed-analogue states [subsequently discovered as the
$Z_c(3900)$ and $Z_c(4020)$]~\cite{Chen:2011xk}.  The relevant
diagrams correspond to (c) in Fig.~\ref{fig:loops2}, where, in the
charm case, the initial meson is taken to be a heavy vector charmonium
state [$\psi(4040)$], the intermediate particles (B,C,D) are $D$ or
$D^*$ states, and the final state is $\pi\pi J/\psi$.

Subsequent work has employed several variants of loop diagrams to
implement the cusp effect.  These can be organized as shown in
Fig.~\ref{fig:loops2}.  At the one-loop level, all diagrams can be
obtained from the box diagram (a) by integrating out various
intermediate mesons (and rescaling appropriately).  Thus, for example,
the ISPE diagram is obtained in the large $m_A$ (or $m_B$) limit.
Other groups have considered diagram (b), which is obtained by taking
$m_C$ (or $m_D$) to be large.  Taking $m_A$ and $m_C$ to infinity
yields the bubble diagram (d).  Of course other combinations can be
taken (but appear to not be employed in the literature); these choices
merge the box diagram to a four-point vertex, or have a bubble with a
single particle on one side and a three-point vertex on the other.
Vertex models are typically built from low-order effective Lagrangians
(they are not effective field theories, in the sense of systematically
including all allowed operators), or with the aid of heavy-quark or
chiral symmetry.  An important take-away point is that the cusp
behavior obtained from all of these diagrams is similar.


The work of Wang {\it et al.}~\cite{Wang:2013cya} is an example of an
alternate analysis of the $Z_c$ states that employs a formalism in
this class of methods.  The authors consider the reaction $Y(4260) \to
\pi\pi J/\psi$ and postulate that the $Y$ is a $D_1\bar D$ bound
state.  This choice leads naturally to the box diagrams of
Fig.~\ref{fig:loops2}a.  The authors also consider the case where
$D^*\bar D$ interactions give rise to a $Z_c(3900)$ and include this
contribution in their analysis of the BESIII data.  The reasonably
good fit to the data leads the authors to claim (i) there is ``strong
evidence that the mysterious $Y(4260)$ is a $\bar DD_1(2420)+ D \bar
D_1(2420)$ molecular state'', and (ii) ``for a more detailed
description of the data the need for an explicit $Z_c(3900)$ pole
seems to be necessary''.

We remark that both of these conclusions seem to be overstatements of
the results.  It is only necessary for the $Y(4260)$ to couple to
strongly to $\bar DD_1$ for their analysis to hold, which is expected
for hybrids as well as molecular states.  With regard to the claimed
existence of a $Z_c$ pole, their own analysis (presented in their
Fig.~2) clearly shows that an explicit $Z_c$ pole appears to {\em
  not\/} be needed to describe the data. Nevertheless, follow-up work
by this group, discussed shortly, found further evidence in support of
the resonance picture of the $Z_c(3900)$.

An application of the simplest nontrivial cusp diagram,
Fig.~\ref{fig:loops2}d, appeared in 2014~\cite{Swanson:2014tra}.  In
this work, it was noted that many of the recently discovered heavy
charged states lie just {\em above\/} nearby open-flavor thresholds.
These are $\bar B B^*$ [$Z_b(10610)$], $\bar B^* B^*$ [$Z_b(10650)$],
$\bar D D^*$ [$Z_c(3900)$], and $\bar D^* D^*$ [$Z_c(4025)$].  While
the proximity to thresholds suggests a molecular interpretation for
states just below these thresholds, it would require unnatural
dynamics to generate such poles above threshold.  Even so,
Ref.~\cite{Guo:2016bjq} argues otherwise.

The exotic bottomonium states were seen in Belle in $e^+e^- \to
\Upsilon(5S) \to \Upsilon(nS)\pi^+\pi^-$ or $\Upsilon(5S) \to
h_b(nP)\pi^+\pi^-$ in the final states $\Upsilon(nS)\pi^\pm$ or
$h_c(nP) \pi^\pm$ (See Sec.~\ref{sec:ee}).  Axial-vector quantum
numbers were heavily favored by an analysis of the angular
distributions.  In modeling these features,
Ref.~\cite{Swanson:2014tra} noted that the $\Upsilon(5S)$ decays
predominantly to $B^{(*)}\bar B^{(*)}$, which leads to an
$\Upsilon(nS)\pi\pi$ amplitude that is roughly constant.  The next
most prolific decay mode of the $\Upsilon(5S)$ is to $B^{(*)}\bar
B^{(*)}\pi$, which can rescatter via Fig.~\ref{fig:loops2}d to yield
the $\Upsilon(nS)\pi\pi$ final state.  Following
tradition~\cite{Tornqvist:1995kr,Bugg:2008wu}, the imaginary part of
the bubble was modeled with an exponential vertex form factor, and the
entire bubble was reconstructed from the dispersion relation.
Couplings and form factors scales were set by data in $\Upsilon(5S)
\to \Upsilon(3S)\pi\pi$, and a surprisingly consistent description of
{\em all\/} the data (13 peaks in 7 invariant mass distributions)
followed.

The following general points were noted:

(i) $Z$ ``states'' have $J^P = 1^+$;

(ii) $Z$ ``states'' lie slightly above open-flavor thresholds;

(iii) Threshold partners produce effects of approximately the same
width if they are observed in the same channel; however, these widths
can differ in different channels. Such behavior is not expected with
$T$-matrix poles;

(iv) $Z_c$ ``states'' may appear in $\bar B^0 \to J/\psi \, \pi^0
\pi^0$ and $B^\pm \to J/\psi \, \pi^\pm\pi^0$;

(v) Similarly, $\bar B_s \to J/\psi \, \varphi \varphi$ and $\bar B_0
\to J/\psi \, \varphi K$ should exhibit cusp effects at $D_s \bar
D_s^*$ and $D_s^*\bar D_s^*$ thresholds, while $\bar B_0 \to J/\psi \,
\eta K$ should display $D\bar D^*$, $D^*\bar D^*$, $D_s\bar D_s^*$, and
$D_s^*\bar D_s^*$ cusp enhancements;

(vi) It should be possible to discern a rich spectrum of exotic
``states'' at higher center-of-mass energy in $J/\psi \, \pi\pi$.
These include a $D_0\bar D_1$ state at 4740 MeV and $D_2\bar D_1$
enhancement at 4880 MeV;

(vii) $\Upsilon(5S) \to K\bar K \Upsilon(nS)$ should show enhancements
at 10695~MeV ($B \bar B_s^*$ and $B^*\bar B_s$) and 10745~MeV
($B^*\bar B_s^*$).

Some aspects of this work were subsequently criticized.
Reference~\cite{Szczepaniak:2015eza} noted that the use of an
exponential form factor in the dispersion relation violates causality.
We remark that, while true, it has little bearing on the dispersion
integral in which it is likely to be used.  Gou {\it et
  al.}~\cite{Guo:2014iya} argued that rescattering effects need to be
summed, and that doing so necessarily forms a $Z_c(3900)$ resonance
pole.  The model employed by the authors includes a four-point $Y
J/\psi \, \pi \pi$ vertex whose strength was fitted to the $Y(4260)
\to D\bar D^* \pi$ distribution at high invariant mass.  Fitting the
same distribution at low invariant mass then required iterating $D\bar
D^*$ bubble diagrams with sufficient strength to generate a
$Z_c(3900)$ pole.

Nevertheless, it is clear the conclusion of Gou {\it et al.} is
contingent upon the model used.  In particular, it is natural to
attribute the enhancement in the $D\bar D^*\pi$ data near threshold to
the threshold opening---as is common in hadronic physics.  The latter
point of view was pursued in Ref.~\cite{Swanson:2015bsa}, where a
simple, causal, and unitarized nonrelativistic model was employed to
describe the $Y\to D \bar D^*\pi$, $Y\to D^*\bar D^* \pi$, $Y \to
J/\psi \, \pi\pi$, and $e^+e^-\to h_c\pi\pi$ data. No $S$-matrix poles
were required to obtain very good agreement with experiment.  This
conclusion is supported by lattice gauge theory computations that
report only weakly repulsive $(D^*\bar D^*)^\pm$ interactions in the
$J^P=1^+$ channel~\cite{Chen:2015jwa}.

This topic has been revisited recently by Zhou and
Xiao~\cite{Zhou:2015jta}, who employed a unitarized coupled-channel
approach similar to that of T{\"o}rnqvist~\cite{Tornqvist:1995kr} to
analyze the same $Z_c$ data as above.  The authors concluded that the
$Z_c(3900)$ signal is related to the combined effect of a pair of
near-threshold ``shadow'' poles and the $D\bar D^*$ threshold, in
which a third-sheet pole might provide a dominant contribution.
Similar work was also carried out by He~\cite{He:2015mja} in a model
that generated relevant interactions with boson exchange and employed
a Bethe-Salpeter formalism to compute amplitudes.

It is disappointing, but not surprising, that conclusions concerning
the dynamical origin of the $Z_b$ and $Z_c$ signals can depend
strongly upon the model assumptions used to generate those
conclusions.  The way forward, of course, is to develop models that
are sufficiently robust and well vetted against experimental results,
such that the conclusions based upon them can be trusted.

Threshold enhancements and openings are generic features of hadronic
systems, and one must therefore be cautious in claiming bound states
where such effects are known to operate.  Near-threshold enhancements
can arise simply because hadrons are soft; thus, the $Y \to D\bar
D^*\pi$ and $Y \to D^*\bar D^*\pi$ data are easily explained.
Similarly, coupled-channel cusps should be regarded as a possible
explanation for bumps seen in rescattering channels slightly above
coupled-channel thresholds.  If the ``widths'' of these enhancements
vary strongly (as they do for the $Z_c$'s) between pure threshold and
rescattering processes, then one has an additional sign that
nonresonant explanations should be considered.  In particular,
threshold bumps arise due to competing effects between form factors
and phase space, whereas a rescattering enhancement width is mediated
by form factors and a rescattering loop.  One sees that
(cusp-dominated) threshold bumps do not exhibit phase motion, while
rescattering enhancements may have phase motion due to the associated
bubble diagrams.

Indeed, both threshold effects and true resonant poles can coexist and
interact; as was shown in Ref.~\cite{Bugg:2008wu}, the presence of a
threshold can, through self-energy diagrams, shift the position of the
fundamental resonance closer to the threshold.  It was noted in
Ref.~\cite{Blitz:2015nra} that this ``pole dragging'' effect does not
depend upon the origin of the pole, whether through meson molecules,
diquark-antidiquark pairs, or intrinsic $q\bar q$ states.


\subsubsection{Further Experimental Considerations}

The LHCb collaboration recently reported the discovery of four states
in the $J/\psi \, \phi$ invariant-mass distribution of $B^+\to J/\psi
\, \phi K^+$~\cite{Aaij:2016iza} (See Sec.~\ref{sec:Y4140}).  Salient
properties of these states are given in Table~\ref{tab:4X}.

\begin{table}[ht]
\centering
\begin{tabular}{l|lll}
\hline\hline
State & Mass (unct.) [MeV] & Width (unct.) [MeV] & $J^{PC}$ \\
\hline
$Y(4140)$ & 4165.5(5,3)  & 83(21,16)  & $1^{++}$ \\
$Y(4274)$ & 4273.3(8,11) & 56(11,10)  & $1^{++}$ \\
$X(4500)$ & 4506(11,13)  & 92(21,21)  & $0^{++}$ \\
$X(4700)$ & 4704(10,19)  & 120(31,35) & $0^{++}$ \\
\hline\hline
\end{tabular}
\caption{Extracted $J/\psi \, \phi$ Breit-Wigner resonance
properties~\cite{Aaij:2016iza}.}
\label{tab:4X}
\end{table}

The resonance parameters of Table~\ref{tab:4X} are based on standard
Breit-Wigner phenomenology.  However, motivated by point (v) above,
the collaboration also fit the data with a cusp model of the
$Y(4140)$.  This amplitude model had one less parameter than the
$s$-wave Breit-Wigner model and was able to fit the data better by
3$\sigma$ (it has five fewer parameters than the full Breit-Wigner
model and in this case fits the data better by 1.6$\sigma$).  This
success motivated the collaboration to construct a $D_s D_{s0}^*$ cusp
model for the $Y(4274)$.  It is perhaps not surprising that this
choice did not perform better than the default amplitude model, since
the $s$-wave quantum numbers are incorrect.

In the bottom sector, the relatively large rate for the reaction
$\Upsilon(5S) \to h_b \pi\pi$ is somewhat mysterious, because a
heavy-quark spin flip is required to make the transition from the
${}^3S_1$ $\Upsilon$ to the ${}^1P_1$ $h_b$ bottomonium state.  It is
tempting to speculate that the spin flip is facilitated by the
presence of light-quark degrees of freedom in the loop-diagram
intermediate state that persist over long time scales.  In effect, the
virtual $B^{(*)}\bar B^{(*)}$ states permit the pions to carry off the
spin component necessary to effect the required $b$-quark spin flip.

Lin {\it et al.}~\cite{Lin:2013mka} have suggested that the coupling
of the $Z_c(3900)$ to $\pi J/\psi$ can be exploited to search for this
exotic state in pho\-to\-pro\-duc\-tion.  The idea is that the virtual
photon converts to a $J/\psi$ via the vector-meson dominance
mechanism, which then interacts with a nucleon by pion exchange, and
creates an $s$-channel $Z_c$, which finally decays to $J/\psi \, \pi$.
The cross section for $\gamma N \to Z N$ was estimated using a
hadronic Lagrangian with dipole form factors, while the $Z_c \pi
J/\psi$ coupling was taken from the measured width of the $Z_c$.  The
resulting cross section was predicted to peak at $\sqrt{s} \approx
7$~GeV, with a readily observable rate.  In spite of these
expectations, a measurement of $\mu N \to \mu J/\psi \, \pi N$ by the
COMPASS collaboration~\cite{Adolph:2014hba} found no evidence for the
$Z_c(3900)$.  This somewhat problematical result may have a resolution
in dynamical effects associated with the high center-of-mass energy
(such as hadronic form factors or Pomeron exchange).

Finally, there appears to be a tension between $e^+e^-$ and
electroweak decay production mechanisms for a subset of exotic states.
In particular, we note that the electroweak decay $\bar B^0 \to J/\psi
\, \pi\pi$ has recently been measured by the LHCb Collaboration, and
the distribution of events in $J/\psi \, \pi$ invariant mass was
published~\cite{Aaij:2014siy}.  A comparison of the analogous
distribution from BESIII reveals a stark difference: Although the
distributions stretch over nearly identical mass ranges, there is no
sign of the $Z_c(3900)$ or $Z_c(4020)$ in the LHCb data.  This result
is difficult to understand, because other than quantum numbers, there
seems to be little difference between $\gamma^* \to c\bar c$
($Y(4260)$ decay where the $Z_c(3900)$ is seen,
Sec.~\ref{sec:eeSubCharmonium}) and $b \to c\bar c d$ ($B$ decay,
where the $Z_c$ is not seen).

\section{Prospects} \label{sec:Prospects}

The search for and discovery of new QCD exotics has been one of the
most productive areas of experimental particle physics in the past two
decades.  For example, of all the particles discovered so far at the
Large Hadron Collider whose existence was not a foregone conclusion,
all are QCD exotics with the exception of the Higgs boson.  The fact
that the rate of appearance of new experimental findings currently
greatly outpaces theoretical explanations in this area [for instance,
even the true structure of the $X(3872)$, the first and best studied
example among the heavy-quark exotics, remains a mystery] is a state
of affairs that has not occurred for many years in particle physics
and provides an excellent indication of intellectual vitality in this
field.

No obvious ceiling to the number of exotics yet to be discovered is
known; there could be many dozens more remaining to be found.
Moreover, the LHC experiments (particularly LHCb) and BESIII will
continue to take data for several more years, the Belle~II upgrade
will soon be coming online, and in the future, detailed processes will
be examined at GlueX (Jefferson Lab), COMPASS (CERN), and
$\bar{\rm P}$ANDA at FAIR (GSI, Darmstadt).  The great majority
of exotics already observed appear in the hidden-charm $c\bar c$ and
hidden-bottom $b\bar b$ tetraquark and $c\bar c$ pentaquark sectors.
The flavor universality of QCD demands that, once the particular
details of hadronic thresholds are taken into account, exotics should
be possible in any combination of quark flavors.  To date, the only
exotic candidate with valence quark content other than $c\bar c$ and
$b\bar b$ is the $X(5568)$ ($\bar b s \bar d u$), and its existence
remains controversial (See Sec.~\ref{sec:confirmation}).  And yet, the
$c\bar c$ sector was the first one in which the first unambiguous
exotics candidates were found, despite decades of dedicated searches
in the lighter-quark sectors.  Is there a sense in which the existence
of exotics, or at least the ability to distinguish them from
backgrounds, requires the presence of heavy quarks?  If it is the
latter, then one can hope to glean some hints of exotics among states
containing some (intermediate-mass) $s$ quarks.  The $D^*_{s0}(2317)$
discovered by BaBar in 2003~\cite{Aubert:2003fg}, for example, is
surprising in having a much lighter mass than predicted for the
expected $c\bar s$ state, and moreover decays through the
isospin-violating mode $D_s \pi^0$.  The state $Y(2175)$ [or
$\phi(2170)$] was discovered by BaBar in 2007~\cite{Aubert:2006bu} as
a $\phi f_0$ resonance in the initial-state radiation process $e^+ e^-
\to \gamma^{\vphantom\dagger}_{\rm ISR} \phi \pi^+ \pi^-$, the strange
analogue of the one in which $Y(4260)$ was originally found (See
Sec.~\ref{sec:eeCharmonium}).  Now that the existence of heavy-quark
exotics has been established, anomalous light-quark systems have come
under greater scrutiny.

In the opposite direction of even more exotic hadrons, one can
anticipate the production of states with manifestly exotic (for $q\bar
q$) quantum numbers such as $J^{PC} = 1^{-+}$.  In that case, if the
state is neutral ($c\bar c q \bar q$), then one faces the enviable
problem of trying to determine whether the states is a
``conventional'' hybrid exotic (See Sec.~\ref{sec:hybrids}) or a
tetraquark exotic.  On the other hand, doubly heavy exotics (such as
$cc \bar q \bar q$) lack heavy-quarkonium decay modes and should be
rather straightforward to identify if they can be produced.  Options
for producing doubly heavy hadrons have been discussed in detail ({\it
  e.g.}, Ref.~\cite{Karliner:2016via,Koshkarev:2016rci}), and while
not even the lowest conventional doubly heavy state [the $\Xi_c (ccq)$
baryon] has yet been confirmed, the prospects for producing such
states are nevertheless considered bright.  A key observation
of~\cite{Karliner:2016via} is that the chief complication in producing
doubly-heavies, to produce two separate heavy $Q\bar Q$ pairs and
induce a quark from each pair to coalesce, is already accomplished in
$B_c$ production.  Copious $B_c$ output (such as at LHCb) is therefore
seen as a promising benchmark for producing doubly heavy hadrons.  At
an even higher level of exoticity are {\it
  hexaquarks\/}~\cite{Jaffe:1976yi,Matveev:1977xt}, which can refer
either to a bound state of 6 $q$'s, or 3 $q$'s plus 3 $\bar q$'s.
Considering the comments on doubly heavies, presumably the latter type
would be easier to produce with heavy quarks; indeed, it is possible
that the $X(4630)$, with its strong coupling to $\Lambda_c {\bar
  \Lambda}_c$ (See Sec.~\ref{sec:eeCharmonium}), is such a state.

The differing status of experiment and theory with respect to our
knowledge of the exotic candidates requires different intellectual
approaches.  Imminent experimental studies present tremendous
opportunities for clarifying multiple issues (discovery of new states,
confirmation of others, measurement of production and decay channels)
in a systematic fashion: One can actually present these goals as a
``to-do'' list, as given below.  Theoretical studies also present
tremendous opportunities, in the sense that all known pictures have
limits to their applicability, and none of them provide a plausible
explanation for the whole set of exotic candidates.  As discussed
below, the direction of progress involves an honest assessment of the
constraints on existing methods (does one really know what being a
molecule means, which fields are truly important near hadronic
thresholds, {\it etc.}), leading both to improvements rendering
existing techniques more flexible and robust, and to the development
entirely new theoretical approaches.

\subsection{Experimental Issues and Prospects}

As mentioned above, the goals for ongoing and future efforts in the 
experimental study of QCD exotica are comparatively straightforward.
Here we simply list ten of the most important.

\begin{enumerate}

\item Search for qualitatively new classes of particles.  There has
  been a growing list of robust experimental discoveries indicating
  potential candidates for QCD exotica.  These include tetraquark
  candidates containing $c\bar{c}$ and light quarks [like the
  $Z_c(3900)$ and $Z_c(4430)$], pentaquark candidates containing
  $c\bar{c}$ and light quarks [the $P_c(4380)$ and $P_c(4450)$], and
  supernumerary states that are potentially hybrid mesons~[like the
  $Y(4260)$].  However, pinning down these interpretations has proven
  difficult.  Finding qualitatively new classes of particles could
  solidify emerging patterns ({\it e.g.}, peaks near thresholds) or
  reveal new ones.  A few examples that could be revealing include:
  hexaquarks~(either dibaryons or tri-mesons); double-heavy open-charm
  exotics~(like $cc\bar{u}\bar{d}$); other tetraquark
  combinations~[the $X(5568)$ is particularly intriguing and should
  obviously be studied further]; and states with exotic $J^{PC}$.

\item Search for new decay modes of the particles that have already
  been discovered.  Once a particle is discovered in some production
  process~({\it e.g.}, $B$ decays or $e^+e^-$ annihilation), that same
  process can be used to search for new decay modes.  So far, many
  particles have only been observed in a single decay mode.  This
  result may, at least in part, be due to experimental commonplaces.
  For example, it is much easier to find decay modes containing a
  $J/\psi$~(which is narrow and has a large dilepton branching
  fraction) than those with an $\eta_c$~(which is wide and decays to
  multiple-hadron final states).  As more data is collected, these
  less experimentally accessible decay modes should be explored.
  Judging from history, besides possibly establishing new decay modes,
  these searches are also likely to lead to the discovery of new
  particles.  Furthermore, none of the discovered exotic states has
  yet to be seen to decay exclusively to light-quark final states, a
  result that might be due to a lack of data and comprehensive
  searches.

\item Identify the conventional mesons in the region between 3.8 and
  4.0 GeV\@.  This region serves as a crucial test for our
  understanding of conventional mesons above open-flavor thresholds,
  which is especially important if we hope to distinguish conventional
  mesons from exotic mesons using their observed properties.  While
  the situation is currently complicated, there are a number of
  experimental measurements yet to be completed.  For example, it
  seems the decay $\chi_{c2}(2P)\to D\bar{D}$ has been established,
  assuming the $Z(3930)$ is the $\chi_{c2}(2P)$.  But can the
  $\chi_{c2}(2P)$ decay to $D\bar{D}^*$ be found?  And what about the
  decay $\chi_{c0}(2P)\to D\bar{D}$?  Besides open-charm decays,
  radiative decays~(despite being much smaller) could also prove
  decisive.

\item Understand $e^+e^-$ cross sections, both exclusive and
  inclusive, as a function of center-of-mass energy.  As discussed in
  Section~\ref{sec:eeCharmonium}, $e^+e^-$ cross sections in the
  charmonium region show a surprisingly diverse range of shapes.
  These shapes should be mapped more finely in the future.  Will
  patterns start to emerge?  Will possibly informative connections
  like the decay $Y(4260)\to\gamma X(3872)$ be established?  Special
  attention should be paid to the open-charm cross sections, which are
  larger than the closed-charm cross sections, and which also display
  surprising structure.  If possible, open-bottom cross sections
  should similarly be mapped.

\item Continue to establish the existence or nonexistence of
  resonances in $B$ decays.  The LHCb experiment has already resolved
  long-standing controversies in the $B\to K\pi \psi(2S)$ and $B\to K
  \phi J/\psi$ channels.  Both revealed the existence of a number of
  states.  Several more channels, however, remain controversial.  The
  decays $B\to K\pi J/\psi$ and $B\to K\pi \chi_{c1}$ should be
  revisited with higher statistics.  Other $B$ and $\Lambda_b$ decays
  should also be explored.  Will the pattern of {\em pairs\/} of peaks
  continue?

\item Explore the differences between the $Z_c$ in $B$ decays and in
  $e^+e^-$ annihilation.  While many $Z_c$ peaks have been seen in
  either $B$ decays or $e^+e^-$ annihilation, one of the most
  conspicuous puzzles is the fact that no $Z_c$ peak has been seen in
  both.  The $Z_c(3900)$, for example, discovered in $e^+e^-\to
  \pi^\mp Z_c$ with $Z_c \to \pi^\pm J/\psi$, is not seen in $B$
  decays to either $K(\pi J/\psi)$ or $\pi(\pi J/\psi)$.  Searches
  with higher statistics should be performed.  Similarly, the
  $Z_c(4430)$, discovered in $B\to K Z_c$ with $Z_c \to \pi^\pm
  \psi(2S)$ should be searched for in $e^+e^-$ annihilation to
  $\pi^\mp Z_c$ with $Z_c \to \pi^\pm \psi(2S)$.  This study requires
  higher statistics at higher center-of-mass energies than currently
  available.

\item Search for previously established particles in hadron
  production.  Of all the candidates for QCD exotica, currently only
  the $X(3872)$, $Y(4140)$, and $X(5568)$ have been reported in prompt
  hadron production.  Several other states, however, might also be
  experimentally accessible.  Besides the $Y(4140)$, the $Y(4274)$,
  $X(4500)$, and $X(4700)$ might also be seen in {\em inclusive\/}
  $\phi J/\psi$ production.  If they are not seen, why not?  Several
  of the $Z_c$ states could also have signatures allowing them to be
  observed in hadron production, such as $Z_c(3900) \to \pi^\pm
  J/\psi$ or $Z_c(4430) \to \pi^\pm \psi(2S)$.

\item Continue mining relatively unexplored production mechanisms,
  such as $\gamma\gamma$ collisions and $e^+e^- \to \psi + X$.  While
  a lot of attention has been paid to $B$ decays and $e^+e^-$
  annihilation, other production mechanisms should continue to be
  explored.  In particular, $\gamma \gamma$ collisions and $e^+e^- \to
  \psi + X$ should be revisited with higher statistics.

\item Explore qualitatively new production mechanisms.
  New production mechanisms are desparately needed and may soon be
  available. At the upcoming ${\bar{\rm P}}$ANDA experiment, a scan of
  $p\bar{p}$ cross sections would be extremely interesting.  How would
  it compare to $e^+e^-$ annihilation?  At GlueX, the photoproduction
  of light-quark hybrid mesons will be studied.  At COMPASS, the
  photoproduction of a variety of charmonium states is feasible and
  studies have just begun~\cite{Adolph:2014hba}.

\item Consider more advanced phenomenological methods, such as coupled
  channels.  While it is relatively straightforward for experiments to
  measure cross sections and to fit peaks, more advanced phenomenology
  becomes important when performing detailed Dalitz plot analyses or
  considering the effects of one channel coupling to another.  There
  has been a lot of success in extending experimental studies in this
  direction.  These efforts need to continue, and cooperation between
  experiment and theory remains essential.

\end{enumerate}

\subsection{Theoretical Issues and Prospects}

Theory finds itself in an interesting predicament because the tools
that have served so well for conventional hadronic states ({\it e.g.},
the quark model, chiral perturbation theory, lattice studies of static
properties) seem to require new insights and extensions to maintain
their applicability to the exotic states.  The best-known theoretical
methods require a clear separation between distinct
distance/time/energy scales, but it is not clear how far apart the
four (or five) quarks reside in these states, or even if they reach an
equilibrium before the exotic state decays, so it is not clear what
substructure degrees of freedom best represent the states.  Moreover,
quantum mechanics allows components of different structures but the
same overall quantum numbers to mix; we have seen this proposal in the
suggestion that the $X(3872)$ is a molecule plus conventional
charmonium (Sec.~\ref{sec:Molecule}).  Similarly, the $J^{PC} =
1^{--}$ states such as the $Y(4260)$ provide an excellent place to
look for the lowest hybrid (Sec.~\ref{sec:hybrids}), but they may mix
with $I=0$ tetraquark states.  Thus far, theoretical pictures tend to
be at their most incisive when they are not as successful in
explaining a given state, because they better indicate the limitations
of the picture.

Indeed, among the generic problems with theory is that it can be
focused too narrowly. For example, quark models may concentrate upon a
specific flavor sector, which may be adequate when sufficient data
exists in that sector, but often it does not; one should investigate
models as broadly as possible to ensure a reasonable level of
reliability.  A related problem occurs for molecular models of states
near thresholds that assume a particular meson pair $AB$ forms a
resonance, but that often do not posit any plausible mechanism for
this binding.  Under these conditions all possible meson pair
thresholds can be thought to form a resonance or bound state!  A
similar situation occurs with threshold-cusp models associated with a
channel $AB$.  In this case, it is incumbent on the proposer to
explain why the resulting cusp should dominate the process under
consideration.  Analogous observations apply to lattice gauge theory
results that have not been obtained with a large interpolating-field
basis (especially in the case of exotic spectroscopy, where
multihadron interpolating fields must be considered), and to QCD
sum rule calculations, which require careful validation against
experiment to ensure robustness.  These problems are compounded when
the technique is applied to incomplete or ambiguous experimental
results, such as is the case in some sectors of exotic spectroscopy.

Lattice QCD could be considered the arbiter of these questions, and
the quality of these calculations is definitely improving ({\it e.g.},
the calculated pion mass is becoming ever closer to its physical
value), but lattice simulations are at their best when dealing with
compact, static, isolated states.  The need to use a suitably large
basis of interpolating operators to properly study exotic states has
already been noted.  Additionally, interpolating operators are
typically pointlike; extended states require the simulation of link
variables, which is computationally expensive.  States with finite
widths like exotics require a lattice technology that was originally
developed for elastic scattering~\cite{Luscher:1990ux} and has been
extended to multiparticle states~\cite{Hansen:2012tf}, but has not yet
been as well developed, particularly for heavy-quark states.
Coupled-channel effects, which are clearly important for the
heavy-quark exotics, have only recently received their first
light-quark lattice simulations~\cite{Dudek:2014qha}, but have not yet
been studied for heavy-quark systems.

From the theoretical perspective, one of many experimental mysteries
is the apparent difficulty of observing some exotic states in $B$
decays, even though they have been seen in $e^+e^-$ collisions.  For
example, the $Z_c(3900)$ and $Z_c(4020)$ are clearly revealed in
$e^+e^- \to J/\psi(h_c) \, \pi \pi$.  Nevertheless, the Dalitz
plot for $\bar B^0 \to J/\psi \, \pi\pi$ has been measured by
LHCb~\cite{Aaij:2014siy}, and shows no hint of these states.  In
contrast, the $X(3823)$ [now believed to be the $\psi_2 (1D)$] has
been observed in both $B\to K \gamma \chi_{c1}$ and in $e^+e^- \to
\pi\pi \gamma \chi_{c1}$.  These findings are difficult to understand
unless the production mechanism depends sensitively on the
environment; if they withstand further experimental scrutiny, it
becomes incumbent upon theorists to explain why these cases are
different.

An analogous situation in the $\phi J/\psi$ channel where the LHCb
collaboration found a quartet of new states in $B\to K Y \to K \phi
J/\psi$~\cite{Aaij:2016iza}.  A notable absence in the list is the
$X(4350)$, which was observed in $\gamma\gamma \to \phi J/\psi$.  One
expects this state to have quantum numbers $J^P = 2^+$ or $0^+$ and
hence show up in $p$- or $s$-wave in $B$ decays, respectively.  Since
the statistical significance of the $X(4350)$ is below $5\sigma$ (See
App.~\ref{app:states}), it may disappear entirely, and models
predicting it to exist become suspect.

There is also a curious dichotomy between exotics that decay to
$J/\psi$ and to $\psi(2S)$.  In particular, the $Z_c(4055)$,
$Y(4360)$, and $Z_c(4430)$ are produced in either $e^+e^-$ or $B$
decays, and so far appear only in $\psi(2S) \, \pi$ or $\psi(2S) \,
\pi\pi$.  In contrast, the $Y(4260)$ and $Z_c(4200)$ were discovered
in the $J/\psi \, \pi\pi$ or $J/\psi \, \pi$ decay modes,
respectively.  Naively, one would expect all the states to couple to
$\psi(nS) \, \pi(\pi)$ with weights approximately given by phase space
and slowly varying couplings.  Perhaps this conundrum is an example,
among many, that points to novel dynamics, such as that proposed in
Sec.~\ref{sect:dyndq}.

\section{Conclusions} \label{sec:Concl}

An entirely new chapter of hadronic physics opened in 2003, with the
discovery of the enigmatic $X(3872)$.  It has since been joined by
about 30 equally mysterious states, many believed to be tetraquarks or
pentaquark resonances, while others might ultimately turn out instead
to be prominent effects due to the opening of hadronic thresholds.

In this review we explored the history and techniques of the
experiments that discovered, confirmed, and measured the mass,
$J^{PC}$, and decay properties of these numerous states.  Very often,
the search to probe a known state led to the discovery of new ones.

We also examined in detail the leading theoretical pictures proposed
for describing these states, finding that no single paradigm yet fits
all of the candidate exotics.  The eventual consensus picture may turn
out to be one that is yet undiscovered, or a hybrid of those already
proposed.  In any case, the rate of theoretical work has not abated,
with a new wave of excitement each time a new exotic candidate is
discovered.

Many of the most prolific experiments
uncovering these new results are still currently active, while a
number of others designed to be sensitive to new production processes
and/or new decay modes will come online in the next few years.  The
directions of both experimental and theoretical discovery can never be
predicted, but only conjectured based upon past experience.  In that
light, the next several years should be just as rich, if not richer,
in the volume of new experimental information and the creation of new
theoretical ideas for these new classes of hadrons.

\section*{Acknowledgments}
This work gratefully acknowledges support from the U.S.\ National
Science Foundation under Grant No.\ 1403891 (R.F.L.) and the U.S.\
Department of Energy under Grant DE-FG02-05ER41374 (R.E.M.).
 
\begin{appendices}

\section{Glossary of Exotic States}
\label{app:states}

\subsection{$X(3823)$ {\rm (}or $\psi_2(1D)${\rm )}}

The $X(3823)$ was discovered by the Belle Collaboration in 2013 in the
reaction $B \to KX$ with $X\to \gamma
\chi_{c1}$~\cite{Bhardwaj:2013rmw}.  The BESIII Collaboration later
found a peak consistent with the $X(3823)$ produced in $e^+e^- \to
\pi^+\pi^- X$, again with $X\to \gamma
\chi_{c1}$~\cite{Ablikim:2015dlj}.  The $X(3823)$ is likely the
$\psi_2(1D)$ state of charmonium.  See Sec.~\ref{sec:X3823} for more
detail.

\subsection{$X(3872)$}

Accidentally discovered by the Belle Collaboration in 2003 in the
reaction $B\to KX$ with $X\to\pi^+\pi^-J/\psi$~\cite{Choi:2003ue}, the
$X(3872)$ was both the first of the $XY\! Z$ states to be discovered
and is the one that has been most studied.  Nevertheless, like most of
the $XY \! Z$ states, there is no interpretation that is universally
agreed upon.  It has been produced in decays of the $B$
meson~\cite{Choi:2003ue,Aaij:2015eva,Choi:2011fc,Aaij:2013zoa,
  Aubert:2008gu,Aubert:2004ns,Aubert:2005zh,Gokhroo:2006bt,
  Aubert:2007rva,Adachi:2008sua,Aaij:2014ala,Aubert:2006aj,
  Aubert:2008ae,Bhardwaj:2011dj,delAmoSanchez:2010jr,Bala:2015wep}, in
hadronic collisions~\cite{Acosta:2003zx,Chatrchyan:2013cld,
  Abazov:2004kp,Aaltonen:2009vj, Aaij:2011sn,Abulencia:2005zc}, and
perhaps in radiative decays of the $Y(4260)$~\cite{Ablikim:2013dyn}.
Besides $\pi^+\pi^-J/\psi$, it has also been seen to decay to $\omega
J/\psi$~\cite{delAmoSanchez:2010jr},
$D^*\bar{D}$~\cite{Gokhroo:2006bt,Aubert:2007rva,Adachi:2008sua},
$\gamma J/\psi$~\cite{Aaij:2014ala,Aubert:2006aj,Aubert:2008ae,
  Bhardwaj:2011dj}, and $\gamma \psi(2S)$~\cite{Aaij:2014ala,
  Aubert:2008ae}.  Its unusual features include a mass that is
currently indistinguishable from the $D^{*0}\bar{D}^0$ threshold~(the
current mass difference is $0.01\pm0.18$~MeV) and a narrow
width~($<1.2$~MeV).  It is has no isospin partners and has
$J^{PC}=1^{++}$.  See Sec.~\ref{sec:X3872} for more discussion of its
experimental properties.

\subsection{$Z_c(3900)$}

The $Z_c(3900)$ was simultaneously discovered in 2013 by the BESIII
and Belle Collaborations in the process $e^+e^- \to \pi^\mp Z_c^\pm$
with $Z_c^\pm \to \pi^\pm J/\psi$.  For the BESIII
observation~\cite{Ablikim:2013mio}, the center-of-mass energy was
fixed to 4.26~GeV\@.  Belle~\cite{Liu:2013dau} used initial-state
radiation to cover the energy region from 4.15 to 4.45~GeV,
corresponding to the region of the $Y(4260)$.  It is not yet clear
whether the production of the $Z_c(3900)$ is associated with the
$Y(4260)$.  The $Z_c(3900)$ has since been seen in decays to $\pi^0
J/\psi$~\cite{Xiao:2013iha,Ablikim:2015tbp} ($Z_c^0$) and in
$D^*\bar{D}$~(both charged and
neutral)~\cite{Ablikim:2015gda,Ablikim:2013xfr,Ablikim:2015swa}.  It
has only been produced in the reaction $e^+e^- \to \pi Z_c$.  See
Sec.~\ref{sec:eeSubCharmonium} for more experimental details.

\subsection{$X(3915)$ {\rm (}or $\chi_{c0}(2P)${\rm )}}

The $X(3915)$ was first seen by the Belle Collaboration in 2010 in the
process $\gamma\gamma \to X$ with $X \to \omega
J/\psi$~\cite{Uehara:2009tx}.  It was later confirmed by the BaBar
Collaboration~\cite{Lees:2012xs}.  It appears as a clear peak with
little background.  Its $J^{PC}$ is likely $0^{++}$, so there is some
possibility that it is the $\chi_{c0}(2P)$ state of charmonium,
although this assignment is controversial.  See Sec.~\ref{sec:XYZ3940}
for more discussion.

\subsection{$Y(3940)$}

The $Y(3940)$ was first observed in 2005 by the Belle Collaboration in
the process $B\to K Y$ with $Y\to \omega J/\psi$~\cite{Abe:2004zs}.
It was the second of the $XY \! Z$ to be discovered [after the
$X(3872)$].  It was later confirmed by the BaBar Collaboration in both
2008~\cite{Aubert:2007vj} and 2010~\cite{delAmoSanchez:2010jr}, but
with a mass around 3915~MeV\@.  Due to its similar mass and width to
the $X(3915)$, and since both states decay to $\omega J/\psi$, the
$Y(3940)$ is usually identified with the $X(3915)$.  See
Sec.~\ref{sec:XYZ3940} for more details.

\subsection{$Z(3930)$ {\rm (}or $\chi_{c2}(2P)${\rm )}}

The $Z(3930)$ was discovered in 2006 by the Belle Collaboration in the
process $\gamma\gamma \to Z$ with $Z \to
D\bar{D}$~\cite{Uehara:2005qd}.  It was confirmed in 2010 by the BaBar
Collaboration in the same process~\cite{Aubert:2010ab}.  Both
collaborations measured $J^{PC} = 2^{++}$.  It is often assumed to be
the $\chi_{c2}(2P)$ state of charmonium.  See Sec.~\ref{sec:XYZ3940}
for more details.

\subsection{$X(3940)$}

The $X(3940)$ was first reported in 2007 by the Belle Collaboration in
$e^+e^-$ collisions near the $\Upsilon(4S)$ resonance.  It appeared in
$e^+e^- \to J/\psi X$, where the $X$ either decayed inclusively or to
$D^*\bar{D}$~\cite{Abe:2007jna}.  A later analysis by Belle using a
similar technique but with more data confirmed the $X(3940)$ decay to
$D^*\bar{D}$~\cite{Abe:2007sya}.  The same analysis also reported a
$X(4160)$ state decaying to $D^*\bar{D}^*$ and a very broad excess of
events near threshold in $D\bar{D}$.  See Sec.~\ref{sec:XYZ3940} for
more details.

\subsection{$Z_c(4020)$}

The $Z_c(4020)$ was discovered in 2013 by the BESIII Collaboration in
the process $e^+e^- \to \pi^\mp Z_c^\pm$ with $Z_c^\pm \to \pi^\pm
h_c$~\cite{Ablikim:2013wzq}.  The center-of-mass energies included
4.23, 4.26, and 4.36~GeV, and the production rate of the $Z_c(4020)$
apparently does not vary greatly over this energy range.  It is thus
unclear whether or not its production is associated with the
$Y(4260)$, $Y(4360)$, or any other state produced in $e^+e^-$
annihilation.  The $Z_c(4020)$ has since been seen to decay to $\pi^0
h_c$~\cite{Ablikim:2014dxl} and to both charged~\cite{Ablikim:2013emm}
and neutral~\cite{Ablikim:2015vvn} combinations of $D^*\bar{D}^*$.
See Sec.~\ref{sec:eeSubCharmonium} for more discussion.

\subsection{$Z_1(4050)$ and $Z_2(4250)$}

The $Z_1(4050)$ and $Z_2(4250)$ were reported in 2008 by the Belle
Collaboration in the process $B\to K Z^\pm$ with $Z^\pm \to \pi^\pm
\chi_{c1}$~\cite{Mizuk:2008me}.  Using a Dalitz plot analysis, the
significance of each structure was reported to be more than $5\sigma$.
BaBar searched for the same structures in the same reaction but used a
phenomenological method to describe the effects of possible resonances
in the $K\pi$ subsystem~\cite{Lees:2011ik}.  They found no need for
the $Z_1(4050)$ or $Z_2(4250)$ states, but with upper limits that are
not inconsistent with the early measurements of Belle.  See
Sec.~\ref{sec:otherB} for more discussion.

\subsection{$Z_c(4055)$}

The $Z_c(4055)$ was reported in 2015 by the Belle Collaboration in the
reaction $e^+e^- \to \pi^\mp Z_c^\pm$ with $Z_c^\pm \to \pi^\pm
\psi(2S)$~\cite{Wang:2014hta}.  A wide range of $e^+e^-$ energies was
covered using the initial-state radiation of the beams.  It appears
that the $Z_c(4055)$ is only produced when the $e^+e^-$ energies are
near the $Y(4360)$~(between 4.0 and 4.5~GeV).  When the energy is in
the $Y(4660)$ region (between 4.5 and 4.9~GeV), no signal is apparent.
See Sec.~\ref{sec:eeSubCharmonium} for more discussion.

\subsection{$Y(4140)$, $Y(4274)$, $X(4500)$, and $X(4700)$}

The $Y(4140)$ was first reported in 2009 by the CDF Collaboration in
the process $B \to K Y$ with $Y \to \phi
J/\psi$~\cite{Aaltonen:2009tz}.  A series of
positive~\cite{Abazov:2013xda,Chatrchyan:2013dma} and
negative~\cite{Aaij:2012pz,Lees:2014lra} searches using the same
process followed, making the status of the $Y(4140)$ uncertain.  In
addition to the $Y(4140)$, the CDF and CMS Collaborations found
evidence for a higher-mass structure, the
$Y(4274)$~\cite{Chatrchyan:2013dma,Aaltonen:2011at}, whose status was
also uncertain.  In 2016, the LHCb Collaboration used a much larger
sample of events to decisively confirm the existence of both the
$Y(4140)$ and the $Y(4274)$~\cite{Aaij:2016iza,Aaij:2016nsc}.  In the
same analysis, LHCb also found evidence for two more structures, the
$X(4500)$ and the $X(4700)$.  See Sec.~\ref{sec:Y4140} for more
detail.

\subsection{$X(4160)$}

The $X(4160)$ was reported by the Belle Collaboration in 2008 in the
reaction $e^+e^- \to J/\psi X$ with $X \to D^*\bar{D}^*$, where the
energy of the $e^+e^-$ collisions was near the $\Upsilon(4S)$
resonance~\cite{Abe:2007sya}.  While the signal is significant
($>5\sigma$), it is yet to be confirmed.  The same reaction with $X
\to D^*\bar{D}$ has so far produced only the $X(3940)$.  See
Sec.~\ref{sec:XYZ3940} for more discussion.

\subsection{$Z_c(4200)$}

The $Z_c(4200)$ was reported in 2014 by the Belle Collaboration in the
process $B\to K Z_c^\pm$ with $Z_c^\pm \to \pi^\pm
J/\psi$~\cite{Chilikin:2014bkk}.  Along with numerous $K^*$ states, an
amplitude analysis of the $K\pi^\pm J/\psi$ system showed evidence for
both the $Z_c(4200)$ and the $Z_c(4430)$ [see the separate entry on
the $Z_c(4430)$].  However, neither the $Z_c(4200)$ nor the
$Z_c(4430)$ have yet been confirmed in this process.  [BaBar reported
a negative search for the $Z_c(4430)$ in $\pi^\pm
J\psi$~\cite{Aubert:2008aa}, but this was before Belle first reported
the $Z_c(4200)$].  See Sec.~\ref{sec:otherB} for more details.

\subsection{$Y(4230)$}

The $Y(4230)$ was reported in 2015 by the BESIII Collaboration in the
reaction $e^+e^- \to Y$ with $Y \to \omega
\chi_{c0}$~\cite{Ablikim:2014qwy}.  Since BESIII only had large data
sets at a few center-of-mass energies, the mass and width of the
$Y(4230)$ could not be measured precisely.  The energy dependence of
the cross section for $e^+e^- \to \omega \chi_{c0}$ was, however,
shown to be decisively different from the $Y(4260)$.  There is some
indication that a narrow peak around 4.23~GeV may also exist in the
$e^+e^- \to \pi^+\pi^-h_c$ cross section~\cite{Chang-Zheng:2014haa}.
See Sec.~\ref{sec:eeCharmonium} for more details.

\subsection{$Y(4260)$ and $Y(4008)$}

The $Y(4260)$ was first observed by the BaBar Collaboration in 2005 in
the reaction $e^+e^- \to Y$ with $Y \to
\pi^+\pi^-J/\psi$~\cite{Aubert:2005rm}.  It was the third of the $XY
\! Z$ states to be discovered.  It was later confirmed by
CLEO-c~\cite{He:2006kg} and by Belle~\cite{Yuan:2007sj} in the same
reaction.  CLEO-c~\cite{Coan:2006rv} and BESIII~\cite{Ablikim:2015tbp}
saw the same peak in $e^+e^- \to \pi^0\pi^0 J/\psi$.  The Belle
Collaboration reported a second peak, called the $Y(4008)$, just below
the much larger $Y(4260)$ peak~\cite{Liu:2013dau,Yuan:2007sj}.  The
$Y(4008)$ peak was not confirmed by a BaBar analysis using the same
method with a data set of similar size~\cite{Lees:2012cn}.
Establishing new decay modes of the $Y(4260)$ involves mapping other
$e^+e^-$ cross sections as a function of center-of-mass energy, which
has proven difficult.  See Sec.~\ref{sec:eeCharmonium} for more
discussion.

\subsection{$X(4350)$}

The $X(4350)$ was reported by the Belle Collaboration in 2010 in the
reaction $\gamma \gamma \to X$ with $X \to \phi
J/\psi$~\cite{Shen:2009vs}.  Its significance is only at the level of
$3.2\sigma$ and is in need of confirmation.  See
Sec.~\ref{sec:confirmation} for more detail.

\subsection{$Y(4360)$ and $Y(4660)$}

The $Y(4360)$ was discovered in 2007 by the BaBar Collaboration in the
process $e^+e^- \to Y$ with $Y \to \pi^+\pi^-
\psi(2S)$~\cite{Aubert:2007zz}.  This unexpected discovery was a
byproduct of a search for the $Y(4260)$ decaying to
$\pi^+\pi^-\psi(2S)$, which was not found.  The Belle Collaboration
soon confirmed the BaBar finding using the same process, but also
found a second peak at higher mass, the $Y(4660)$~\cite{Wang:2007ea}.
Both peaks have been confirmed with higher statistics by both the
Belle~\cite{Wang:2014hta} and BaBar Collaborations~\cite{Lees:2012pv}.
See Sec.~\ref{sec:eeCharmonium} for more details.

\subsection{$Z_c(4430)$ and $Z_c(4240)$}

The $Z_c(4430)$ was first reported in 2008 by the Belle Collaboration
in the process $B \to K Z_c^\pm$ with $Z_c^\pm \to \pi^\pm
\psi(2S)$~\cite{Choi:2007wga}.  As the first of the electrically
charged $XY \! Z$ states to be reported, it was the subject of much
scrutiny.  
The initial report by Belle was based on a one-dimensional fit
to the $\pi^\pm \psi(2S)$ invariant mass, but the data was
later reanalyzed using more sophisticated
amplitude analyses~\cite{Mizuk:2009da,Chilikin:2013tch}.
The BaBar
Collaboration, however, using a technique that allowed for $K\pi^\pm$
resonances in a model-independent way, could not confirm the
$Z_c(4430)$ in either its reported decay of $\pi^\pm \psi(2S)$ or in
the possible decay $\pi^\pm J/\psi$~\cite{Aubert:2008aa}.  Belle later
also reported evidence for the $Z_c(4430)$ in $B\to K Z_c^\pm$ with
$Z_c^\pm \to \pi^\pm J/\psi$ [also see the entry for the
$Z_c(4200)$]~\cite{Chilikin:2014bkk}.  In 2014, the LHCb Collaboration
confirmed the existence of the $Z_c(4430)$ in $\pi^\pm \psi(2S)$ using
a larger data set, and also observed a lighter and wider structure
named the $Z_c(4240)$~\cite{Aaij:2014jqa,Aaij:2015zxa}.  See
Sec.~\ref{sec:Z4430} for more details.

\subsection{$X(4630)$}

The $X(4630)$ was seen in 2008 by the Belle Collaboration as a
threshold enhancement in $e^+e^- \to X$ with $X \to \Lambda_c
\bar{\Lambda}_c$~\cite{Pakhlova:2008vn}.  The $X(4630)$ is one of
several peaks seen in exclusive $e^+e^-$ cross sections, such as the
$Y(4260)$ in $e^+e^-\to \pi\pi J/\psi$ and the $Y(4360)$ in $e^+e^-\to
\pi\pi \psi(2S)$.  The fact that it is named $X$ rather than $Y$ is
thus somewhat of an anomaly.  See Sec.~\ref{sec:eeCharmonium} for more
details.

\subsection{$P_c(4380)$ and $P_c(4450)$}

The $P_c(4380)$ and $P_c(4450)$ pentaquark candidates were reported in
2015 by the LHCb Collaboration in an amplitude analysis of the decay
$\Lambda_b \to K^-J/\psi \, p$~\cite{Aaij:2015tga}.  In addition to a
number of excited resonant $\Lambda$ decays to $K^- p$, amplitudes
corresponding to $P_c \to \, J/\psi p$ resonances were found to be
necessary to describe the data.  See Sec.~\ref{sec:Pc} for more
details.

\subsection{$X(5568)$}

The $X(5568)$ was reported in 2016 by the D0 Collaboration in
inclusive hadronic~($p\bar{p}$) production of
$B_s\pi^\pm$~\cite{D0:2016mwd}.  It was not confirmed by the LHCb
Collaboration in $pp$ collisions at the LHC\@.  While the data sets at
LHCb are larger, the production mechanisms and energies are different.
The state needs confirmation.  See Sec.~\ref{sec:confirmation} for
more details.

\subsection{$Z_b(10610)$ and $Z_b(10650)$}

The $Z_b(10610)$ and $Z_b(10650)$ were discovered in 2012 by the Belle
Collaboration in $e^+e^-$ annihilations near the $\Upsilon(5S)$
mass~\cite{Belle:2011aa}.  They were produced in the process $e^+e^-
\to \pi^\mp Z_b^\pm$ and were found to decay through the five channels
$\pi^\pm \Upsilon(1S,2S,3S)$ and $\pi^\pm h_b(1P,2P)$, with consistent
properties in each.  Their $J^P$ were measured to be
$1^+$~\cite{Garmash:2014dhx}. A neutral version of the $Z_b(10610)$
was later found in the process $e^+e^- \to \pi^0 Z_b$ with $Z_b \to
\pi^0 \Upsilon(2S,3S)$~\cite{Krokovny:2013mgx}.  The open-bottom
decays $Z_b(10610)\to B^*\bar{B}$ and $Z_b(10650)\to B^*\bar{B}^*$
were also reported by the Belle Collaboration~\cite{Garmash:2015rfd}.
See Sec.~\ref{sec:eeSubBottomonium} for more discussion.

\subsection{$Y_b(10888)$}

The $Y_b(10888)$ was originally reported in 2010 by the Belle
Collaboration in the process $e^+e^- \to Y_b$, with $Y_b \to
\pi^+\pi^- \Upsilon(1S,2S,3S)$~\cite{Chen:2008xia}.  This initial
report found evidence for a deviation in mass between the peak in the
$e^+e^- \to \pi^+\pi^-\Upsilon(1S,2S,3S)$ cross section and the peak
in the inclusive $e^+e^-$ cross section, the latter thought to be the
$\Upsilon(5S)$.  A later analysis found that the two peaks could be
described with the same parameterization~\cite{Santel:2015qga}.  It is
thus not clear that there is a $Y_b(10888)$ distinct from the
$\Upsilon(5S)$, although the large closed-bottom cross sections are
still a mystery.  The energy dependences of the $e^+e^- \to
\pi^+\pi^-h_b(1P,2P)$ cross sections are consistent with those of the
$e^+e^- \to \pi^+\pi^-\Upsilon(1S,2S,3S)$ cross
sections~\cite{Abdesselam:2015zza}.  See Sec.~\ref{sec:eeBottomonium}
for more details.









\end{appendices}

\clearpage

\section*{References}

\bibliographystyle{elsarticle-num}
\bibliography{REVIEW_REFS}

\end{document}